\documentclass[12pt]{article}
\catcode`\@=11
\@addtoreset{equation}{section}

\def\@seccntformat#1{\csname the#1\endcsname.~~}
\makeatother

\baselineskip=\normalbaselineskip

\usepackage{mathrsfs,amsbsy,amssymb,latexsym,amsfonts,amsmath}
\usepackage{graphicx,color}

\allowdisplaybreaks[4]
\renewcommand{\thefootnote}{\arabic{footnote}}
\def\cM{\mathcal M}
\newcommand{\tr}{\operatorname{tr}}
\def\rmd{{\mathrm d}}
\newcommand{\Lie}{\pounds}
\newcommand{\barg}{\bar{g}}
\newcommand{\barh}{\bar{h}}
\newcommand{\barT}{\bar{T}}
\newcommand{\barK}{\bar{K}}
\newcommand{\barN}{\bar{N}}
\newcommand{\nn}{\nonumber}
\newcommand{\tc}{{\tilde{c}}}
\newcommand{\te}{{\tilde{e}}}
\newcommand{\tn}{{\tilde{n}}}
\newcommand{\tp}{{\tilde{p}}}
\newcommand{\ts}{{\tilde{s}}}
\newcommand{\tsigma}{{\tilde{\sigma}}}
\newcommand{\eq}[1]{(\ref{#1})}
\newcommand{\Exp}[1]{\operatorname{e}^{#1}}
\newcommand{\ii}{{\mathrm i}}
\newcommand{\Stot}{{\hat{S}}}
\def\cG{\mathcal G}
\def\cH{\mathcal H}
\def\cK{\mathcal K}
\newcommand{\bartr}{\overline{\tr}}
\newcommand{\ord}{{\mathcal O}}
\newcommand{\NS}{{\scriptscriptstyle \mathrm{NS}}}
\newcommand{\bx}{{\boldsymbol{x}}}\newcommand{\sfT}{{\mathsf T}}
\newcommand{\sfe}{{\mathsf e}}
\newcommand{\by}{{\mathbf{y}}}

\begin{document}

\begin{titlepage}
\renewcommand{\thefootnote}{\fnsymbol{footnote}}

\begin{flushright}
\parbox{3.5cm}
{KUNS-2329
% \\ {\tt paper11}
}
\end{flushright}

\vspace*{1.0cm}

\begin{center}
{\Large \bf Relativistic viscoelastic fluid mechanics}%
\end{center}
\vspace{1.0cm}

\centerline{
{Masafumi Fukuma}%
\footnote{E-mail address: 
fukuma@gauge.scphys.kyoto-u.ac.jp} ~and~ 
{Yuho Sakatani}%
\footnote{E-mail address: 
yuho@gauge.scphys.kyoto-u.ac.jp}
}

\vspace{0.2cm}

\begin{center}
{\it Department of Physics, Kyoto University \\ 
Kyoto 606-8502, Japan\\}

% \vspace*{1cm}
\end{center}
\vspace*{1cm}
\begin{abstract}

A detailed study is carried out for the relativistic theory of viscoelasticity 
which was recently constructed 
on the basis of Onsager's linear nonequilibrium thermodynamics. 
After rederiving the theory using a local argument with the entropy current, 
we show that this theory universally 
reduces to the standard relativistic Navier-Stokes fluid mechanics 
in the long time limit. 
Since effects of elasticity are taken into account, 
the dynamics at short time scales is modified 
from that given by the Navier-Stokes equations, 
so that acausal problems intrinsic to 
relativistic Navier-Stokes fluids are significantly remedied. 
We in particular show that
the wave equations for the propagation of disturbance 
around a hydrostatic equilibrium in Minkowski spacetime
become symmetric hyperbolic for some range of parameters, 
so that the model is free of acausality problems. 
This observation suggests that 
the relativistic viscoelastic model with such parameters
can be regarded as a causal completion of 
relativistic Navier-Stokes fluid mechanics. 
By adjusting parameters to various values, 
this theory can treat a wide variety of materials 
including elastic materials, Maxwell materials, Kelvin-Voigt materials, 
and (a nonlinearly generalized version of) 
simplified Israel-Stewart fluids, 
and thus we expect the theory to be the most universal description of 
single-component relativistic continuum materials. 
We also show that the presence of strains and the corresponding change in temperature 
are naturally unified through the Tolman law 
in a generally covariant description of continuum mechanics.

\end{abstract}
\thispagestyle{empty}
\end{titlepage}

\tableofcontents
\thispagestyle{empty}

\setcounter{page}{0}
\setcounter{footnote}{0}
\newpage

%%%%%%%%%%%%%%%%%%%%%%%%%%%%%%%%%%%%%%%%%%%%%%%%%%%%%% 
\section{Introduction}
%%%%%%%%%%%%%%%%%%%%%%%%%%%%%%%%%%%%%%%%%%%%%%%%%%%%%% 

The dynamics of fluids at large scales is universally described 
by the Navier-Stokes equations, 
which represent the regression to a global equilibrium 
with transfers of conserved quantities 
(such as energy-momentum and particle number) among fluid particles \cite{LL_fluid}. 
This can be formulated in a generally covariant way, 
but it is known that there arises a problem of acausality. 
In fact, the obtained equations for the propagation of disturbance 
are basically parabolic 
and thus predict infinitely large speed of propagation 
for infinitely high frequency modes, leaving light cones. 
One should note here that this does not imply the breakdown of 
the internal consistency of the description 
because the Navier-Stokes equations are simply an effective description 
at large spacetime scales 
and need not describe high frequency modes correctly. 
However, this is still troublesome 
when adopting the equations in numerical simulations; 
the initial value problems are ill posed, 
and unacceptable numerical solutions can be obtained easily.

To remedy the problem, M\"uller, Israel, and Stewart 
\cite{Mueller,Israel:1976tn,Israel:1979wp} extended the theory 
by treating the dissipative part of stress tensor, $\tau^{\mu\nu}_{\rm (d)}$\,, 
and the heat flux $q^\mu$ (for the Eckart frame) 
or the particle diffusion current $\nu^\mu$ (for the Landau-Lifshitz frame)
as additional thermodynamic variables 
on which the entropy density can depend. 
This prescription is based on the so-called extended thermodynamics
and corresponds to taking into account higher derivative corrections
to the effective theory. 
It has been shown that such modified theories have a good causal behavior 
and that linear perturbations around a hydrostatic equilibrium  
obey hyperbolic differential equations. 
This is now regarded as a fundamental framework 
for the numerical study of relativistic viscous fluids.

Meanwhile, modifications of the Navier-Stokes equations 
have also been studied in the area of \emph{rheology}, 
and the materials treated there are generically called 
\emph{viscoelastic materials} or \emph{viscoelastic fluids}.
Historically, viscoelasticity was defined by Maxwell in the 19th century 
as the characteristic property of such continuum materials 
that behave as elastic solids at short time scales 
and as viscous fluids at long time scales \cite{LL_fluid,LL_elasticity}. 
In 1948, Eckart proposed a theory of elasticity and anelasticity \cite{Eckart:1948}, 
which describes the nonrelativistic dynamics of single-component 
viscoelastic materials and was reinvented recently \cite{afky} 
in the light of the covariance 
under foliation preserving diffeomorphisms. 
In this description, elastic strains 
(or equivalently, the ``intrinsic metric'' defined below)
are introduced as additional thermodynamic variables, 
as in the theory of elasticity. 
As explicitly shown in \cite{afky}, this theory of viscoelasticity
contains the theory of elasticity and the theory of fluids 
as special limiting cases, 
and correctly reproduces the Navier-Stokes equations in the fluid limit.
Furthermore, as was pointed out in \cite{afky2}, 
since the dynamics at short time scales is dominated by elasticity, 
shear modes of linear perturbations around a hydrostatic equilibrium 
obey differential equations with second-order time derivatives
(in contrast to the equations obtained from the Navier-Stokes equations
that contain only a first-order time derivative), 
so that causal behaviors for large frequencies are significantly improved.

Recently, on the basis of Onsager's linear regression theory 
on nonequilibrium thermodynamics 
\cite{Onsager:I,Onsager:II,Casimir,LL_stat}, 
the present authors proposed a relativistic theory of viscoelasticity \cite{FS} 
which generalizes the theory of elasticity and anelasticity 
\cite{Eckart:1948,afky} in a generally covariant form. 
In the present paper, after rederiving the theory 
relying on a local argument with the entropy current, 
we study the detailed properties of relativistic viscoelasticity. 
We show that fluidity is universally realized in the long time limit 
and also that acausal problems disappear for a wide region of parameters. 
Thus, the relativistic theory of viscoelasticity with such parameters
can be regarded as a causal completion 
of relativistic Navier-Stokes fluid mechanics, 
and we expect that it could be used as another basis 
in the numerical study of relativistic viscous fluids.

This paper is organized as follows. 
In Sec.\ \ref{Relativistic_viscoelastic_mechanics} 
we rederive the viscoelastic model of \cite{FS} 
using a local argument with the entropy current. 
We also show that the presence of strains and the corresponding change in temperature 
are naturally unified in a generally covariant description of continuum mechanics.
In Sec.\ \ref{fluid_elastic_limits} 
we consider the long and short time limits of our viscoelastic model.
We prove that the model universally gives relativistic Navier-Stokes fluids 
in the long time limit. 
In Sec.\ \ref{Israel_model} 
we show that when some parameters take specific values, 
our viscoelastic model reduces to (a higher-dimensional extension of) 
the nonlinear generalization of the simplified Israel-Stewart model 
\cite{Denicol:2008ua}. 
In Sec.\ \ref{hyperbolic_dispersion} 
we consider linear perturbations around a hydrostatic equilibrium  
in Minkowski spacetime. 
The dispersion relations show that the evolutions are certainly stable. 
Although the wave equations for the linear perturbations 
are not always hyperbolic,  
if some parameters are chosen appropriately 
(including the parametrizations for the simplified Israel-Stewart model)
they become symmetric hyperbolic and thus free of acausality problems. 
Section \ref{conclusion} is devoted to conclusion and discussions.

%%%%%%%%%%%%%%%%%%%%%%%%%%%%%%%%%%%%%%%%%%%%%%%%%%%%%% 
\section{Relativistic viscoelastic mechanics}
%%%%%%%%%%%%%%%%%%%%%%%%%%%%%%%%%%%%%%%%%%%%%%%%%%%%%% 
\label{Relativistic_viscoelastic_mechanics}

In this section, we rederive the fundamental equations 
for relativistic viscoelastic mechanics 
using a local argument with the entropy current.
In Appendix~\ref{appendix:Review} 
we show that the present formulation is equivalent 
to the ``entropic formulation'' proposed in our previous paper \cite{FS} 
which is based on Onsager's linear regression theory.

%%%%%%%%%%%%%%%%%%%%%%%%%%%%%%%%%%%%%%%%%%%%%%%%%%%%%% 
\subsection{Definitions}
%%%%%%%%%%%%%%%%%%%%%%%%%%%%%%%%%%%%%%%%%%%%%%%%%%%%%% 
\label{Definitions}

We start by giving a brief review 
on the generally covariant definitions 
of viscoelastic materials \cite{FS}. 

%%%%%%%%%%%%%%%%%%%%%%%%%%%%%%%%%%%%%%%%%%%%%%%%%%%%%% 
\paragraph{\underline{{\bf Geometrical setup}}\\}
%%%%%%%%%%%%%%%%%%%%%%%%%%%%%%%%%%%%%%%%%%%%%%%%%%%%%% 

We consider a single-component continuum material 
living in a $(D+1)$-dimensional Lorentzian manifold $\cM$\,. 
The local coordinates are denoted by $x^{\mu}$ $(\mu=0,1,\cdots,D)$\,, 
and the background Lorentzian metric with signature $(-,+,\cdots,+)$ 
by $g_{\mu\nu}(x)$\,.
Following the convention of Landau and Lifshitz \cite{LL_fluid}, 
we define the velocity field $u=u^\mu(x) \partial_\mu$ 
from the momentum $(D+1)$-vector $p_\mu$ as
\begin{align}
 u^\mu(x) \equiv g^{\mu\nu}(x)\,p_\nu(x)/e(x) = p^\mu(x)/e(x)\,,
\end{align}
where $e(x)\equiv \sqrt{-g^{\mu\nu}(x)\,p_\mu(x)\,p_\nu(x)}$ 
is the proper energy density. 
Note that $u^\mu(x)$ is normalized as $g_{\mu\nu}(x)u^\mu(x)u^\nu(x)=-1$\,.
Here and hereafter indices are subscripted (or superscripted) always with $g_{\mu\nu}$ 
(or with its inverse $g^{\mu\nu}$).

Assuming that the velocity field is hypersurface orthogonal, 
we introduce a foliation of $\cM$ 
consisting of spatial hypersurfaces (timeslices) orthogonal to $u^\mu$\,. 
We parametrize the timeslices with a real parameter $t$ 
and denote them by $\Sigma_t$\,. 
We exclusively (except for Sec.\ \ref{hyperbolic_dispersion})
use a coordinate system $x=(x^\mu)=(x^0,\,\bx)$ such that $x^0=t$\,, 
$\bx=(x^i)$ $(i=1\,,\cdots,D)$\,, 
for which the shape of the material at time $t$ is given by 
the induced metric on $\Sigma_t$\,:
\begin{align}
 h_{\mu\nu}(x)\equiv g_{\mu\nu}(x) + u_{\mu}(x)\,u_{\nu}(x) 
  = g_{\mu\nu}(x) + \frac{p_\mu(x)\,p_\nu(x)}{e^2(x)}\,.
\label{hypersurface_metric}
\end{align}
We also define the extrinsic curvature $K_{\mu\nu}$ of the hypersurface 
as half the Lie derivative of $h_{\mu\nu}$ 
with respect to the velocity field $u=u^\mu\partial_\mu$\,: 
\begin{align}
 K_{\mu\nu}\equiv \frac{1}{2}\,\Lie_u h_{\mu\nu}
  = \frac{1}{2}\,h_\mu^{~\rho} h_\nu^{~\sigma}
  \bigl(\nabla_{\rho} u_{\sigma} + \nabla_{\sigma} u_{\rho} \bigr)\,.
\end{align}
This measures the rate of change in the induced metric $h_{\mu\nu}$ 
as material particles flow along $u^\mu$. 
Note that this tensor is symmetric and orthogonal to $u^\mu$, 
$K_{\mu\nu}\,u^\nu=0$\,.

In the ADM parametrization, 
the metric and the velocity are represented
with the lapse $N(x)$ and the shifts $N^i(x)$ $(i=1,\cdots,D)$ as
\begin{align}
 \rmd s^2 &=g_{\mu\nu}(x)\,\rmd x^{\mu}\,\rmd x^{\nu} 
  = -N^2(x)\, \rmd t^2 
  + h_{ij}(x)\,\bigl(\rmd x^i-N^i(x)\,\rmd t\bigr)\,
   \bigl(\rmd x^j-N^j(x)\,\rmd t \bigr)\,,
\label{ADM_decomp}\\
 u &= u^\mu(x)\,\partial_\mu = \frac{1}{N(x)}\,\partial_0 
  + \frac{N^i(x)}{N(x)}\,\partial_i 
 \quad \bigl( \Leftrightarrow~ u_\mu(x)\,\rmd x^\mu = -N(x)\,\rmd t\bigr)\,.
\label{velocity}
\end{align}
The volume element on the hypersurface is given by 
the $D$-form $\sqrt{h}\,\rmd^D\bx 
 \equiv \sqrt{\det(h_{ij})}\,\rmd^D\bx = N^{-1}\sqrt{-g}\, \rmd^D\bx$\,.

With a given foliation, we still have the symmetry of 
foliation preserving diffeomorphisms  
that give rise to transformations only among the points on each timeslice. 
Using this residual gauge symmetry 
we can impose the {\it synchronized gauge}, $N^i(x)\equiv 0$\,, 
so that the background metric and the velocity field are expressed as
\begin{align}
 \rmd s^2 &= g_{\mu\nu}(x)\,\rmd x^\mu\,\rmd x^\nu
  \equiv -N^2(x)\,\rmd t^2 + h_{ij}(x)\,\rmd x^i\,\rmd x^j\,,\\
 u&=u^\mu(x)\,\partial_\mu 
  = \frac{1}{N(x)}\,\frac{\partial}{\partial t} = \frac{\partial}{\partial \tau}\,,
\end{align}
where $\tau$ is the local proper time defined by $\rmd \tau=N\,\rmd t$\,.
In this gauge, 
due to the relation $\partial/\partial t=N(x)\,\partial/\partial\tau$\,,  
the proper energy density $e(x)$ measured with the proper time $\tau$ 
is related to the energy density $\sfe(x)$ measured with time $t$ as 
\begin{align}
 \sfe(x) = N(x)\,e(x)\,.
\end{align}
Note that $\sfe(x)$ includes the gravitational potential through the factor $N(x)$\,.
Accordingly, 
the local temperature $T$ measured with $\tau$ 
% $\bigl(T\equiv (\partial \ts/ \partial \te)^{-1}\bigr)$
is related to the temperature $\sfT$ measured with $t$ 
% $\bigl(\sfT \equiv (\partial \ts/ \partial \tilde{\sfe})^{-1}\bigr)$ 
through the following Tolman law:
\begin{align}
 \sfT(x) = N(x)\,T(x)\,.
\label{Tolman}
\end{align}

%%%%%%%%%%%%%%%%%%%%%%%%%%%%%%%%%%%%%%%%%%%%%%%%%%%%%% 
\paragraph{\underline{{\bf Definition of (relativistic) viscoelastic materials}}\\}
%%%%%%%%%%%%%%%%%%%%%%%%%%%%%%%%%%%%%%%%%%%%%%%%%%%%%% 

\begin{figure}[htbp]
\begin{quote}
\begin{center}
\includegraphics[scale=0.6]{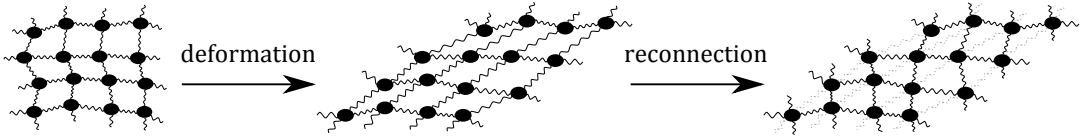}
\caption{Processes of deformation and stress relaxation \cite{afky,afky2}.
\label{fig:reconnection}}
\end{center}
\end{quote}
\vspace{-2ex}
\end{figure}

According to the definition of Maxwell, 
viscoelastic materials behave as elastic solids at short time scales 
and as viscous fluids at long time scales 
(see, e.g., Sec.\ 36 in \cite{LL_elasticity}). 
In order to understand how such materials evolve in time, 
we consider a material consisting of many molecules bonding each other 
and assume that the molecules first stay at their equilibrium positions 
in the absence of strains (as in the leftmost illustration of Fig.\ \ref{fig:reconnection}) 
\cite{afky,afky2}. 
%%%% 
%%%% 
We now suppose that an external force is applied to deform the material.
An internal strain is then produced in the body, and according to the definition, 
the accompanied internal stress can be treated as an elastic force 
at least during short intervals of time.
However, if we keep the deformation much longer than the relaxation times 
(characteristic to each material), 
then the bonding structure changes to maximize the entropy, 
and the internal strain vanishes eventually 
as in the rightmost of Fig.\ \ref{fig:reconnection}.
The point is that two figures (the central and the rightmost) 
have the same shape (same induced metric) $h_{\mu\nu}$\,, 
but different bonding structures.

The internal bonding structure can be specified 
by the \textit{intrinsic metric} $\bar{h}_{\mu\nu}$\,, 
which measures the shape that the material would take 
when all the internal strains are removed virtually \cite{Eckart:1948,afky}.
For the example given in Fig.\ \ref{fig:reconnection}, 
the intrinsic metric for the center illustration is given by the induced metric for the leftmost illustration,
while the intrinsic metric for the rightmost illustration agrees with the induced metric for itself.
Thus, the plastic (i.e., nonelastic) deformation from the center illustration 
to the rightmost illustration is described 
as the evolution of the intrinsic metric.%
\footnote{%%
$\barh_{\mu\nu}$ is also called the ``strain metric'' 
and was first introduced by Eckart to embody 
``the principle of relaxability-in-the-small'' in anelasticity \cite{Eckart:1948}.
Some examples of the explicit form of $h_{\mu\nu}$ and $\barh_{\mu\nu}$ 
under various deformations can be found in \cite{afky,afky2}. 
} %%%%%%%%%%

\begin{figure}[htbp]
\begin{quote}
\begin{center}
\includegraphics[scale=0.35]{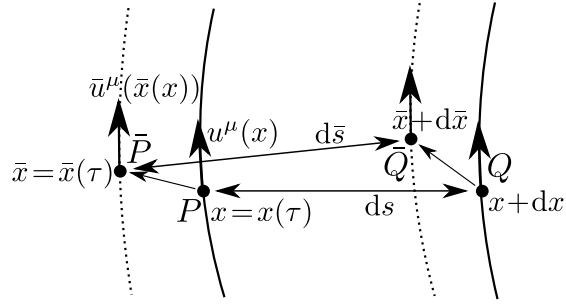}
\caption{Real ($x^\mu(\tau)$) and virtual ($\bar{x}^\mu(\tau)$) trajectories 
of material particles. 
The distance between $\bar{P}$ and $\bar{Q}$ 
gives the definition of the intrinsic metric $\bar{g}_{\mu\nu}$\,.}
\label{fig:u_bar}
\end{center}
\end{quote}
\vspace{-2ex}
\end{figure}

Its generally covariant generalization can be defined in the following way. 
Suppose that we have two adjacent, spatially separated spacetime points 
$P$ and $Q$\,, 
each of which represents a point on the trajectory of a material particle
(see Fig.\ \ref{fig:u_bar}).
%%% 
%%% 
By denoting their coordinates by $x=(x^\mu)$ and $x+\rmd x=(x^\mu+\rmd x^\mu)$\,, 
respectively, the distance between $P$ and $Q$ in the real configuration 
is of course given with the metric $g_{\mu\nu}$ as (the square root of)
\begin{align}
 \rmd s^2 = g_{\mu\nu}(x)\,\rmd x^\mu\,\rmd x^\nu\,.
\end{align}
We now virtually remove all the strains
in a sufficiently small spacetime region including the two points. 
Then $P$ and $Q$ would move to other positions $\bar{P}$ and $\bar{Q}$, 
whose coordinates we denote by $\bar{x}=(\bar{x}^\mu)$ 
and $\bar{x}+\rmd\bar{x}=(\bar{x}^\mu+\rmd\bar{x}^\mu)$\,, respectively. 
This correspondence defines a local map 
$f:\,x\mapsto\bar{x}=\bar{x}(x)$\,, 
with which we define the intrinsic metric $\bar{g}_{\mu\nu}(x)$  
as the metric measuring the virtual distance between $\bar{P}$ and $\bar{Q}$ 
(or, as the pullback of the metric $g_{\mu\nu}$ for the map; 
$\bar{g}_{\mu\nu}\equiv f^\ast g_{\mu\nu}$):%
\footnote{%%%
As in the standard theory of elasticity \cite{LL_elasticity}, 
there may be an arbitrariness in defining $\bar{x}^\mu$, 
but the intrinsic metric $\bar{g}_{\mu\nu}$ can still be defined uniquely. 
} %%%%%%%%%%%
\begin{align}
 \rmd \bar{s}^2 
  &\equiv g_{\rho\sigma}(\bar{x})\,\rmd \bar{x}^\rho\,\rmd \bar{x}^\sigma
  =g_{\rho\sigma}\bigl(\bar{x}(x)\bigr)\,\frac{\partial\bar{x}^\rho}{\partial x^\mu}\,
  \frac{\partial\bar{x}^\sigma}{\partial x^\nu}\,\rmd x^\mu\,\rmd x^\nu\nn\\
 &\equiv \bar{g}_{\mu\nu}(x)\,\rmd x^\mu\,\rmd x^\nu\,.
\end{align}
With the velocity vector $u=u^\mu(x)\,\partial_\mu$\,, 
we parametrize $\barg_{\mu\nu}$ as
\begin{align}
 \barg_{\mu\nu} &= - (1+2\theta)\,u_\mu u_\nu
  - \varepsilon_\mu u_\nu - \varepsilon_\nu\, u_\mu
  + \barh_{\mu\nu} \nn\\
 &\bigl( 
  \varepsilon_\mu u^\mu=0\,,\quad h_{\mu\nu}\, u^\nu=0\,,
  \quad \barh_{\mu\nu}\, u^\nu=0\bigr)\,. 
\end{align}
The \emph{strain tensor} is then introduced as
\begin{align}
 E_{\mu\nu}(x) 
  &\equiv \frac{1}{2}\,\bigl(g_{\mu\nu}(x)-\barg_{\mu\nu}(x)\bigr) \nn\\
 &= \,\theta\, u_\mu u_\nu 
 + \frac{1}{2}\,\bigl(\varepsilon_\mu u_\nu
 + \varepsilon_\nu\, u_\mu \bigr)+ \varepsilon_{\mu\nu}\,,
\label{strain_general}
\end{align}
where 
\begin{align}
 \varepsilon_{\mu\nu}(x)
  \equiv \frac{1}{2}\,\bigl(h_{\mu\nu}(x)-\barh_{\mu\nu}(x) \bigr)
\label{spatial_strain}
\end{align}
is the spatial strain tensor. 
Note that if we define the extrinsic curvature 
associated with the spatial intrinsic metric $\barh_{\mu\nu}$ as 
\begin{align}
 \barK_{\mu\nu} \equiv \frac{1}{2}\,\Lie_u \barh_{\mu\nu}
  =\frac{1}{2}\,\bigl(u^\lambda\, \partial_\lambda \barh_{\mu\nu}
  + \partial_\mu u^\lambda\, \barh_{\lambda\nu}
  + \partial_\nu u^\lambda\, \barh_{\mu\lambda}\bigr)\,, 
\end{align}
the following identity holds:
\begin{align}
 \Lie_u\varepsilon_{\mu\nu} = K_{\mu\nu}-\barK_{\mu\nu}\,.
\end{align}

A viscoelastic material is a thermodynamic system 
consisting of material particles as its subsystems. 
While the system regresses to a thermodynamic equilibrium, 
one can imagine that the virtual trajectory of each material particle 
approaches its real trajectory, 
so that the strain tensor $E_{\mu\nu}$ approaches zero. 
Such an irreversible process is called \emph{plastic} (i.e., nonelastic), 
and thus we see that the dynamics of $E_{\mu\nu}$ 
includes plastic evolutions 
(in addition to reversible, elastic evolutions).
In the following discussions, we assume that 
$E_{\mu\nu}=(\varepsilon_{\mu\nu}\,,\,\varepsilon_\mu\,,\,\theta)$ 
are all small quantities, 
such that their nonlinear effects can be neglected. 
We shall denote the contraction of a spatial tensor%
\footnote{%%%
By spatial we mean that $A_{\mu\nu}$ is orthogonal to $u^\mu$\,,
$A_{\mu\nu} u^\nu=0=A_{\mu\nu}u^\mu$\,.
Recall that $g^{\mu\nu}=-u^\mu u^\nu+h^{\mu\nu}$\,.
} %%%%%%%%%%%
 $A_{\mu\nu}$ 
with $g^{\mu\nu}$ by $\tr A$\,, so that
\begin{align}
 \tr\varepsilon\equiv g^{\mu\nu}\varepsilon_{\mu\nu}=h^{\mu\nu}\varepsilon_{\mu\nu}\,, 
 \quad \tr K\equiv g^{\mu\nu}K_{\mu\nu}=h^{\mu\nu}K_{\mu\nu}\,. 
\end{align}

We close this section by explaining the physical meaning of the strain tensor
$E_{\mu\nu}=(\varepsilon_{\mu\nu}\,,\,\varepsilon_\mu\,,\,\theta)$\,.
The spatial strain tensor $\varepsilon_{\mu\nu}$ 
stands for the standard strains, 
measuring the difference between the induced metric $h_{\mu\nu}$ 
and the spatial induced metric $\barh_{\mu\nu}$\,.
One can easily see that 
the quantity $\varepsilon^{\mu}$ represents 
the relative velocity of a material particle in its real trajectory 
with respect to that in its virtual trajectory, 
$\varepsilon^\mu= u^\mu-\bar{u}^\mu
\equiv \rmd x^\mu/\rmd \tau - \rmd \bar{x}^\mu/\rmd \tau$\,, 
where $\tau$ is a common proper time (see Fig.\ \ref{fig:u_bar}). 
In order to understand the meaning of $\theta$, 
we first recall that the covariant vector $u_\mu$ is expressed as 
$u_\mu\,\rmd x^\mu=-N\,\rmd x^0$\,.
We can then rewrite $\rmd s^2$ and $\rmd \bar{s}^2$ as
\begin{align}
 \rmd s^2 &= -\,N^2(x)\,(\rmd x^0)^2 + h_{\mu\nu}(x)\,\rmd x^\mu\,\rmd x^\nu\,,\\
 \rmd \bar{s}^2 &=
  -\,\bigl(1+2\theta(x)\bigr)\,N^2(x)\,(\rmd x^0)^2 
  + 2 N(x)\,\varepsilon_\mu(x)\,\rmd x^\mu\,\rmd x^0
  + \bar{h}_{\mu\nu}(x)\,\rmd x^\mu\,\rmd x^\nu\,,
\end{align}
with $h_{\mu\nu}\,\rmd x^\mu\,\rmd x^\nu=h_{ij}\,(\rmd x^i-N^i\,\rmd x^0)\,
(\rmd x^j-N^j\,\rmd x^0)$ 
and a similar (but a bit more complicated) expression 
for $2N\,\varepsilon_\mu\,\rmd x^\mu\,\rmd x^0
  + \bar{h}_{\mu\nu}\,\rmd x^\mu\,\rmd x^\nu$\,.
These equations mean that 
$\bar{N}\equiv \sqrt{1+2\theta}\,N \simeq (1+\theta)\,N$
represents the lapse function for the intrinsic metric. 
Then, through the Tolman law, 
we can relate the virtual temperature $\barT$ observed in the absence of strains  
to the actual temperature $T$ as $N\,T=\barN\,\barT\,(=\sfT)$\,.
We thus obtain the relation 
$\theta=(\bar{N}^2/N^2-1)/2=(T^2/\bar{T}^2-1)/2\simeq (T-\barT)/\barT$, 
and conclude that the scalar $\theta$ expresses 
the increase of the temperature due to strains. 
This conclusion shows that the presence of strains and the corresponding change 
in temperature 
are naturally unified in a generally covariant description of continuum mechanics.

%%%%%%%%%%%%%%%%%%%%%%%%%%%%%%%%%%%%%%%%%%%%%%%%%%%%%% 
\subsection{Entropy production rate}
%%%%%%%%%%%%%%%%%%%%%%%%%%%%%%%%%%%%%%%%%%%%%%%%%%%%%% 
\label{Entropy_production}

As was adopted in \cite{FS}, 
in order to develop thermodynamics in a generally covariant manner, 
it is convenient to distinguish density quantities from other intensive quantities, 
and, by multiplying them with the spatial volume element $\sqrt{h}$\,, 
we construct new quantities which are spatial densities on each timeslice. 
For example, the entropy density $s$, the energy-momentum density $p_\mu$\,, 
and the number density $n$ are density quantities, 
and for them we introduce the following spatial densities:
\begin{align}
 \ts\equiv\sqrt{h}\,s\,,\quad \tp_\mu\equiv \sqrt{h}\,p_\mu\,,\quad
  \tn\equiv \sqrt{h}\,n\,.
\end{align}
We assume that each material particle is in its local thermodynamic equilibrium, 
and that the local entropy $\ts$ is a function of 
$\tp_\mu$\,, $\tn$\,, and $g_{\mu\nu}$ as well as
of the strain tensor $E_{\mu\nu}=(\varepsilon_{\mu\nu}\,, \, \varepsilon_\mu\,, \, 
\theta)$\,: 
\begin{align}
 \ts(x) = \ts\bigl(E_{\mu\nu}(x),\tp_\mu(x),\tn(x), g_{\mu\nu}(x)\bigr) \,. 
\label{local_entropy}
\end{align}
We further assume that $\ts$ depends on $\tp_\mu$ 
only through the local proper energy $\te(\tp_\mu,\,g_{\mu\nu})
\equiv\sqrt{-g^{\mu\nu}\,\tp_\mu\,\tp_\nu}$\,, 
so that $\ts$ can also be expressed as
\begin{align}
 \ts(x)&=\tsigma\bigl( E_{\mu\nu}(x)\,,\,\te(x)\,,\,\tn(x)\,,\,g_{\mu\nu}(x)\bigr)\nn\\
 &=\tsigma\bigl( \varepsilon_{\mu\nu}(x)\,,\,\varepsilon_\mu(x)\,,\,
  \theta(x)\,,\,\te(\tp_\mu(x),g_{\mu\nu}(x))\,,\tn(x)\,,\,g_{\mu\nu}(x)\bigr)\,.
\end{align}
Since we are only interested in linear nonequilibrium thermodynamics, 
we only need to expand $\ts$ in $E_{\mu\nu}$ to second order:%
\footnote{%%
For a tensor $A_{\mu\nu}$\,, we define
$A_{\langle\mu\nu\rangle}\equiv (1/2)\,h_\mu^{~\rho}\,h_\nu^{~\sigma}\,
  \bigl[ A_{\rho\sigma}+A_{\sigma\rho}
  -(2/D)\,h^{\alpha\beta}\,A_{\alpha\beta}\,h_{\rho\sigma}\bigr]$\,.
} %%%%%%%%%% 
\begin{align}
 \ts&=\mbox{(terms independent of $E_{\mu\nu}$)} \nn\\
 &~~~~- \frac{1}{2T}\bigl[
     2\lambda_1\varepsilon_{\langle\mu\nu\rangle}\varepsilon^{\langle\mu\nu\rangle} 
   + \lambda_2 \,\varepsilon_{\mu}\varepsilon^{\mu} 
   + \gamma_1 \,(\tr\varepsilon)^2 
   + 2\gamma_2\,(\tr\varepsilon)\, \theta 
   + \gamma_3\,\theta^2 
 \bigr] \,.
\end{align}
We require the stability of the system under the change in strains $E_{\mu\nu}$\,,
so that the constants $\lambda_1$ and $\lambda_2$ are non-negative, 
and the matrix 
${\boldsymbol\gamma}
 =\bigl(\begin{smallmatrix}\gamma_1
 &\gamma_2\cr \gamma_2&\gamma_3\end{smallmatrix}\bigr)$ 
is positive semidefinite.

Then the fundamental thermodynamic relation can be written as
\begin{align}
 \delta \ts =&~ 
  - \frac{u^\nu}{T} \,\delta \tp_\nu
  - \frac{\mu}{T}\,\delta \tn
  + \frac{\sqrt{h}}{2T}\,T^{\mu\nu}_\mathrm{(q)} \,\delta g_{\mu\nu}\nn\\
 &~ - \frac{\sqrt{h}}{T} \, 2\lambda_1\,\varepsilon^{\langle\mu\nu\rangle}
                                 \,\delta \varepsilon_{\langle\mu\nu\rangle}
  - \frac{\sqrt{h}}{T}\,\bigl(\gamma_1\,\tr\varepsilon+\gamma_2\,\theta\bigr)\,
  \delta (\tr\varepsilon) 
  \nn\\
 &~ - \frac{\sqrt{h}}{T} \,\lambda_2\,\varepsilon^\mu\,\delta\varepsilon_\mu
  - \frac{\sqrt{h}}{T}\,\bigl(\gamma_3\,\theta +\gamma_2\,\tr\varepsilon\bigr)\, 
  \delta\theta \,.
\label{delta_s_visc}
\end{align}
Here the temperature $T$\,, the chemical potential $\mu$ 
and the quasiconservative part of the stress tensor, $\tau_\mathrm{(q)}^{\mu\nu}$\,, 
are defined as%
\footnote{%%
We here use a convention that the quasiconservative stress tensor 
$\tau_\mathrm{(q)}^{\mu\nu}$ does not include stresses 
originated from strains.
} %%%%%%%%%%
\begin{align}
 \frac{\partial\tsigma}{\partial \te}=\frac{1}{T}\,,\quad
  \frac{\partial\tsigma}{\partial \tn}=-\frac{\mu}{T}\,,\quad
  \frac{\partial\tsigma}{\partial g_{\mu\nu}}
  =\frac{\sqrt{h}}{2T}\,\tau_\mathrm{(q)}^{\mu\nu}\,,
\end{align}
where we require that $\tau^{\mu\nu}_\mathrm{(q)}$ be orthogonal to $u^\mu$\,, 
$\tau^{\mu\nu}_\mathrm{(q)}\,u_\nu=0$\,.
The quasiconservative part of the energy-momentum tensor is then defined as 
\begin{align}
 T^{\mu\nu}_\mathrm{(q)} &\equiv e\, u^\mu u^\nu + \tau^{\mu\nu}_\mathrm{(q)} \,.
\end{align}
In deriving Eq.\ \eq{delta_s_visc}, 
we have used the relations 
\begin{align}
 \frac{\partial\te(\tp_\mu, g_{\mu\nu})}{\partial \tp_\nu}=-u^\nu\,,\qquad
  \frac{\partial\te(\tp_\mu, g_{\mu\nu})}{\partial g_{\mu\nu}}
  = \frac{\te}{2}\,u^\mu u^\nu\,.
\end{align}

We now set the variation in Eq.\ \eq{delta_s_visc} to be $\delta=\Lie_u$\,.
We then obtain
\begin{align}
 \sqrt{h}\,\nabla_\mu (s\,u^\mu) = \sqrt{h}\,\Big[&
  - \frac{u^\nu}{T} \,\nabla_\mu (p_\nu\, u^\mu)
  - \frac{\mu}{T}\,\nabla_\mu (n\,u^\mu) 
  + \frac{1}{T}\,\tau^{\mu\nu}_\mathrm{(q)}\, K_{\mu\nu}\nn\\
  & - \frac{2\lambda_1}{T}\,\varepsilon^{\langle\mu\nu\rangle}
                                 \,\Lie_u \varepsilon_{\langle\mu\nu\rangle}
  - \frac{1}{T}\bigl(\gamma_1\,\tr\varepsilon+\gamma_2\,\theta\bigr)\,
  \Lie_u(\tr\varepsilon) \nn\\
  & - \frac{\lambda_2}{T}\,\varepsilon^\mu\,\Lie_u\varepsilon_\mu
  - \frac{1}{T}\bigl(\gamma_3\,\theta +\gamma_2\,\tr\varepsilon \bigr)\,
  \Lie_u\theta \Big]\,.
\label{div_entropy}
\end{align}
Here we have used the identities for Lie derivatives:
\begin{align}
 \Lie_u\,\ts
   = \sqrt{h}\,\nabla_\mu\bigl(s\, u^\mu)\,,\quad 
 \Lie_u\,\tp_\nu
   = \sqrt{h}\,\nabla_\mu\bigl(p_\nu\, u^\mu\bigr)\,, \quad
 \Lie_u\,\tn
   = \sqrt{h}\,\nabla_\mu\bigl(n\, u^\mu)\,,
\end{align}
which can be shown by using the identities 
$\Lie_u \sqrt{h} = \sqrt{h}\, \nabla_\mu u^\mu$ 
and $p_{\mu}\nabla_\nu u^{\mu} =0$\,.
Note that 
$\tr\bigl(\Lie_u\varepsilon_{\mu\nu}\bigr)
 =h^{\mu\nu}\,\Lie_u\varepsilon_{\mu\nu}$ 
can be replaced by $\Lie_u(\tr\varepsilon)$ in our approximation 
because the difference 
$\Lie_u(\tr\varepsilon) -\tr\bigl(\Lie_u\varepsilon_{\mu\nu}\bigr)
=(\Lie_u\barh^{\mu\nu})\,\varepsilon_{\mu\nu}=-2 K^{\mu\nu}\,\varepsilon_{\mu\nu}$ 
is of higher orders.

The full energy-momentum tensor $T^{\mu\nu}$ and the full number current $n^\mu$ 
are given by
\begin{align}
 T^{\mu\nu} &\equiv e\, u^\mu u^\nu + \tau^{\mu\nu}\,, \qquad
       n^\mu \equiv n\,u^\mu + \nu^\mu\,,\nn\\
 &\bigl( \tau^{\mu\nu} u_\nu=0= \nu^\mu u_\mu \bigr)
\end{align}
where $\tau^{\mu\nu}$ and $\nu^\mu$ are 
the stress tensor and the diffusion current, respectively. 
Then, by introducing the entropy current 
\begin{align}
 s^\mu \equiv s\, u^\mu - \frac{\mu}{T}\,\nu^\mu \,,
\end{align}
and by using Eq.\ \eq{div_entropy} together with the current conservation laws
\begin{align}
 \nabla_\nu T^{\mu\nu} = 0\,,\qquad \nabla_\mu n^\mu =0\,,
\end{align}
the local entropy production rate can be evaluated as
\begin{align}
 \nabla_\mu s^\mu 
  =&~ -\frac{1}{T}\,\bigl(\tau^{\mu\nu}-\tau_{\rm(q)}^{\mu\nu}\bigr)\, K_{\mu\nu}\,
  + \nu^\mu\, \partial_\mu \Bigl(-\frac{\mu}{T}\Bigr)\nn\\
  &~ - \frac{2\lambda_1}{T}\,\varepsilon^{\langle\mu\nu\rangle}\,
  \Lie_u \varepsilon_{\mu\nu}
  -\frac{1}{T}\,\bigl(\gamma_1\,\tr\varepsilon+\gamma_2\,\theta\bigr)\,
  \Lie_u \tr\varepsilon \nn\\
  &~ - \frac{\lambda_2}{T}\,\varepsilon^\mu\,\Lie_u\varepsilon_\mu
  - \frac{1}{T}\bigl(\gamma_3\,\theta + \gamma_2\,\tr\varepsilon \bigr)\, \Lie_u\theta 
\nn\\
  =&~ 
  \begin{pmatrix}
   \varepsilon^{\langle\mu\nu\rangle} & (-1/T)\,K^{\langle\mu\nu\rangle} 
  \end{pmatrix}
  \begin{pmatrix}
   (-2\lambda_1/T)\,\Lie_u\varepsilon_{\langle\mu\nu\rangle} \cr 
   \tau_{\langle\mu\nu\rangle}-\tau^{\rm(q)}_{\langle\mu\nu\rangle}
  \end{pmatrix}\nn\\
  &~ 
  + \begin{pmatrix}
    \varepsilon^\mu 
   &\nabla^\mu (-\mu/T) 
  \end{pmatrix}
  \begin{pmatrix}
   (-\lambda_2/T)\,\Lie_u\varepsilon_\mu \cr
   \nu_\mu
  \end{pmatrix}\nn\\
 &~+
  \begin{pmatrix}
   \tr \varepsilon & \theta & (-1/T)\,\tr K 
  \end{pmatrix}
  \begin{pmatrix}
   -(1/T)\,{\boldsymbol\gamma}\,
  \begin{pmatrix} \Lie_u(\tr\varepsilon) \cr
   \Lie_u\theta \end{pmatrix} \cr
   (1/D)\,\bigl(\tr\tau-\tr\tau_\mathrm{(q)}\bigr)
  \end{pmatrix} \,.
\label{entropy_prod_rate}
\end{align}
Thus, if we require that each term be separately positive definite, 
we obtain the following equations:
\begin{align}
  \begin{pmatrix}
   -(2\lambda_1/T)\,\Lie_u\varepsilon_{\langle\mu\nu\rangle} \cr 
   \tau_{\langle\mu\nu\rangle}-\tau_{\langle\mu\nu\rangle}^\mathrm{(q)}
  \end{pmatrix}
  &= 2\,({\boldsymbol\cG}+{\boldsymbol\eta})
  \begin{pmatrix}
   \varepsilon_{\langle\mu\nu\rangle} \cr (-1/T)\,K_{\langle\mu\nu\rangle} 
  \end{pmatrix} \,,
\label{2nd_law_1}\\
  \begin{pmatrix}
   -(\lambda_2/T)\,\Lie_u\varepsilon_\mu \cr \nu_\mu
  \end{pmatrix}
  &= ({\boldsymbol\cH}+{\boldsymbol\sigma})
  \begin{pmatrix}
    \varepsilon_\mu \cr
    h_{\mu}^{~\nu} \partial_\nu (-\mu/T) 
  \end{pmatrix} \,,
\label{2nd_law_2}\\
   \begin{pmatrix}
   - (1/T)\,{\boldsymbol\gamma}\,
  \begin{pmatrix} \Lie_u(\tr\varepsilon) \cr \Lie_u\theta \end{pmatrix} \cr
  (1/D)\bigl(\tr\tau-\tr\tau_\mathrm{(q)}\bigr)
  \end{pmatrix} 
  &= ({\boldsymbol\cK}+{\boldsymbol\zeta})
  \begin{pmatrix}
   \tr \varepsilon \cr \theta \cr (-1/T)\,\tr K 
  \end{pmatrix} \,.
\label{2nd_law_3}
\end{align}
Here ${\boldsymbol\cG}$\,, ${\boldsymbol\cH}$\,, and ${\boldsymbol\cK}$ 
are antisymmetric matrices,
\begin{align}
 {\boldsymbol\cG} = 
  \begin{pmatrix} 
   0 & \cG \cr -\cG & 0
  \end{pmatrix} \,,\quad
 {\boldsymbol\cH} =
  \begin{pmatrix} 
   0 & \cH \cr 
   -\cH & 0 \cr
  \end{pmatrix} \,,\quad
 {\boldsymbol\cK} =
  \begin{pmatrix} 
   0 & \cK' & \cK \cr 
   -\cK' & 0  & -\cK a \cr 
   -\cK & \cK a & 0
  \end{pmatrix}  \,,
\label{2nd_law_4}
\end{align}
and ${\boldsymbol\eta}$\,, ${\boldsymbol\sigma}$\,, and ${\boldsymbol\zeta}$ 
are positive semidefinite symmetric matrices,
\begin{align}
 {\boldsymbol\eta} = 
  \begin{pmatrix} 
   \eta_1 & \eta_2 \cr \eta_2 & \eta_3
  \end{pmatrix} \,,\quad
 {\boldsymbol\sigma} =
  \begin{pmatrix} 
   \sigma_1 & \sigma_2 \cr 
   \sigma_2 & \sigma_3 \cr
  \end{pmatrix} \,,\quad
 {\boldsymbol\zeta} =
  \begin{pmatrix} 
   \zeta_1 & \zeta_2  & \zeta_4 \cr 
   \zeta_2 & \zeta_3  & \zeta_5 \cr 
   \zeta_4 & \zeta_5  & \zeta_6
  \end{pmatrix}  \,.
\label{2nd_law_5}
\end{align}
Note that only the symmetric matrices contribute 
when substituted to the entropy production rate \eq{entropy_prod_rate}. 
This means that the matrices 
${\boldsymbol\eta}$\,, ${\boldsymbol\sigma}$\,, and ${\boldsymbol\zeta}$ 
are associated with irreversible processes, 
while the matrices 
${\boldsymbol\cG}$\,, ${\boldsymbol\cH}$\,, and ${\boldsymbol\cK}$ 
are with reversible ones.

The relationship between the equations given above 
and the corresponding ones given in \cite{FS} 
is summarized in Appendix \ref{appendix:Review}.

%%%%%%%%%%%%%%%%%%%%%%%%%%%%%%%%%%%%%%%%%%%%%%%%%%%%%% 
\subsection{Fundamental equations}
%%%%%%%%%%%%%%%%%%%%%%%%%%%%%%%%%%%%%%%%%%%%%%%%%%%%%% 
\label{Fundamental_equations}

Using Eqs.\ \eq{2nd_law_1}--\eq{2nd_law_5} 
at each point $x=(x^0=t,\,\bx)$ on timeslice $\Sigma_t$\,, 
we can express 
(A) the currents $\tau^{\mu\nu}$ and $\nu^\mu$ 
% in the energy-momentum tensor $T^{\mu\nu}$ and the number current $n^\mu$\,, 
% respectively, 
and (B) the evolution of strains, 
$\Lie_u\varepsilon_{\langle\mu\nu\rangle}\,, \ \Lie_u\varepsilon_\mu\,, \ 
\Lie_u\tr\varepsilon$\,, and $\Lie_u\theta$\,, 
only in terms of local thermodynamic quantities on $\Sigma_t$\,.

We thus conclude that the dynamics of relativistic viscoelastic materials
is described by the following two sets of equations \cite{FS,afky}: 

\noindent
\underline{(A) current conservation laws:}
\begin{align}
 \nabla_\mu T^{\mu\nu}&=\nabla_\mu \bigl(e\,u^\mu u^\nu + \tau^{\mu\nu}\bigr)= 0\,,
\label{conservation1}\\
 \nabla_\mu n^\mu&=\nabla_\mu \bigl(n\,u^\mu + \nu^\mu\bigr) =0\,, 
\label{conservation2}
\end{align}
with the constitutive equations
\begin{align}
 \tau^{\mu\nu} 
 =&~ \tau^{\mu\nu}_\mathrm{(q)} -2\,(\cG-\eta_2)\,\varepsilon^{\langle\mu\nu\rangle} 
      - \frac{2\eta_3}{T}\,K^{\langle\mu\nu\rangle} \nn\\
  &~ -\Bigl[(\cK-\zeta_4)\, \tr \varepsilon -(\cK a+\zeta_5)\,\theta
   + \frac{\zeta_6}{T}\,\tr K\Bigr]\, h^{\mu\nu} \,,
\label{stress_final}\\
 \nu^\mu 
 =&~ -(\cH-\sigma_2)\,\varepsilon^\mu
     + \sigma_3\,h^{\mu\nu}\partial_\nu \Bigl(-\frac{\mu}{T}\Bigr) \,.
\label{diffusion_final}
\end{align}

\noindent
\underline{(B) rheology equations:}
\begin{align}
 \Lie_u\varepsilon_{\langle\mu\nu\rangle}
  &= -\frac{\eta_1\,T}{\lambda_1}\, \varepsilon_{\langle\mu\nu\rangle}
     +\frac{\cG + \eta_2}{\lambda_1} \, K_{\langle\mu\nu\rangle} \,,
\label{rheology1}\\
 \Lie_u\varepsilon_\mu
  &= -\frac{\sigma_1\,T}{\lambda_2} \varepsilon_\mu
     -\frac{(\cH+\sigma_2)\,T}{\lambda_2}
  h_\mu^{~\nu} \partial_\nu \Bigl(-\frac{\mu}{T}\Bigr) \,,
\label{rheology2}\\
  \begin{pmatrix}
  \Lie_u(\tr\varepsilon) \cr
  \Lie_u\theta
  \end{pmatrix}
  &= 
  \begin{pmatrix}
   \gamma_1 & \gamma_2 \cr \gamma_2 & \gamma_3
  \end{pmatrix}^{-1}
  \begin{pmatrix}
   -\zeta_1\,T\,\tr \varepsilon - (\cK'+\zeta_2)\,T\,\theta + (\cK  +\zeta_4)\,\tr K \cr
   (\cK'-\zeta_2)\,T\,\tr \varepsilon - \zeta_3\,T\,\theta  - (\cK a-\zeta_5)\,\tr K 
  \end{pmatrix}\nn\\
  &= 
  \begin{pmatrix}
  -\frac{(\gamma_3\,\zeta_1+\gamma_2\,(\cK'-\zeta_2))\,T}
        {\det{\boldsymbol\gamma}}\,\tr \varepsilon
  +\frac{(\gamma_2\,\zeta_3- \gamma_3\,(\cK'+\zeta_2))\,T}
        {\det{\boldsymbol\gamma}}\,\theta 
  +\frac{\gamma_3\,(\cK  +\zeta_4)+ \gamma_2\,(\cK a-\zeta_5)}
        {\det{\boldsymbol\gamma}}\,\tr K \cr
   \frac{(\gamma_2\,\zeta_1+\gamma_1(\cK'-\zeta_2))\,T}
        {\det{\boldsymbol\gamma}}\,\tr \varepsilon
  -\frac{(\gamma_1\zeta_3-\gamma_2\,(\cK'+\zeta_2))\,T}
        {\det{\boldsymbol\gamma}}\,\theta
  -\frac{\gamma_2\,(\cK  +\zeta_4)+\gamma_1\,(\cK a-\zeta_5)}
        {\det{\boldsymbol\gamma}}\,\tr K
  \end{pmatrix}
\label{rheology3}\,.
\end{align}
The former set of equations describes the dynamics 
of $D+2$ conserved quantities $(p_\mu=e\,u_\mu,\,n)$\,, 
while the latter that of $D(D+1)/2$ dynamical variables 
$E_{\mu\nu}=(\varepsilon_{\mu\nu},\varepsilon_\mu,\theta)$\,.

It is  convenient to introduce the following parameters:
\begin{align}
 \tau_\mathrm{s}&\equiv \frac{\lambda_1}{\eta_1\,T}\,,\qquad 
  \tau_{\sigma}\equiv \frac{\lambda_2}{\sigma_1\,T}\,,
\\
 \tau_{\pm}&\equiv 
  \frac{2\det{\boldsymbol\gamma}}{T\,\bigl(P_{\zeta\gamma}\mp\sqrt{P_{\zeta\gamma}^2
  -4\det{\boldsymbol\gamma}\,(\det{\boldsymbol\zeta}_\mathrm{s}+\cK^{\prime 2})}\bigr)}\,,
\\
 a_{\pm}&\equiv
  \frac{-2\bigl(\zeta_3\,\gamma_2- (\cK'+\zeta_2)\, \gamma_3\bigr)}
  {\zeta_3\,\gamma_1-\zeta_1\,\gamma_3- 2 \cK'\,\gamma_2\pm 
  \sqrt{P_{\zeta\gamma}^2-4\det{\boldsymbol\gamma}\, (\det{\boldsymbol\zeta}_\mathrm{s}
  +\cK^{\prime 2})}} \,,
\end{align}
where 
$P_{\zeta\gamma} \equiv \zeta_3 \gamma_1+\zeta_1 \gamma_3-2 \zeta_2 \gamma_2\geq 0$\,, 
and 
${\boldsymbol\zeta}_\mathrm{s}$ is the principal submatrix of ${\boldsymbol\zeta}$ 
defined by
${\boldsymbol\zeta}_\mathrm{s}\equiv 
\bigl(\begin{smallmatrix} \zeta_1 & \zeta_2 \cr \zeta_2 & \zeta_3
\end{smallmatrix}\bigr)$\,.
Since ${\boldsymbol\zeta}_\mathrm{s}$ is positive semidefinite, 
$\det{\boldsymbol\zeta}_\mathrm{s}$ is non-negative. 
Note that $\tau_\mathrm{s}$\,, $\tau_{\sigma}$\,, and $\mathrm{Re}\,\tau_\pm$ 
are all non-negative. 
We further introduce the scalar variables 
\begin{align}
 \varepsilon_{\pm} \equiv \frac{1}{2}\,(\tr\varepsilon - a_{\pm} \, \theta) \,.
\end{align}
Then the rheology equations \eq{rheology1}--\eq{rheology3} can be rewritten 
in a more compact form:

\noindent
\underline{(B${}^{\,\prime}$) rheology equations:}
\begin{align}
 \Lie_u\varepsilon_{\langle\mu\nu\rangle} 
  =&~ -\frac{1}{\tau_\mathrm{s}}\, \varepsilon_{\langle\mu\nu\rangle}
     +\frac{\cG + \eta_2}{\lambda_1} \, K_{\langle\mu\nu\rangle} \,, 
\label{rheology1a}
\\
 \Lie_u\varepsilon_\mu
  =&~ -\frac{1}{\tau_\sigma}\, \varepsilon_\mu
     -\frac{(\cH+\sigma_2)\,T}
           {\lambda_2}\, h_\mu^{~\nu}\, \partial_\nu \Bigl(-\frac{\mu}{T}\Bigr) \,,
\label{rhelogy2a}\\
 \Lie_u\varepsilon_{\pm} 
  =&~ -\frac{1}{\tau_{\pm}}\,\varepsilon_{\pm}
      +\frac{(\cK a -\zeta_5)\,(a_{\pm}\,\gamma_1+\gamma_2)
  +(\cK+\zeta_4)\,(a_{\pm}\,\gamma_2+\gamma_3)}{2\det{\boldsymbol\gamma}} \,\tr K 
\label{rheology3a}\,.
\end{align}
From these, we see that $\tau_\mathrm{s}$\,, $\tau_\sigma$\,, and $\mathrm{Re}\,\tau_\pm$ 
give the typical time scales for the relaxation of strains.

The relation between the viscoelastic models and a few well-known rheological models 
(such as the Maxwell model and the Kelvin-Voigt model) 
is discussed in Appendix \ref{constitutive_eq}.

%%%%%%%%%%%%%%%%%%%%%%%%%%%%%%%%%%%%%%%%%%%%%%%%%%%%%% 
\section{Fluid and elastic limits}
%%%%%%%%%%%%%%%%%%%%%%%%%%%%%%%%%%%%%%%%%%%%%%%%%%%%%% 
\label{fluid_elastic_limits}

In this section, 
we discuss the limits of elasticity and fluidity 
in the relativistic theory of viscoelasticity.
We first identify the properties that characterize a given material 
as a fluid or as an elastic material. 
We then consider the long-time and short-time limits of our dynamical equations 
and show that fluidity is universally realized in the long time limit. 
We also make a comment on the subtlety existing 
in Maxwell's definition of viscoelasticity.

%%%%%%%%%%%%%%%%%%%%%%%%%%%%%%%%%%%%%%%%%%%%%%%%%%%%%% 
\subsection{Fluidity and elasticity}
%%%%%%%%%%%%%%%%%%%%%%%%%%%%%%%%%%%%%%%%%%%%%%%%%%%%%% 

Fluidity is characterized by the property that the relaxation of 
the strains $E_{\mu\nu}=(\varepsilon_{\mu\nu},\,\varepsilon_\mu,\,\theta)$
proceeds instantaneously. 
Thus, their rheology equations are expressed as
\begin{align}
 \Lie_u\varepsilon_{\mu\nu} = 0\,,\quad
  \Lie_u\varepsilon_{\mu} = 0\,,\quad
  \Lie_u\theta = 0\,, \quad\mbox{(fluids)}
\end{align}
or equivalently, 
\begin{align}
 \Lie_u\varepsilon_{\langle\mu\nu\rangle} = 0\,,\quad
  \Lie_u\varepsilon_{\mu} = 0\,,\quad
  \Lie_u\varepsilon_\pm = 0\,. \quad\mbox{(fluids)}
\end{align}
This situation can also be realized in the long time limit, 
and we show in the next section 
that the constitutive equations for our viscoelastic model 
universally reduces to those for the Navier-Stokes fluids 
in the long time limit.

On the other hand, 
elastic materials by definition do not undergo any plastic deformations, 
and thus their intrinsic metric $\barh_{\mu\nu}$ does not evolve for any processes. 
Thus, a given viscoelastic material is regarded as being elastic 
when its rheology equations are expressed as \cite{Eckart:1948,Carter:1977qf,afky}
\begin{align}
 \barK_{\mu\nu}=\frac{1}{2}\,\Lie_u \barh_{\mu\nu} = 0\,.
  \quad\mbox{(elastics)}
\end{align}

%%%%%%%%%%%%%%%%%%%%%%%%%%%%%%%%%%%%%%%%%%%%%%%%%%%%%% 
\subsection{Long time limit as a fluid limit}
%%%%%%%%%%%%%%%%%%%%%%%%%%%%%%%%%%%%%%%%%%%%%%%%%%%%%% 

Let the time scale of observation be $T_\mathrm{obs}$\,.
If the observation is made much longer than the relaxation times 
(i.e., $T_\mathrm{obs}\gg \tau_\mathrm{s}\,,\, \tau_\sigma\,,\,\mathrm{Re}\,\tau_{\pm}$)\,, 
then we can neglect the terms 
$\Lie_u \varepsilon_{\langle\mu\nu\rangle}$\,, $\Lie_u \varepsilon_\mu$\,, 
and $\Lie_u \varepsilon_\pm$ in Eqs.\ \eq{rheology1a}--\eq{rheology3a} 
because, for example, 
$\Lie_u \varepsilon_{\langle\mu\nu\rangle}\sim
T_\mathrm{obs}^{-1}\,\varepsilon_{\langle\mu\nu\rangle}
\ll \tau_\mathrm{s}^{-1}\,\varepsilon_{\langle\mu\nu\rangle}$\,.
We thus obtain
\begin{align}
 \varepsilon_{\langle\mu\nu\rangle}
  &\simeq 
  \tau_\mathrm{s}\,\frac{\cG + \eta_2}{\lambda_1} \, K_{\langle\mu\nu\rangle} 
  =\frac{\cG+\eta_2}{\eta_1\,T}\,K_{\langle\mu\nu\rangle} \,, \\
 \varepsilon_\mu 
  &\simeq 
     \tau_\sigma\,\frac{(\cH + \sigma_2)T}{\lambda_2} \, h_\mu^{~\nu}\,
     \partial_\nu \Bigl(-\frac{\mu}{T}\Bigr) 
  =\frac{\cH + \sigma_2}{\sigma_1} \, h_\mu^{~\nu}\,
     \partial_\nu \Bigl(-\frac{\mu}{T}\Bigr) \,, \\
 \varepsilon_{\pm} &\simeq 
   \tau_{\pm}\,\frac{(\cK a -\zeta_5)\,(a_{\pm}\,\gamma_1+\gamma_2)
  +(\cK+\zeta_4)\,(a_{\pm}\,\gamma_2+\gamma_3)}
        {2\det{\boldsymbol\gamma}} \,\tr K \nn\\
  &\hspace{-3ex}\left[\,
   \mbox{or}\quad
   \begin{pmatrix}\tr\varepsilon \cr \theta\end{pmatrix}
   \simeq \frac{\tr K}{(\det {\boldsymbol\zeta}_\mathrm{s}+\cK^{\prime 2})\,T}\,
   \begin{pmatrix}
    \zeta_3(\zeta_4+\cK)-(\zeta_2+\cK')(\zeta_5-\cK a) \cr
    -(\zeta_2-\cK')(\zeta_4+\cK) + \zeta_1(\zeta_5-\cK a)
   \end{pmatrix}\,\,
  \right]
  \,.
\end{align}
By substituting these equations to Eqs.\ \eq{stress_final} and \eq{diffusion_final}, 
the constitutive equations take the following form:
\begin{align}
 \tau^{\mathrm{(long)}}_{\mu\nu} &=
  \tau_{\mu\nu}^\mathrm{(q)} - 2\eta_{\NS} \, K_{\langle\mu\nu\rangle}
                          - \zeta_{\NS}\, (\tr K) \, h_{\mu\nu} \,,\\
 \nu^{\mathrm{(long)}}_\mu&=
  \sigma_{\NS}\,h_\mu^{~\nu}\,\partial_\nu \Bigl(-\frac{\mu}{T}\Bigr)\,,
\end{align}
where we have defined viscosity and diffusion coefficients by
\begin{align}
 \eta_{\NS} &\equiv \frac{\det{\boldsymbol\eta}+\cG^2}{\eta_1\,T} \,,
\label{eta_NS}\\
 \zeta_{\NS}&\equiv 
 \frac{\det{\boldsymbol\zeta} + \cK^2\,(a^2\,\zeta_1+2a\,\zeta_2+\zeta_3)
  -2\cK\, \cK'(a\zeta_4+\zeta_5)
  +\cK^{\prime 2}\,\zeta_6}{(\det{\boldsymbol\zeta}_\mathrm{s}+\cK^{\prime 2})\,T} \,,
\label{zeta_NS}\\
 \sigma_{\NS}&\equiv \frac{\det{\boldsymbol\sigma}+\cH^2}{\sigma_1}\,.
\label{sigma_NS}
\end{align}
Note that they are always non-negative, as can be seen from the inequality
\begin{align}
 &\cK^2\,(a^2\,\zeta_1+2a\,\zeta_2+\zeta_3)
  -2\cK\, \cK'(a\zeta_4+\zeta_5) +\cK^{\prime 2}\,\zeta_6
  = \begin{pmatrix} \cK a & \cK & -\cK'\end{pmatrix}
 {\boldsymbol\zeta} \begin{pmatrix} \cK a \cr \cK \cr -\cK'\end{pmatrix} \geq 0 \,.
\end{align}
In particular, when the material is locally isotropic,
we can take $\tau^{\mu\nu}_\mathrm{(q)} = P\, h^{\mu\nu}$\,, 
with $P$ the pressure,
and thus the stress tensor certainly gives the constitutive equations 
for a relativistic Navier-Stokes fluid:
\begin{align}
 \tau_{\mathrm{(long)}}^{\mu\nu} = -2\,\eta_{\NS} \,K^{\langle\mu\nu\rangle} 
  + (P- \zeta_{\NS}\,\tr K)\,h^{\mu\nu} \,.
\end{align}
We thus confirm that our viscoelastic model 
always exhibits fluidity in the long time limit.

%%%%%%%%%%%%%%%%%%%%%%%%%%%%%%%%%%%%%%%%%%%%%%%%%%%%%% 
\subsection{Short time limit as an elastic limit}
%%%%%%%%%%%%%%%%%%%%%%%%%%%%%%%%%%%%%%%%%%%%%%%%%%%%%% 
\label{short_elastic}

In contrast, at short time scales 
($T_\mathrm{obs}\ll \tau_\mathrm{s}\,,\,\mathrm{Re}\,\tau_{\pm}$)\,,
we have 
\begin{align}
  \Lie_u\varepsilon_{\langle\mu\nu\rangle} \gg 
  -\frac{1}{\tau_\mathrm{s}}\, \varepsilon_{\langle\mu\nu\rangle}\,,\qquad 
 \Lie_u\varepsilon_\pm  \gg -\frac{1}{\tau_{\pm}}\,\varepsilon_\pm \,, 
\end{align}
so that Eqs.\ \eq{rheology1a}--\eq{rheology3a} 
can be approximated as
\begin{align}
 \Lie_u\varepsilon_{\langle\mu\nu\rangle}
  &\simeq \frac{\cG + \eta_2}{\lambda_1} \, K_{\langle\mu\nu\rangle} \,, 
\label{short_time_limit1}
\\
 \Lie_u\varepsilon_{\pm} 
  &\simeq \frac{(\cK a -\zeta_5)\,(a_{\pm}\,\gamma_1+\gamma_2)
 +(\cK+\zeta_4)\,(a_{\pm}\,\gamma_2+\gamma_3)}{2\det{\boldsymbol\gamma}} \,\tr K \,,\\
 \Bigl(\Rightarrow \Lie_u(\tr\varepsilon) 
  &\simeq
  \frac{(\cK a-\zeta_5)\,\gamma_2+(\cK+\zeta_4)\,\gamma_3}
  {\det{\boldsymbol\gamma}} \,\tr K \Bigr)
\label{short_time_limit2}\,.
\end{align}
By substituting Eqs.\ \eq{short_time_limit1}--\eq{short_time_limit2} 
into Eq.\ \eq{stress_final}, 
the stress tensor can be rewritten in the following form:
\begin{align}
 \tau^{\mathrm{(short)}}_{\mu\nu} 
 =&~ \tau^\mathrm{(q)}_{\mu\nu}
      -2\,(\cG-\eta_2) \,\varepsilon_{\langle\mu\nu\rangle} 
      - \frac{2\cG\,\eta_3}
             {(\cG+\eta_2)\,T}\,\Lie_u\varepsilon_{\langle\mu\nu\rangle} \nn\\
  &~ -\Bigl[(\cK-\zeta_4)\, \tr \varepsilon 
             -(\cK a+\zeta_5)\,\theta \nn\\
  &~ \qquad  + \frac{\zeta_6\,\det{\boldsymbol\gamma}}
                    {\bigl[(\cK a-\zeta_5)\,\gamma_2+(\cK+\zeta_4)\,\gamma_3\bigr]\,T}\,
    \Lie_u(\tr\varepsilon) \Bigr]\, h_{\mu\nu} \,.
\label{stress_short_limit}
\end{align}
These constitutive equations have the same form as those of a Kelvin-Voigt material 
(see Appendix \ref{constitutive_eq}). 
However, one cannot yet identify the material at short time scales 
with a Kelvin-Voigt material, 
because they generically obey a different type of rheology equations.

As discussed in the first subsection, 
elasticity is characterized by the condition that the intrinsic metric $\barh_{\mu\nu}$ 
does not evolve, 
and the rheology equations for elastic materials are given by 
$\barK_{\mu\nu}=0$\,, or equivalently by $\Lie_u\varepsilon_{\mu\nu}=K_{\mu\nu}$ 
\cite{Eckart:1948,Carter:1977qf,afky}.
However, this is realized only when the conditions 
$\cG+\eta_2=\lambda_1$ and 
$(\cK a-\zeta_5)\,\gamma_2+(\cK+\zeta_4)\,\gamma_3
 =\det{\boldsymbol\gamma}$ are satisfied. 
That is, for generic values of parameters, 
even if the observation time is sufficiently shorter than the relaxation times,
the intrinsic metric $\barh_{\mu\nu}$ evolves 
when the induced metric $h_{\mu\nu}$ does 
(i.e., $\barK_{\mu\nu}\neq 0$ if $K_{\mu\nu}\neq 0$). 
Thus, Maxwell's original definition of viscoelasticity 
(considered only for the situations 
where the induced metric is static, $K_{\mu\nu}=(1/2)\,\Lie_u h_{\mu\nu}=0$)
needs to be modified for generic values of parameters, 
such that $\barh_{\mu\nu}$ is allowed to evolve when $h_{\mu\nu}$ does.

%%%%%%%%%%%%%%%%%%%%%%%%%%%%%%%%%%%%%%%%%%%%%%%%%%%%%% 
\section{Simplified Israel-Stewart fluids}
%%%%%%%%%%%%%%%%%%%%%%%%%%%%%%%%%%%%%%%%%%%%%%%%%%%%%% 
\label{Israel_model}

In this section, as an interesting example,
we consider the case where $\cK'=\eta_3=\sigma_3=\zeta_6=0$ and 
$\tau_\mathrm{(q)}^{\mu\nu}=P\,h^{\mu\nu}$\,. 
In this case, from the positivity of matrices ${\boldsymbol\eta}$\,, 
${\boldsymbol\sigma}$\,, and ${\boldsymbol\zeta}$\,, 
the conditions $\eta_2=\sigma_2=\zeta_4=\zeta_5=0$  
also must be imposed. 
Then the conserved currents take the following form:%
\footnote{%%
From this form of the bulk stress 
and the relation $\theta\simeq (T-\bar{T})/\bar{T}$\,,
we see that $a/\bar{T}$ can be identified with 
the thermal expansion coefficient. 
} %%%%%%%%%%
\begin{align}
 T^{\mu\nu} &= e\,u^\mu u^\nu + P\, h^{\mu\nu} 
     -2\cG\,\varepsilon^{\langle\mu\nu\rangle} 
  - \cK\,\bigl(\tr\varepsilon - a \,\theta\bigr)\, h^{\mu\nu} \,, \\
 n^\mu &= n\,u^\mu - \cH \,\varepsilon^\mu \,,
\end{align}
and the rheology equations become
\begin{align}
 \Lie_u\varepsilon_{\langle\mu\nu\rangle}
  &= -\frac{1}{\tau_\mathrm{s}}\, \varepsilon_{\langle\mu\nu\rangle}
     +\frac{\cG}{\lambda_1} \, K_{\langle\mu\nu\rangle} \,, \\
 \Lie_u\varepsilon_\mu
  &= -\frac{1}{\tau_\sigma}\, \varepsilon_\mu
     -\frac{\cH \,T}{\lambda_2}\,
  h_\mu^{~\nu}\, \partial_\nu \Bigl(-\frac{\mu}{T}\Bigr) \,,\\
  \Lie_u(\tr\varepsilon) 
  &=   -\frac{(\gamma_3\,\zeta_1-\gamma_2\,\zeta_2)\,T}{\det{\boldsymbol\gamma}}\,
  \tr \varepsilon
  + \frac{(\gamma_2\,\zeta_3- \gamma_3\,\zeta_2)\,T}{\det{\boldsymbol\gamma}}\,
  \theta 
  +\frac{\cK\,(a\,\gamma_2+\gamma_3)}{\det{\boldsymbol\gamma}}\,\tr K \,, \\
  \Lie_u\theta
  &= \frac{(\gamma_2\,\zeta_1-\gamma_1\,\zeta_2)\,T}{\det{\boldsymbol\gamma}}\,
  \tr \varepsilon
  -\frac{(\gamma_1\,\zeta_3-\gamma_2\,\zeta_2)\,T}{\det{\boldsymbol\gamma}}\,\theta
  -\frac{\cK\,(a\,\gamma_1+\gamma_2)}{\det{\boldsymbol\gamma}}\,\tr K\,.
\end{align}
By using the relations
\begin{align}
 \tau_{\langle\mu\nu\rangle}
   = -2 \cG \,\varepsilon_{\langle\mu\nu\rangle}\,,\quad 
 \nu^\mu = - \cH \,\varepsilon^\mu \,, \quad 
 \Pi\equiv \frac{1}{D}\,\bigl(\tr\tau-\tr\tau_\mathrm{(q)}\bigr)
          = -\cK\,\bigl(\tr\varepsilon - a \,\theta\bigr) \,,
\end{align}
the rheology equations can be rewritten as
\begin{align}
 \Lie_u\tau_{\langle\mu\nu\rangle}
  =&~ -\frac{1}{\tau_\mathrm{s}}\, \tau_{\langle\mu\nu\rangle}
     -\frac{2\cG^2}{\lambda_1} \, K_{\langle\mu\nu\rangle} \,, \\
 %%%%
 \Lie_u\nu_\mu
  =&~ -\frac{1}{\tau_\sigma}\, \nu_\mu
     +\frac{\cH^2 \,T}{\lambda_2}\, 
  h_\mu^{~\nu}\, \partial_\nu \Bigl(-\frac{\mu}{T}\Bigr) \,,\\
 %%%%
  \Lie_u\Pi 
  =&~ - \frac{\bigl[(a\,\gamma_2+\gamma_3)\,\zeta_1
  -(a\,\gamma_1+\gamma_2)\,\zeta_2\bigr]\,T}{\det{\boldsymbol\gamma}} \, \Pi 
  -\frac{\cK^2\,(a^2 \gamma_1+2a\gamma_2+\gamma_3)}{\det{\boldsymbol\gamma}}\,\tr K 
\nn\\
  &~ 
  + \frac{\cK\,T\,\bigl[a\,\zeta_1\,(a\,\gamma_2+\gamma_3)
  -\zeta_2\,(a^2\,\gamma_1-\gamma_3) 
  -\zeta_3\,(a\,\gamma_1+\gamma_2) \bigr]}{\det{\boldsymbol\gamma}}
  \,\theta \,, \\
 %%%%
  \Lie_u\theta
  =&~
   \frac{(\gamma_1\,\zeta_2-\gamma_2\,\zeta_1)\,T}{\cK\,\det{\boldsymbol\gamma}}\,\Pi 
  -\frac{\cK\,(a\,\gamma_1+\gamma_2)}{\det{\boldsymbol\gamma}}\,\tr K \nn\\
   &~ -\frac{\bigl[\gamma_1\,(a\,\zeta_2+\zeta_3)
   -\gamma_2\,(a\,\zeta_1+\zeta_2)\bigr]\,T}{\det{\boldsymbol\gamma}} \,\theta \,.
\end{align}
This model gives hyperbolic differential equations for small perturbations 
around a hydrostatic equilibrium, 
as is shown in Sec.\ \ref{hyperbolic_dispersion}.

For brevity, 
we here consider the case when $\theta$ is decoupled from other variables.
This can be realized by setting $a=\gamma_2=\zeta_2=0$ in the above equations, 
and the rheology equations become 
\begin{align}
 \tau_\mathrm{s}\,\Lie_u\tau_{\langle\mu\nu\rangle}
  =&~ -\tau_{\langle\mu\nu\rangle}
  - \eta_{\NS}\, K_{\langle\mu\nu\rangle} \,, \\
 \tau_{\sigma}\,\Lie_u\nu_\mu
  =&~ - \nu_\mu
     +\sigma_{\NS}\, 
  h_\mu^{~\nu}\, \partial_\nu \Bigl(-\frac{\mu}{T}\Bigr) \,,\\
 \tau_\mathrm{b}\,\Lie_u\Pi =&~ - \Pi -\zeta_{\NS}\,\tr K \,, 
\label{nonlinear}\\
  \Lie_u\theta =&~ -\frac{\zeta_3\,T}{\gamma_3}\,\theta \,.
\end{align}
Here we have introduced $\tau_\mathrm{b}\equiv \gamma_1/(\zeta_1\,T)$\,,
and the viscosity and diffusion coefficients are given in this case by 
$\eta_{\NS} = \tau_\mathrm{s}\,\cG^2/\lambda_1 = \cG^2/(\eta_1 T)$\,,
$\zeta_{\NS} = \tau_\mathrm{b}\,\cK^2/\gamma_1 = \cK^2/(\zeta_1 T)$\,, 
and $\sigma_{\NS}=\cH^2/\sigma_1$\,.
These equations look like the \emph{nonlinear causal dissipative hydrodynamics} 
proposed in \cite{Denicol:2008ua}.
Although the nonlinear terms in \cite{Denicol:2008ua} 
(e.g., $h_\mu^{~\rho}\,\nu_\nu\,\nabla_\rho u^\nu$) 
are important for numerical simulations of ultra-relativistic dynamics, 
these terms, in principle, cannot be treated properly 
in our first-order formalism. 
However, 
if we do not make the approximation 
$\Lie_u(\tr\varepsilon)\simeq \tr(\Lie_u\varepsilon_{\mu\nu})$\,, 
then Eq.\ \eq{nonlinear} becomes 
$-\tau_\mathrm{b}\,\cK\,\tr(\Lie_u\varepsilon_{\mu\nu}) 
 = \tau_\mathrm{b}\,\bigl(\Lie_u\Pi + (1/D)\,\tr K\,\Pi 
 -\cK\,K^{\langle\mu\nu\rangle}\,\varepsilon_{\langle\mu\nu\rangle}\bigr)
 = - \Pi -\zeta_{\NS}\,\tr K$ 
and coincides with Eq.\ (14) in \cite{Denicol:2008ua} 
where the spatial dimension is set to be $D=1$\,.

If we neglect the nonlinear terms, we then get relations of Maxwell-Cattaneo type:
\begin{align}
\begin{aligned}
 \pi^{\mu\nu}
 &= - 2\,\eta_{\NS}\,K^{\langle\mu\nu\rangle}
  - \tau_\mathrm{s}\, h^{\mu\gamma}\,h^{\nu\delta}\,u^\rho\, 
  \nabla_\rho \pi_{\gamma\delta} \,,\\
 \Pi &= - \zeta_{\NS}\, \tr K 
  - \tau_\mathrm{b}\,u^\gamma\, \nabla_\gamma \Pi \,,\\
 \nu^\mu&= \sigma_{\NS} \,h^{\mu\nu}\,\partial_\nu\Bigl(-\frac{\mu}{T}\Bigr)
  - \tau_\sigma\, h^\mu_{~\nu}\, u^\gamma\, \nabla_\gamma \nu^\nu \,, 
\end{aligned}
\label{ISsimple}
\end{align}
where $\pi^{\mu\nu}\equiv \tau^{\langle\mu\nu\rangle}
-\tau_\mathrm{(q)}^{\langle\mu\nu\rangle}$\,.
They are the constitutive equations for the simplified version 
of the Israel-Stewart model.% 
\footnote{%%
The constitutive equations for a simplified Israel-Stewart fluid is obtained 
by setting the viscous-heat coupling coefficients to be zero 
in those for an Israel-Stewart fluid 
(i.e., $\alpha_0=\alpha_1=0$ in Eqs.\ (8a)--(8c) in \cite{Israel:1976tn}).
} %%%%%%%%%%

Thus, in this case the rheology equations are equivalent 
to the constitutive equations 
for the simplified Israel-Stewart model \eq{ISsimple},
and the $\bigl[D+1+1+D(D+1)/2+D\bigr]$ dynamical variables (excluding $\theta$) 
can be determined from 
the $D+2$ conservation laws ($\nabla_\mu n^\mu=\nabla_\mu T^{\mu\nu}=0$) and 
the $D(D+1)/2+D$ equations \eq{ISsimple}.

%%%%%%%%%%%%%%%%%%%%%%%%%%%%%%%%%%%%%%%%%%%%%%%%%%%%%% 
\section{Hyperbolicity and dispersion relations}
%%%%%%%%%%%%%%%%%%%%%%%%%%%%%%%%%%%%%%%%%%%%%%%%%%%%%% 
\label{hyperbolic_dispersion}

In this section, 
we study linear perturbations around a hydrostatic equilibrium 
in Minkowski spacetime. 
We exclusively take a coordinate system $(x^\mu)=(x^0,\,x^i)$ 
in which the background metric is written as 
$g_{\mu\nu}=\eta_{\mu\nu}\equiv \mathrm{diag} (-1, 1,\cdots, 1)$\,.
A hydrostatic equilibrium is then specified 
by the velocity $u_{(0)}=u_{(0)}^\mu \partial_\mu \equiv \partial_0$ 
(i.e., $u_{(0)}^\mu=\delta^\mu_0$), 
the proper energy density $e_{(0)}$\,, the number density $n_{(0)}$\,, 
and the vanishing strain tensor $E^{(0)}_{\mu\nu}\equiv0$\,. 
The induced metric is then given by 
$h^{(0)}_{\mu\nu}=\eta_{\mu\nu}+u^{(0)}_\mu u^{(0)}_\nu
=\mathrm{diag} (0,1,\cdots,1)$\,. 
Note that from the fundamental relation for the hydrostatic equilibrium, 
$\ts_{(0)}=\tsigma_{(0)}\bigl(\te_{(0)},\tn_{(0)},\sqrt{h_{(0)}}\bigr)
\equiv\sqrt{h_{(0)}}\,s_{(0)}( e_{(0)},n_{(0)} )$\,, 
other thermodynamic quantities such as 
the temperature $T_{(0)}$\,, the chemical potential $\mu_{(0)}$ 
and the pressure $P_{(0)}$ are determined as 
\begin{align}
 \delta\ts_{(0)}=\frac{1}{T_{(0)}}\,\delta\te_{(0)} 
  -\frac{\mu_{(0)}}{T_{(0)}}\,\delta\tn_{(0)}
  +\frac{P_{(0)}}{T_{(0)}}\,\delta\sqrt{h_{(0)}}\,,
\end{align}
or
\begin{align}
 \delta s_{(0)}=\frac{1}{T_{(0)}}\,\delta e_{(0)} 
  -\frac{\mu_{(0)}}{T_{(0)}}\,\delta n_{(0)}
\end{align}
with the Euler-Gibbs-Duhem relation 
\begin{align}
 s_{(0)}=\frac{e_{(0)}}{T_{(0)}} - \frac{\mu_{(0)}}{T_{(0)}}
  +\frac{P_{(0)}}{T_{(0)}}\,.
\end{align}

%%%%%%%%%%%%%%%%%%%%%%%%%%%%%%%%%%%%%%%%%%%%%%%%%%%%%% 
\subsection{Linear perturbations around a hydrostatic equilibrium}
%%%%%%%%%%%%%%%%%%%%%%%%%%%%%%%%%%%%%%%%%%%%%%%%%%%%%% 
\label{linear_perturbations}

We now consider linear perturbations around the hydrostatic equilibrium,
\begin{align}
 g_{\mu\nu}&= \eta_{\mu\nu}+0\,, & u^\mu &= \delta_0^{\mu} + \delta u^\mu \,,& 
  h_{\mu\nu} &= \rlap{$\displaystyle  h^{(0)}_{\mu\nu}
  + \eta_{0\mu}\,\delta u_\nu + \delta u_\mu \,\eta_{0\nu}\,,$} & &\\
 e&=e_{(0)}+\delta e \,, & n&=n_{(0)}+\delta n\,, & E_{\mu\nu}&= 0 + E_{\mu\nu}\,, & &
\end{align}
and denote their conjugate thermodynamic variables by 
\begin{align}
 T =T_{(0)} +\delta T \,,\quad \mu=\mu_{(0)}+\delta \mu \,,\quad
  P=P_{(0)}+\delta P\,.
\end{align}
We only consider the locally isotropic case: 
$\tau^{\mu\nu}_{\mathrm{(q)}}=P\,h^{\mu\nu}$\,.
Using the identity $-1=u^\mu\, u_\mu= -1 + 2\delta u_0= - 1 - 2\delta u^0$\,, 
we can show that $\delta u_0=\delta u^0=0$\,, 
and the acceleration vector 
$a^\mu = u^\nu\, \partial_\nu u^\mu= \partial_0 \delta u^\mu$ 
has only spatial components: 
$a^0 = \partial_0 \delta u^0=0$ and $a^i = \partial_0 \delta u^i$\,.
Moreover, from $0=\varepsilon_{\mu\nu}\, u^\nu = \varepsilon_{\mu 0}$\,, 
$\varepsilon_{\mu\nu}$ also has only spatial components, $\varepsilon_{ij}$\,,  
in this linear approximation.
Similarly, since $0=K_{\mu\nu}\, u^\nu = K_{\mu 0}$\,,
the extrinsic curvature also has only spatial components,  
which are expressed as
\begin{align}
 K_{ij}
  = \frac{1}{2}\, h_i^\mu h_j^\nu\, 
  \bigl(\partial_\mu u_\nu + \partial_\nu u_\mu \bigr)
  =\frac{1}{2}\, \bigl(\partial_i \delta u_j + \partial_i \delta u_j \bigr)\,,
\end{align}
or
\begin{align}
 \tr K=\partial_i \delta u^i\,, \qquad
  K_{\langle ij\rangle}=\frac{1}{2}\, \Bigl(\partial_i \delta u_j
  + \partial_j \delta u_i
  - \frac{2}{D}\,(\partial_k \delta u^k)\, h^{(0)}_{ij}\Bigr)\,.
\end{align}
As for the stress tensor \eq{stress_final}, 
by decomposing it as $\tau_{\mu\nu}=\tau^{(0)}_{\mu\nu}+\delta \tau_{\mu\nu}$\,, 
the zeroth part is given by
$\tau^{(0)}_{\mu i}= P_{(0)}\, h^{(0)}_{\mu i}$\,, 
and from 
$0=\tau_{\mu\nu}\,u^\nu=\tau^{(0)}_{\mu i}\,\delta u^i + \delta \tau_{\mu0}$ 
we can show that
$\delta \tau_{00}=0$\,,\, 
$\delta \tau_{i0}=-\tau^{(0)}_{ij}\,\delta u^j=-P_{(0)}\,\delta u_i$\,,
and the spatial components are written as
\begin{align}
 \delta \tau_{ij} &=\delta P\, h^{(0)}_{ij}
  -2\,(\cG-\eta_2)\,\varepsilon_{\langle ij\rangle}
  - \frac{\eta_3}{T_{(0)}}\,\Bigl[\partial_i \delta u_j
  + \partial_j \delta u_i
  - \frac{2}{D}\,(\partial_k \delta u^k)\, h^{(0)}_{ij}\Bigr] \nn\\
  &\quad -\bigl[(\cK-\zeta_4)\,\tr\varepsilon
  - (\cK a+\zeta_5)\,\theta\bigr]\, h^{(0)}_{ij}
  -\frac{\zeta_6}{T_{(0)}}\,(\partial_k \delta u^k)\, h^{(0)}_{ij} \,.
\end{align}
The diffusion current is written as 
\begin{align}
 \nu^\mu = -\,(\cH-\sigma_2)\,\varepsilon^\mu
  + \sigma_3\,h_{(0)}^{\mu\nu}\,\partial_\nu\,\delta\Bigl(-\frac{\mu}{T}\Bigr)\,.
\end{align}

We now substitute the above expressions 
to the set of fundamental equations, 
consisting of (A) the conservation laws 
\eq{conservation1}--\eq{diffusion_final} 
and (B) the rheology equations \eq{rheology1}--\eq{rheology3} 
(or \eq{rheology1a}--\eq{rheology3a}).

\noindent
\underline{(A)} 
As for the conservation laws of energy-momentum tensor, 
the component along $u^\mu$ is given by 
$0=u_\nu\,\partial_\mu T^{\mu\nu}
=\partial_\mu(T^{\mu\nu} u_\nu)-T^{\mu\nu}\,\partial_\mu u_\nu
=-\partial_\mu(e u^\mu) - \tau^{\mu\nu}\,\partial_\mu u_\nu$\,.
From this we obtain 
\begin{align}
 \partial_\mu \bigl(eu^\mu\bigr)&= \partial_0 \delta e
  + e_{(0)}\, \partial_i \delta u^i
\nn\\
 &= -\tau^{\mu\nu}\,\partial_\mu u_\nu 
  = -\tau_{(0)}^{\mu\nu}\,\partial_\mu \delta u_\nu
  = - P_{(0)}\, \partial_i \delta u^i\,,
\end{align}
or
\begin{align}
 \partial_0 \delta e = - w_{(0)}\, \partial_i \delta u^i\,.
\end{align}
Here $w_{(0)}\equiv e_{(0)}+P_{(0)}$ is the enthalpy density 
at the hydrostatic equilibrium. 
As for the components orthogonal to $u^\mu$\,, 
from the equations $0=h_{\lambda\nu}\,\partial_\mu T^{\mu\nu}
=h_{\lambda\nu}\,\partial_\mu(e u^\mu u^\nu)
+h_{\lambda\nu}\,\partial_\mu\tau^{\mu\nu}
=e\,u^\mu\,\partial_\mu u_\lambda + h_{\lambda\nu}\,\partial_\mu\tau^{\mu\nu}
=e\,a_\lambda+ h_\lambda^{~\nu}\,\partial^\mu\tau_{\mu\nu}$\,,
we obtain
\begin{align}
 e\, a_i&=e_{(0)}\,\partial_0\delta u_i \nn\\
 &= -h_i^\nu\, \partial^\mu \tau_{\mu\nu}=- \partial^\mu \delta \tau_{\mu i}
  =- \partial^0 \delta \tau_{0i} - \partial^k \delta \tau_{ik}
  = P_{(0)}\,\partial^0 \delta u_i - \partial^k \delta \tau_{ik}\,,
\end{align}
or
\begin{align}
 w_{(0)}\, \partial_0\delta u_i &= -\,\partial^k \delta \tau_{ik} \nn\\
 &= 
  -\,\partial_i\delta P
  +2\,(\cG-\eta_2)\,\partial^k\varepsilon_{\langle ik\rangle}
  +\Bigl( \frac{(D-2)\eta_3}{D\, T_{(0)}}+\frac{\zeta_6}{T_{(0)}}\Bigr)\,
  \partial_i\partial_k\delta u^k + \frac{\eta_3}{T_{(0)}}\,\triangle\delta u^i
\nn\\
 &\quad\, + (\cK-\zeta_4)\,\partial_i(\tr\varepsilon)
  - (\cK a+\zeta_5)\,\partial_i\theta \,,
\end{align}
where $\triangle$ is the spatial Laplacian, 
$\triangle\equiv \delta^{ij}\,\partial_i\,\partial_j$\,.
The conservation law of particle number current becomes
\begin{align}
 0= \partial_\mu \bigl(nu^\mu +\nu^\mu\bigr)
  = \partial_0 \delta n + n_{(0)}\, \partial_i\delta u^i
    + \sigma_3\,\triangle\, \delta\Bigl(-\frac{\mu}{T}\Bigr)
    - (\cH-\sigma_2)\, \partial_i \varepsilon^i \,. 
\end{align}

\noindent
\underline{(B)} 
The rheology equations are linearized as 
\begin{align}
 \partial_0 \varepsilon_{\langle ij\rangle}
  =&~ \frac{\cG + \eta_2}{2\lambda_1} \, 
  \Bigl(\partial_i \delta u_j
        + \partial_j \delta u_i 
        - \frac{2}{D}\,\bigl(\partial_k \delta u^k\bigr)\, h^{(0)}_{ij}\Bigr) 
  -\frac{1}{\tau_\mathrm{s}}\, \varepsilon_{\langle ij\rangle} \,, \\
 \partial_0 \varepsilon_i
  =&~ -\frac{1}{\tau_\sigma}\, \varepsilon_i
     -\frac{(\cH+\sigma_2)\, T_{(0)}}{\lambda_2}\, 
  \partial_i \,\delta\Bigl(-\frac{\mu}{T}\Bigr) \,,\\
  \partial_0 (\tr\varepsilon) 
  =&~ -\frac{\bigl(\gamma_3\,\zeta_1+\gamma_2\,(\cK'-\zeta_2)\bigr)\,T_{(0)}}
   {\det{\boldsymbol\gamma}}\,\tr \varepsilon
     +\frac{\bigl(\gamma_2\,\zeta_3- \gamma_3\,(\cK'+\zeta_2)\bigr)\,T_{(0)}}
           {\det{\boldsymbol\gamma}}\,\theta \nn\\
   &~ +\frac{\gamma_2\,(\cK a-\zeta_5)+\gamma_3\,(\cK +\zeta_4)}
           {\det{\boldsymbol\gamma}}\, \partial_i\delta u^i \,,\\
  \partial_0 \theta
  =&~ \frac{\bigl(\gamma_1\,(\cK'-\zeta_2)+\gamma_2\,\zeta_1\bigr)\,T_{(0)}}
          {\det{\boldsymbol\gamma}} \,\tr \varepsilon
    -\frac{(\gamma_1\,\zeta_3-\gamma_2\,\bigl(\cK'+\zeta_2)\bigr)\,T_{(0)}}
     {\det{\boldsymbol\gamma}}\,\theta \nn\\
   &~ -\frac{\gamma_2\,(\cK +\zeta_4)+\gamma_1\,(\cK a-\zeta_5)}
          {\det{\boldsymbol\gamma}}\, \partial_i\delta u^i \,,
\end{align}
where we have used the approximation 
$\Lie_u\varepsilon_{ij}\simeq \partial_0 \varepsilon_{ij}$\,, 
$\Lie_u\varepsilon_{i}\simeq \partial_0 \varepsilon_i$\,, 
$\Lie_u(\tr\varepsilon)\simeq \partial_0 (\tr\varepsilon)$\,, 
and
$\Lie_u\theta \simeq \partial_0 \theta$\,.

Since we are considering locally isotropic materials, 
the fundamental thermodynamic relation \eq{delta_s_visc} can be rewritten
with the use of the Euler relation \eq{Euler} as
\begin{align}
 \delta s = \frac{1}{T}\, \delta e -\frac{\mu}{T}\, \delta n 
  & - \frac{1}{T}\, 2\lambda_1\,
      \varepsilon^{\langle\mu\nu\rangle} \,\delta \varepsilon_{\langle\mu\nu\rangle}
  - \frac{1}{T}\,\bigl(\gamma_1\,\tr\varepsilon
                       +\gamma_2\,\theta\bigr)\,\delta (\tr\varepsilon) 
  \nn\\
  & - \frac{1}{T} \,\lambda_2\,\varepsilon^\mu\,\delta\varepsilon_\mu
  - \frac{1}{T}\,\bigl(\gamma_3\,\theta
                       +\gamma_2\,\tr\varepsilon\bigr)\, \delta\theta \,.
\end{align}
If we denote the thermodynamic variables %for material particles per unit volume 
collectively by 
$(a^r)=
(e,n,\varepsilon_{\langle\mu\nu\rangle},\varepsilon_\mu,\tr\varepsilon,\theta)$\,, 
the matrix 
${\bf A} \equiv -\bigl(\partial^2s/\partial a^r\partial a^s\bigr)\bigr\rvert_{(0)}$ 
is positive definite from the convexity of entropy.
Here $\vert_{(0)}$ means that the matrix is evaluated at the hydrostatic state.
In the following discussions, we assume for brevity 
that the matrix takes the following form:
\begin{align}
{\bf A} = 
\begin{pmatrix}
 {\bf A}_1 & {\bf A}_2 &     0     &     0     &     0     &     0     \cr
 {\bf A}_2 & {\bf A}_3 &     0     &     0     &     0     &     0     \cr
 0 & 0 & {\bf A}_4^{\langle\mu\nu\rangle,\langle\rho\sigma\rangle} &  0 & 0 & 0 \cr
    0    &    0    &    0    & {\bf A}_5^{\mu\nu} &     0     &    0    \cr
     0     &     0     &     0     &     0     & {\bf A}_6 & {\bf A}_7 \cr
     0     &     0     &     0     &     0     & {\bf A}_7 & {\bf A}_8
\end{pmatrix} \,,
\end{align}
where the principal submatrix
\begin{align}
 {\bf A}_\mathrm{s}\equiv
\begin{pmatrix}
 {\bf A}_1&{\bf A}_2\cr {\bf A}_2&{\bf A}_3
\end{pmatrix}
 = \begin{pmatrix}
 -\frac{\partial^2s}{\partial e^2}\Bigr\rvert_{(0)} 
 & -\frac{\partial^2s}{\partial e\partial n}\Bigr\rvert_{(0)} \\
 -\frac{\partial^2s}{\partial e\partial n}\Bigr\rvert_{(0)} 
 & -\frac{\partial^2s}{\partial n^2}\Bigr\rvert_{(0)}
   \end{pmatrix}
 = \begin{pmatrix}
 -\frac{\partial (1/T)}{\partial e}\Bigr\rvert_{(0)} 
 & \frac{\partial (\mu/T)}{\partial e}\Bigr\rvert_{(0)} \\
 -\frac{\partial (1/T)}{\partial n}\Bigr\rvert_{(0)} 
 & \frac{\partial (\mu/T)}{\partial n}\Bigr\rvert_{(0)}
   \end{pmatrix}
\end{align}
is positive definite. 
Then the Gibbs-Duhem equation \eq{Gibbs-Duhem} can be written as%
\footnote{%%
Note that the right-hand side of Eq.\ \eq{Gibbs-Duhem} can be set to zero 
for the linear perturbations around a hydrostatic equilibrium.
} %%%%%%%%%%
\begin{align}
 \partial_i\delta P &= s_{(0)}\, \partial_i\delta T + n_{(0)}\, \partial_i\delta \mu \nn\\
 &= s_{(0)}\, \partial_i\,\delta \bigl[(1/T)^{-1}\bigr]
  + n_{(0)}\, \partial_i\,\delta \bigl[(1/T)^{-1}\,(\mu/T)\bigr] \nn\\
  &= \bigl(w_{(0)}\, {\bf A}_1+n_{(0)}\, {\bf A}_2\bigr)\, T_{(0)}\, \partial_i\delta e
  +\bigl(w_{(0)}\, {\bf A}_2+n_{(0)}\, {\bf A}_3\bigr)\, T_{(0)}\, \partial_i\delta n\,, \\
 \partial_i\,\delta\Bigl(-\frac{\mu}{T}\Bigr)
  &= -\bigl({\bf A}_2\,\partial_i\delta e+{\bf A}_3\,\partial_i\delta n \bigr)\,,
\end{align}
and we finally obtain the following set of linearized equations of motion:
\begin{align}
 \partial_0 \delta e 
 =&~ -w_{(0)}\, \partial_i \delta u^i\,,\\
 w_{(0)}\, \partial_0\delta u_i 
 =&~ 
  2\,(\cG-\eta_2)\,\partial^k\varepsilon_{\langle ik\rangle}
  +\Bigl( \frac{(D-2)\eta_3}{D\, T_{(0)}}+\frac{\zeta_6}{T_{(0)}}\Bigr)\,
  \partial_i\partial_k\delta u^k + \frac{\eta_3}{T_{(0)}}\,\triangle\delta u_i
\nn\\
 &~ + (\cK-\zeta_4)\,\partial_i(\tr\varepsilon) - (\cK a+\zeta_5)\,\partial_i\theta 
\nn\\
 &~ 
  -\bigl(w_{(0)}\, {\bf A}_1+n_{(0)}\, {\bf A}_2\bigr)\, T_{(0)}\, \partial_i\delta e
  -\bigl(w_{(0)}\, {\bf A}_2+n_{(0)}\, {\bf A}_3\bigr)\, T_{(0)}\, \partial_i\delta n \,,\\
 \partial_0 \delta n 
  =&~ - n_{(0)}\, \partial_i\delta u^i
    + \sigma_3\,\bigl({\bf A}_2\,\triangle \delta e
                      +{\bf A}_3\, \triangle \delta n\bigr)
    + (\cH-\sigma_2)\, \partial_i \varepsilon^i  \,,\\
 \partial_0 \varepsilon_{\langle ij\rangle}
  =&~ \frac{\cG + \eta_2}{2\lambda_1} \, 
  \Bigl(\partial_i \delta u_j
        + \partial_j \delta u_i
        - \frac{2}{D}\,(\partial_k \delta u^k)\,h^{(0)}_{ij}\Bigr) 
  -\frac{1}{\tau_\mathrm{s}}\, \varepsilon_{\langle ij\rangle} \,, \\
 \partial_0 \varepsilon_i
  =&~ -\frac{1}{\tau_\sigma}\, \varepsilon_i
     +\frac{(\cH+\sigma_2)\,T_{(0)}}{\lambda_2} \,
      \bigl({\bf A}_2\,\partial_i\delta e+{\bf A}_3\,\partial_i\delta n \bigr) \,,\\
  \partial_0 (\tr\varepsilon) 
  =&~ -\frac{\bigl(\gamma_3\,\zeta_1+\gamma_2\,(\cK'-\zeta_2)\bigr)\,T_{(0)}}
  {\det{\boldsymbol\gamma}}\,\tr \varepsilon
  +\frac{\bigl(\gamma_2\, \zeta_3- \gamma_3\,(\cK'+\zeta_2)\bigr)\,T_{(0)}}
        {\det{\boldsymbol\gamma}}\,\theta
 \nn\\
  &~
  +\frac{\gamma_2\,(\cK a-\zeta_5)+\gamma_3\,(\cK +\zeta_4)}
        {\det{\boldsymbol\gamma}}\, \partial_i\delta u^i \,,\\
  \partial_0 \theta
  =&~
 \frac{\bigl(\gamma_1\,(\cK'-\zeta_2)+\gamma_2\,\zeta_1\bigr)\,T_{(0)}}
      {\det{\boldsymbol\gamma}} \,\tr \varepsilon
 -\frac{\bigl(\gamma_1\,\zeta_3-\gamma_2\,(\cK'+\zeta_2)\bigr)\,T_{(0)}}
  {\det{\boldsymbol\gamma}}\,\theta \nn\\
   &~ -\frac{\gamma_2\, (\cK +\zeta_4)+\gamma_1\,(\cK a-\zeta_5)}
            {\det{\boldsymbol\gamma}}\, \partial_i\delta u^i \,.
\end{align}

We now consider wave propagations in the $x^D$ direction, 
demanding that perturbations depend only on $x^0$ and $x^D$:
\begin{align}
 \delta u_i = \delta u_i(x^0,x^D) \,,\quad
  \varepsilon_{ij}= \varepsilon_{ij} (x^0,x^D) \,,\quad 
  \delta e = \delta e(x^0,x^D) \,,\quad 
  \delta n = \delta n(x^0,x^D) \,.
\end{align}
Then the above equations can be rewritten as follows:
\begin{align}
 \partial_0 \varepsilon_{\langle II\rangle}
  =&~ - \frac{\cG + \eta_2}{D\lambda_1} \, \partial_D \delta u_D 
     - \frac{1}{\tau_\mathrm{s}}\, \varepsilon_{\langle II\rangle} \,, \\
 \partial_0 \varepsilon_{\langle IJ\rangle}
  =&~ -\frac{1}{\tau_\mathrm{s}}\, \varepsilon_{\langle IJ\rangle}
  \qquad (I\neq J) \,, \\
 \partial_0 \varepsilon_{\langle ID\rangle}
  =&~ \frac{\cG + \eta_2}{2\lambda_1} \, \partial_D \delta u_I
  -\frac{1}{\tau_\mathrm{s}}\, \varepsilon_{\langle ID\rangle} \,, \\
 w_{(0)}\,\partial_0\delta u_I 
  =&~  
   2\,(\cG-\eta_2)\,\partial_D\varepsilon_{\langle ID\rangle}
  +\frac{\eta_3}{T_{(0)}}\, \partial_D^2 \delta u_I \,,\\
 \partial_0 \varepsilon_I =&~ -\frac{1}{\tau_\sigma}\, \varepsilon_I \,,\\
 \partial_0 \delta e =&~ -w_{(0)}\, \partial_D \delta u_D\,,\\
 w_{(0)}\, \partial_0\delta u_D 
  =&~ 
   2\,(\cG-\eta_2)\,\partial_D\varepsilon_{\langle DD\rangle}
  +\Bigl(\frac{2(D-1)\,\eta_3}{D\,T_{(0)}}
         +\frac{\zeta_6}{T_{(0)}}\Bigr)\,\partial_D^2 \delta u_D \nn\\
  &~ + (\cK-\zeta_4)\,\partial_D(\tr\varepsilon) - (\cK a+\zeta_5)\,\partial_D\theta \nn\\
  &~ 
  -\bigl(w_{(0)}\,{\bf A}_1+n_{(0)}\,{\bf A}_2\bigr)\, T_{(0)} \partial_D\delta e
  -\bigl(w_{(0)}\,{\bf A}_2+n_{(0)}\,{\bf A}_3\bigr)\, T_{(0)} \partial_D\delta n \,,\\
 \partial_0 \delta n 
  =&~ - n_{(0)}\, \partial_D\delta u_D
    + \sigma_3\,\bigl({\bf A}_2\,\partial_D^2 \delta e+{\bf A}_3\,
  \partial_D^2 \delta n\bigr)
    + (\cH-\sigma_2)\, \partial_D \varepsilon_D  \,,\\
 \partial_0 \varepsilon_{\langle DD\rangle}
  =&~ \frac{(D-1)\,(\cG + \eta_2)}{D\,\lambda_1} \, \partial_D \delta u_D 
  -\frac{1}{\tau_\mathrm{s}}\, \varepsilon_{\langle DD\rangle} \,, \\
 \partial_0 \varepsilon_D
  =&~ -\frac{1}{\tau_\sigma}\, \varepsilon_D
     +\frac{(\cH+\sigma_2)\,T_{(0)}}{\lambda_2} \bigl({\bf A}_2\,\partial_D\delta e 
           + {\bf A}_3\,\partial_D\delta n\bigr) \,,\\
  \partial_0 (\tr\varepsilon) 
  =&~ -\frac{\bigl(\gamma_3\,\zeta_1+\gamma_2\,(\cK'-\zeta_2)\bigr)\,T_{(0)}}
   {\det{\boldsymbol\gamma}}\,\tr \varepsilon
     +\frac{\bigl(\gamma_2\,\zeta_3- \gamma_3\,(\cK'+\zeta_2)\bigr)\,T_{(0)}}
           {\det{\boldsymbol\gamma}}\,\theta \nn\\
   &~ +\frac{\gamma_2\,(\cK a-\zeta_5)+\gamma_3\,(\cK +\zeta_4)}
            {\det{\boldsymbol\gamma}}\, \partial_D\delta u_D\,,\\
  \partial_0 \theta
  =&~ \frac{\bigl(\gamma_1\,(\cK'-\zeta_2)+\gamma_2\,\zeta_1\bigr)\, T_{(0)}}
          {\det{\boldsymbol\gamma}}\,\tr\varepsilon
    -\frac{\bigl(\gamma_1\,\zeta_3-\gamma_2\,(\cK'+\zeta_2)\bigr)\,T_{(0)}}
   {\det{\boldsymbol\gamma}}\,\theta \nn\\
   &~ -\frac{\gamma_2\,(\cK +\zeta_4)+\gamma_1\,(\cK a-\zeta_5)}
            {\det{\boldsymbol\gamma}}\, \partial_D\delta u_D\,,
\end{align}
where $I,J= 1,\cdots,D-1$\,.
This set of equations can be further decomposed 
according to the transformation properties 
under the little group $\mathrm{SO}(D-1)$\,: 
\begin{enumerate}
\item tensor modes: 
 $(\varepsilon_{\langle II\rangle},\,\varepsilon_{\langle IJ\rangle})$\,;
\item shear modes: 
 $(\varepsilon_{\langle ID\rangle},\,\delta u_I,\,\varepsilon_I)$\,;
\item sound modes: 
 $(\tr\varepsilon,\,\varepsilon_{\langle DD\rangle},\,
   \delta u_D,\,\delta e,\,\delta n,\,\varepsilon_D,\,\theta)$\,.
\end{enumerate}
In the remainder of this section, 
we study hyperbolicity and dispersion relations 
for each type of perturbation modes.

%%%%%%%%%%%%%%%%%%%%%%%%%%%%%%%%%%%%%%%%%%%%%%%%%%%%%% 
\subsection{Tensor modes}
%%%%%%%%%%%%%%%%%%%%%%%%%%%%%%%%%%%%%%%%%%%%%%%%%%%%%% 
\label{Tensor_modes}

For tensor modes, the set of equations can be written as
\begin{align}
 \partial_0 \varepsilon_{\langle II\rangle}
  &= - \frac{\cG + \eta_2}{D\lambda_1} \, \partial_D \delta u_D 
     - \frac{1}{\tau_\mathrm{s}}\, \varepsilon_{\langle II\rangle} \,, \\
 \partial_0 \varepsilon_{\langle IJ\rangle}
  &= -\frac{1}{\tau_\mathrm{s}}\, \varepsilon_{\langle IJ\rangle} \,.
\end{align}
From the identity 
$\sum_I\varepsilon_{\langle II\rangle}+\varepsilon_{\langle DD\rangle}=0$\,, 
the number of independent variables of $\varepsilon_{\langle II\rangle}$ is $D-2$\,.
If we define the variables $E_{IJ}$ by
\begin{align}
 E_{IJ}\equiv 
 \begin{cases} \varepsilon_{\langle II\rangle}
 -\varepsilon_{\langle (D-1)(D-1)\rangle} & (\text{for } I=J) \cr
 \varepsilon_{\langle IJ\rangle} & (\text{for } I\neq J)
\end{cases}\,,
\end{align}
then the number of independent $E_{IJ}$ is 
$D(D-1)/2-1=(D-2)(D+1)/2$ 
because $E_{(D-1)(D-1)}=0$\,,
and the equations for $E_{IJ}$ become
\begin{align}
 0 = \partial_0 E_{IJ} + \frac{1}{\tau_\mathrm{s}} E_{IJ} \,.
\label{tensor_linear_eom}
\end{align}
Thus, if we consider plane waves propagating in the $x^D$ direction,
\begin{align}
 \delta u_i = \widetilde{\delta u}_i(\omega,k) \Exp{\ii k\,x^D-\ii \omega x^0}\,,\qquad 
 \varepsilon_{ij}
  = \widetilde{\varepsilon}_{ij}(\omega,k) \Exp{\ii k\,x^D-\ii \omega x^0}\,,
\end{align}
we obtain the dispersion relation 
$\omega= - \ii/\tau_\mathrm{s}$ 
which represents nonpropagating, purely dissipating modes. 
Since $\tau_\mathrm{s}$ is positive, the imaginary part of $\omega$ is 
always negative, 
and thus we find that the tensor modes are always stable. 
Such \emph{relaxation modes} correspond to stress relaxations 
observed at rheological time scales 
($T_\mathrm{obs}\sim \tau_\mathrm{s}$), 
and will disappear at hydrodynamic time scales 
($T_\mathrm{obs}\gg \tau_\mathrm{s}$).

%%%%%%%%%%%%%%%%%%%%%%%%%%%%%%%%%%%%%%%%%%%%%%%%%%%%%% 
\subsection{Shear modes}
%%%%%%%%%%%%%%%%%%%%%%%%%%%%%%%%%%%%%%%%%%%%%%%%%%%%%% 
\label{Shear_modes}

For shear modes, we have the equations
\begin{align}
 0&=\bigl(\partial_0 + \tau_\mathrm{s}^{-1}\bigr)\, \varepsilon_{\langle ID\rangle} 
    - \frac{\cG + \eta_2}{2\lambda_1} \, \partial_D \delta u_I \,, 
\label{shear_linear_eom_1}\\
  0&=\partial_0\delta u_I -\frac{2\,(\cG-\eta_2)}{w_{(0)}}\,
  \partial_D\varepsilon_{\langle ID\rangle}
     -\frac{\eta_3}{w_{(0)}\,T_{(0)}}\, \partial_D^2 \delta u_I \,,
\label{shear_linear_eom_2}\\
 0&= \bigl(\partial_0+\tau_\sigma^{-1}\bigr)\,\varepsilon_I \,.
\label{shear_linear_eom_3}
\end{align}
Note that $\varepsilon_I$ is decoupled from the other variables, 
and Eq.\ \eq{shear_linear_eom_3} represents its pure relaxation 
with relaxation time $\tau_\sigma\,(\geq 0)$\,.

If we set $\eta_3=0$\,, by redefining the variables by
\begin{align}
 s_{I\pm} 
  \equiv 
 \pm \sqrt{\frac{\lambda_1}{w_{(0)}}}\,\varepsilon_{\langle ID\rangle}
 + \frac{1}{2}\, \delta u_I \,,
\end{align}
the set of linearized equations for $s_{I\pm}$ can be written as
\begin{align}
 \Biggl(\partial_0 \mp c_\mathrm{shear}\,\partial_D\Biggr)\, s_{I\pm}
     \pm \frac{s_{I+}-s_{I-}}{2\tau_\mathrm{s}} =0\,,
\label{shear_linearized}
\end{align}
for $I=1,2,\dotsc,D-2$\,. 
These are hyperbolic equations and the characteristic velocity 
is given by
\begin{align}
 c_\mathrm{shear}\equiv \sqrt{\frac{\eta_{\NS}}{w_{(0)}\,\tau_\mathrm{s}}} \,.
\end{align}

For generic cases, 
from Eqs.\ \eq{shear_linear_eom_1} and \eq{shear_linear_eom_2}, 
we obtain telegrapher's equations with Kelvin-Voigt damping
\begin{align}
 \Bigl(\partial_0^2 + \frac{1}{\tau_\mathrm{s}}\, \partial_0 
        - \frac{\eta_3}{w_{(0)}\,T_{(0)}}\, \partial_D^2\partial_0
        - c_\mathrm{shear}^2\, \partial_D^2\Bigr)\, 
\begin{pmatrix} \varepsilon_{\langle ID\rangle}\cr \delta u_I \end{pmatrix} 
=0\,.
\label{telegraphers}
\end{align}
Although they are generically nonhyperbolic 
and have infinite wave-front velocity 
as in the standard relativistic fluid mechanics,
they can be made into hyperbolic telegrapher's equations
by setting $\eta_3=0$\,.%
\footnote{%%
In this case, from the non-negativeness of the matrix $\boldsymbol{\eta}$\,,
$\eta_2$ (and thus $\det\boldsymbol{\eta}$) must vanish.
However, this still gives a positive shear viscosity if $\cG\neq 0$\,,
as can be seen from Eq.\ \eq{eta_NS}. 
}%%%%%%%%%%%

If we consider the short time limit ($\tau_\mathrm{s}\to \infty$), 
the differential equations become
\begin{align}
 \Bigl(\partial_0^2 - \frac{\eta_3}{w_{(0)}\,T_{(0)}}\, \partial_D^2\partial_0
        - c_\mathrm{shear}^2\, \partial_D^2\Bigr) \,
 \begin{pmatrix} \varepsilon_{\langle ID\rangle}\cr \delta u_I \end{pmatrix} =0\,.
\end{align}
The wave equations in this form also appear 
for viscous solids such as Kelvin-Voigt materials, 
%and sometimes this is called the Stokes's equation,
and reduce to the wave equations when $\eta_3=0$\,.

Finally, for plane waves 
\begin{align}
 \delta u_i = \widetilde{\delta u}_i(\omega,k) \Exp{\ii k\,x^D-\ii \omega x^0}\,,\qquad
 \varepsilon_{ij} = \widetilde{\varepsilon}_{ij}(\omega,k)
  \Exp{\ii k\,x^D-\ii \omega x^0}\,,
\end{align}
from \eq{telegraphers}, we obtain the dispersion relation
\begin{align}
 \Gamma^2 
 + \Bigl(\frac{1}{\tau_\mathrm{s}} + \frac{\eta_3}{w_{(0)}\,T_{(0)}} \,k^2\Bigr)\, \Gamma
 + c_\mathrm{shear}^2\, k^2 =0\,,
\label{dispersion_shear}
\end{align}
where $\Gamma\equiv -\ii\, \omega$\,.
Since all the coefficients are positive, 
the real part of $\Gamma$ (or the imaginary part of $\omega$) always 
takes negative values, 
and thus we see that there are no unstable growing modes in the shear modes. 
Equation \eq{dispersion_shear} has two solutions, 
which are expanded around $k=0$ as
\begin{align}
 \omega = \Bigg\{\begin{array}{l}
  -\frac{\ii}{\tau_\mathrm{s}} 
  + \ii\, (1-r_\mathrm{s})\, c_\mathrm{shear}^2\,\tau_\mathrm{s} k^2
  + \ii\, (1-r_\mathrm{s})\, c_\mathrm{shear}^4\, \tau_\mathrm{s}^3\, k^4 + \ord(k^6) \cr
  - \ii\, c_\mathrm{shear}^2\, \tau_\mathrm{s}\, k^2
  - \ii\, (1-r_\mathrm{s})\, c_\mathrm{shear}^4\, \tau_\mathrm{s}^3\, k^4 + \ord(k^6)\,,
  \end{array} 
\label{shear_dispersion}
\end{align}
with $r_\mathrm{s}\equiv \eta_3/(\eta_{\NS}\,T_{(0)})$\,. 
The former represents the relaxation modes 
which are not observed 
at hydrodynamic time scales ($T_\mathrm{obs}\gg \tau_\mathrm{s}$).
The latter represents the \emph{hydrodynamic modes} 
where $\omega\to 0$ in the limit $k^2\to 0$\,, 
and from the coefficients of $k^2$\,, 
the diffusion coefficient is found to be
$c_\mathrm{shear}^2\, \tau_\mathrm{s}=\eta_{\NS}/w_{(0)}$\,. 
Moreover, by the comparison with the dispersion relation of Maxwell-Cattaneo type, 
the effective relaxation time associated with the hydrodynamic modes 
is read off from the coefficients of $k^4$ as 
$(1-r_\mathrm{s})\, \tau_\mathrm{s}$\,. 
Indeed, if we set $r_\mathrm{s}=1$\,, 
the effective relaxation time becomes zero 
and the dispersion relation becomes purely diffusive;
$\omega=-\ii\,(\eta_{\NS}/w_{(0)})\,k^2$\,.

If we are interested only in the hydrodynamic modes, 
the dispersion relation coincides with that of the Israel-Stewart model 
up to $\ord(k^4)$ by identifying $(1-r_\mathrm{s})\, \tau_\mathrm{s}$ 
with the relaxation time $\tau_{\pi}$ in the Israel-Stewart model. 
However, if the relaxation modes are also taken into account, 
our viscoelastic model has a richer structure than the Israel-Stewart model, 
which is the special case ($r_\mathrm{s}=0$) of the viscoelastic model.

%%%%%%%%%%%%%%%%%%%%%%%%%%%%%%%%%%%%%%%%%%%%%%%%%%%%%% 
\subsection{Sound modes}
%%%%%%%%%%%%%%%%%%%%%%%%%%%%%%%%%%%%%%%%%%%%%%%%%%%%%% 
\label{Sound_modes}

Finally, for sound modes, we have the following set of differential equations:
\begin{align}
\label{soundmodes_linear1}
 0=&~\partial_0 \delta e + w_{(0)}\, \partial_D \delta u_D\,,\\
 0=&~ \partial_0\delta u_D - \frac{2\,(\cG-\eta_2)}{w_{(0)}}\,
  \partial_D\varepsilon_{\langle DD\rangle}
  -\Bigl(\frac{2(D-1)\,\eta_3}{D\,w_{(0)}\,T_{(0)}}
         +\frac{\zeta_6}{w_{(0)}\,T_{(0)}}\Bigr)\,\partial_D^2 \delta u_D \nn\\
  &~ - \frac{\cK-\zeta_4}{w_{(0)}}\,\partial_D(\tr\varepsilon)
          + \frac{\cK a+\zeta_5}{w_{(0)}}\,\partial_D\theta \nn\\
  &~ 
  +\frac{\bigl(w_{(0)}\,{\bf A}_1
  +n_{(0)}\,{\bf A}_2\bigr)\,T_{(0)}}{w_{(0)}}\, \partial_D\delta e
  +\frac{\bigl(w_{(0)}\,{\bf A}_2+n_{(0)}\,{\bf A}_3\bigr)\,T_{(0)}}{w_{(0)}}\, 
  \partial_D\delta n \,,\\
 0=&~ \partial_0 \delta n + n_{(0)}\, \partial_D\delta u_D
    - \sigma_3\,\bigl({\bf A}_2\,\partial_D^2 \delta e+{\bf A}_3\,\partial_D^2 \delta n\bigr)
    - (\cH-\sigma_2)\, \partial_D \varepsilon_D  \,,\\
 0=&~ \partial_0 \varepsilon_{\langle DD\rangle}
  - \frac{(D-1)\,(\cG + \eta_2)}{D\,\lambda_1} \, \partial_D \delta u_D 
  + \frac{1}{\tau_\mathrm{s}}\, \varepsilon_{\langle DD\rangle} \,, \\
 0=&~ \partial_0 \varepsilon_D
  +\frac{1}{\tau_\sigma}\, \varepsilon_D
  -\frac{(\cH+\sigma_2)\,T_{(0)}}{\lambda_2}\, \bigl({\bf A}_2\,\partial_D\delta e 
           + {\bf A}_3\,\partial_D\delta n\bigr) \,,\\
 0=&~ \partial_0 (\tr\varepsilon) 
    +\frac{\bigl(\gamma_3\,\zeta_1+\gamma_2\,(\cK'-\zeta_2)\bigr)\,T_{(0)}}
    {\det{\boldsymbol\gamma}}\,\tr \varepsilon  \nn\\
   &~ -\frac{\gamma_2\,(\cK a-\zeta_5)+\gamma_3\,(\cK +\zeta_4)}
          {\det{\boldsymbol\gamma}}\, \partial_D\delta u_D
      -\frac{\bigl(\gamma_2\,\zeta_3- \gamma_3\,(\cK'+\zeta_2)\bigr)\,T_{(0)}}
            {\det{\boldsymbol\gamma}}\,\theta \,,\\
 0=&~ \partial_0 \theta
    +\frac{\gamma_2\,(\cK  +\zeta_4)+\gamma_1\,(\cK a-\zeta_5)}
          {\det{\boldsymbol\gamma}}\, \partial_D\delta u_D \nn\\
  &~
   -\frac{\bigl(\gamma_1\,(\cK'-\zeta_2)+\gamma_2\,\zeta_1\bigr)\,T_{(0)}}
         {\det{\boldsymbol\gamma}}\,\tr \varepsilon
  +\frac{\bigl(\gamma_1\,\zeta_3-\gamma_2\,(\cK'+\zeta_2)\bigr)\,T_{(0)}}
  {\det{\boldsymbol\gamma}}\,\theta \,.
\label{soundmodes_linear7}
\end{align}
In particular, if we consider the case where $\eta_3=\zeta_6=\sigma_3=0$\,, 
the set of equations reduces to the following linear differential equations:
\begin{align}\label{linear_eom_sound}
 \bigl(\partial_0& + B_0\, \partial_D + B_1 \bigr)\,\vec{Y}=0\,,\\
 B_0 &\equiv\left(\begin{smallmatrix}
 0 & 0 & -\frac{\cK\,(a\,\gamma_2+\gamma_3)}{\det{\boldsymbol\gamma}} & 0 & 0 & 0 & 0 \\
 0 & 0 & -\frac{(D-1)\,\cG}{D\,\lambda_1} & 0 & 0 & 0 & 0 \\
 -\frac{\cK}{w_{(0)}} & -\frac{2\cG}{w_{(0)}} & 0 & 
   \frac{T_{(0)}}{w_{(0)}}\,(w_{(0)}\,{\bf A}_1+ n_{(0)}\,{\bf A}_2) & 
   \frac{T_{(0)}}{w_{(0)}}\, (w_{(0)}\,{\bf A}_2+n_{(0)}\,{\bf A}_3) & 
   0 & \frac{\cK a}{w_{(0)}} \\
 0 & 0 & w_{(0)} & 0 & 0 & 0 & 0 \\
 0 & 0 & n_{(0)} & 0 & 0 & -\cH & 0 \\
 0 & 0 & 0 & -\frac{{\bf A}_2\,T_{(0)}\,\cH}{\lambda_2} &
  -\frac{{\bf A}_3\,T_{(0)}\,\cH}{\lambda_2} & 0 & 0 \\
 0 & 0 & \frac{\cK\,(a\,\gamma_1+\gamma_2)}{\det{\boldsymbol\gamma}} & 0 & 0 & 0 & 0
\end{smallmatrix}\right)\,,\\
 B_1 &\equiv\left(\begin{smallmatrix}
   \frac{(\gamma_3\,\zeta_1+\gamma_2\,(\cK'-\zeta_2))\,T_{(0)}}{\det{\boldsymbol\gamma}} &
   0 & 0 & 0 & 0 & 0 &
   -\frac{(\gamma_2\,\zeta_3-\gamma_3\,(\cK'+\zeta_2))\,T_{(0)}}{\det{\boldsymbol\gamma}} \\
 0 & \frac{1}{\tau_\mathrm{s}} & 0 & 0 & 0 & 0 & 0 \\
 0 & 0 & 0 & 0 & 0 & 0 & 0 \\
 0 & 0 & 0 & 0 & 0 & 0 & 0 \\
 0 & 0 & 0 & 0 & 0 & 0 & 0 \\
 0 & 0 & 0 & 0 & 0 & \frac{1}{\tau_{\sigma}} & 0 \\
 -\frac{(\gamma_1\,(\cK'-\zeta_2)+\gamma_2\,\zeta_1)\,T_{(0)}}{\det{\boldsymbol\gamma}} &
  0 & 0 & 0 & 0 & 0 & 
  \frac{(\gamma_1\,\zeta_3-\gamma_2\,(\cK'+\zeta_2))\,T_{(0)}}{\det{\boldsymbol\gamma}}
 \end{smallmatrix}\right) \,,\qquad
  \vec{Y} \equiv\left(\begin{smallmatrix}
   \tr\varepsilon \\ \varepsilon_{\langle DD\rangle}\\ \delta u_D\\ 
   \delta e\\ \delta n\\ \varepsilon_D\\ \theta \end{smallmatrix}\right)\,.
\end{align}
Here we have defined
\begin{align}
 c_\mathrm{s}^2\equiv \frac{T_{(0)}}{w_{(0)}}\,\bigl(w_{(0)}^2\,{\bf A}_1
  + 2 w_{(0)}\,n_{(0)}\, {\bf A}_2 + n_{(0)}^2\,{\bf A}_3\bigr)
  =\frac{\partial p}{\partial e}\Bigr\rvert_{\frac{s}{n}}\,,
\end{align}
and 
\begin{align}
 M&\equiv
  \left(\begin{smallmatrix}
 \sqrt{\frac{w_{(0)}\,(a\, \gamma_2+ \gamma_3)}{\det{\boldsymbol\gamma}}} & 
 0 & 0 & 0 & 0 & 0 & 0 \\
 0 & \sqrt{\frac{w_{(0)}\,(D-1)}{2 D\,\lambda_1}} & 0 & 0 & 0 & 0 & 0 \\
 0 & 0 & 1 & 0 & 0 & 0 & 0 \\
 0 & 0 & 0 & \frac{w_{(0)}}{c_\mathrm{s}} & 
 -\frac{w_{(0)}\,{\bf A}_2+n_{(0)}\,{\bf A}_3}{\sqrt{\det{\bf A}_\mathrm{s}}\,c_\mathrm{s}} &
 0 & 0 \\
 0 & 0 & 0 & \frac{n_{(0)}}{c_\mathrm{s}} & 
 \frac{w_{(0)}\,{\bf A}_1+n_{(0)}\,{\bf A}_2}{\sqrt{\det{\bf A}}_{(1)}\,c_\mathrm{s}} 
  & 0 & 0 \\
 0 & 0 & 0 & 0 & 0 & \sqrt{\frac{w_{(0)}}{\lambda_2}} & 0 \\
 0 & 0 & 0 & 0 & 0 & 0 &
  \sqrt{\frac{w_{(0)}\,(a \gamma_1+ \gamma_2)}{a\,\det{\boldsymbol\gamma}}} 
  \end{smallmatrix}\right)\,,\\
  \vec{Y}'&\equiv M^{-1} \vec{Y}\,.
\end{align}
We then have
\begin{align}
 \bigl(\partial_0 + M^{-1}B_0 M\, \partial_D + M^{-1}B_1 M\,\bigr)\, \vec{Y}'=0\,,
\end{align}
with
\begin{align}
 &M^{-1}B_0 M =\left(\begin{smallmatrix}
  0 &  0 & -M_1 & 0 & 0 & 0 & 0 \\
  0 &  0 & -M_2 & 0 & 0 & 0 & 0\\
  -M_1 & -M_2 &  0 & c_\mathrm{s} & 0 & 0 & M_3 \\
  0 &  0 &  c_\mathrm{s} & 0 & 0 & -M_4 & 0 \\
  0 &  0 &  0 & 0 & 0 & -M_5 & 0\\
  0 &  0 &  0 & -M_4 & -M_5 & 0 & 0\\
  0 &  0 &  M_3 & 0 & 0 & 0 & 0\\
  \end{smallmatrix}\right) \,,\\
 &M_1\equiv \sqrt{\frac{\cK^2\,(a\,\gamma_2+\gamma_3)}{w_{(0)}\,\det{\boldsymbol\gamma}}}
 \,,\quad 
 M_2\equiv \sqrt{\frac{2\,(D-1)\,\eta_{\NS}}{D\,w_{(0)}\,\tau_\mathrm{s}}}\,,\quad 
 M_3\equiv \sqrt{\frac{\cK^2\,a\,(a\gamma_1+\gamma_2)}
  {w_{(0)}\,\det{\boldsymbol\gamma}}} \nn\\
 &M_4= \frac{\cH \,\bigl(w_{(0)}\,{\bf A}_2+n_{(0)}\,{\bf A}_3\bigr)}{c_\mathrm{s}}
  \sqrt{\frac{T_{(0)}}{\tau_{\sigma}\,w_{(0)}\,\sigma_1}}\,,\quad 
 M_5= \frac{\cH}{c_\mathrm{s}}\,
 \sqrt{\frac{\det{\bf A}_\mathrm{s}\,T_{(0)}\,w_{(0)}}{\tau_{\sigma}\,\sigma_1}}\,,\\
 &M^{-1}B_1 M =
  \left(\begin{smallmatrix}
 \frac{(\gamma_3\,\zeta_1+\gamma_2\,(\cK'-\zeta_2))\,T_{(0)}}{\det{\boldsymbol\gamma}} &
  0 & 0 & 0 & 0 & 0 & 
  \frac{((\cK'+\zeta_2)\,\gamma_3-\zeta_3\,\gamma_2)\,T_{(0)}}{\det{\boldsymbol\gamma}}\,
 \sqrt{\frac{a\gamma_1+\gamma_2}{a(a \gamma_2+\gamma_3)}} \\
 0 & \frac{1}{\tau_\mathrm{s}} & 0 & 0 & 0 & 0 & 0 \\
 0 & 0 & 0 & 0 & 0 & 0 & 0 \\
 0 & 0 & 0 & 0 & 0 & 0 & 0 \\
 0 & 0 & 0 & 0 & 0 & 0 & 0 \\
 0 & 0 & 0 & 0 & 0 & \frac{1}{\tau_{\sigma}} & 0 \\
 -\frac{((\cK'-\zeta_2)\,\gamma_1+\zeta_1\,\gamma_2)\,T_{(0)}}{\det{\boldsymbol\gamma}}\,
 \sqrt{\frac{a (a \gamma_2+\gamma_3)}{a \gamma_1+\gamma_2}} & 0 & 0 & 0 & 0 & 0 &
 \frac{(\gamma_1\,\zeta_3-\gamma_2\,(\cK'+\zeta_2))\,T_{(0)}}{\det{\boldsymbol\gamma}}
 \end{smallmatrix}\right) \,.
\end{align}
The real matrix $M^{-1}B_0 M$ is symmetric and can be diagonalized. 
The eigenvalues are calculated to be $\{0,0,0,\pm v_\pm\}$\,,
where
\begin{align}
 v_\pm^2 &\equiv{\scriptstyle 
 \frac{1}{2} \Bigl(c_\mathrm{s}^2
 +\frac{\zeta_{\NS}} {w_{(0)}\,\tau_\mathrm{b}}
 +\frac{2\,(D-1)\,\eta_{\NS}}{D\,w_{(0)}\,\tau_\mathrm{s}}
 +\frac{{\bf A}_3\,\sigma_{\NS}}{\tau_{\sigma}} \Bigr)
 }\nn\\
 &\quad \ {\scriptstyle 
 \pm \frac{1}{2} 
 \sqrt{\Bigl(c_\mathrm{s}^2
 +\frac{\zeta_{\NS}}{w_{(0)}\,\tau_\mathrm{b}}
 +\frac{2\,(D-1)\,\eta_{\NS}}{D\,w_{(0)}\,\tau_\mathrm{s}}
 +\frac{{\bf A}_3\,\sigma_{\NS}}{\tau_{\sigma}} \Bigr)^2
 -\frac{4{\bf A}_3\,\sigma_{\NS}}{\tau_{\sigma}} \Bigl(
  \frac{\zeta_{\NS}}{w_{(0)}\,\tau_\mathrm{b}}
 +\frac{2\,(D-1)\,\eta_{\NS}}{D\,w_{(0)}\,\tau_\mathrm{s}}
 +\frac{\det{\bf A}_\mathrm{s}\,w_{(0)}\,T_{(0)}}{{\bf A}_3}\Bigr)}
}
\end{align}
give the characteristic velocities.
Since all the eigenvalues are real, 
we see that the system of differential equations \eq{linear_eom_sound}
is hyperbolic.

If we particularly set $\cH=0$ (and thus $\sigma_{\NS}=0$), 
the characteristic velocity reduces to 
\begin{align}
v_\pm =
\sqrt{c_\mathrm{s}^2+\frac{\zeta_{\NS}}{w_{(0)}\,\tau_\mathrm{b}}
+\frac{2\,(D-1)\,\eta_{\NS}}{D\,w_{(0)}\,\tau_\mathrm{s}}} 
\end{align}
and agrees with the large wave-number limit of the group velocity 
(which in our case coincides with the front velocity 
and the characteristic velocity) 
in the M\"uller-Israel-Stewart theory (see, e.g., Eq.\ (49) in \cite{Romatschke:2009im}).
If we take the long time limit, 
$\tau_\mathrm{b}\,,\, \tau_\mathrm{s}\to 0$\,, 
the characteristic velocity becomes infinitely large, 
and thus causality gets violated.

For generic cases (i.e., when we do not impose the conditions 
$\eta_3=\zeta_6=\sigma_3=0$), 
from Eqs.\ \eq{soundmodes_linear1}--\eq{soundmodes_linear7}, 
the dispersion relation for the plane wave
\begin{align}
 \delta u_i &= \widetilde{\delta u}_i(\omega,k) \Exp{\ii k\,x^D-\ii \omega x^0}\,,
  &\varepsilon_{ij}
  &= \widetilde{\varepsilon}_{ij}(\omega,k) \Exp{\ii k\,x^D-\ii \omega x^0}\,,\\
 \delta e &= \widetilde{\delta e}(\omega,k) \Exp{\ii k\,x^D-\ii \omega x^0}\,, &
  \delta n&=\widetilde{\delta n}(\omega,k) \Exp{\ii k\,x^D-\ii \omega x^0}\,,
\end{align}
is obtained as 
\begin{align}
 &{\textstyle \Gamma^7 + \bigl(c_{60}+c_{62}\,k^2\bigr)\,\Gamma^6
 + \bigl(c_{50}+c_{52}\,k^2+c_{54}\,k^4\bigr)\,\Gamma^5
 + \bigl(c_{40}+c_{42}\,k^2+c_{44}\,k^4\bigr)\,\Gamma^4} \nn\\
 &{\textstyle
 + \bigl(c_{30}+c_{32}\,k^2+c_{34}\,k^4\bigr)\,\Gamma^3
 + \bigl(c_{22}\,k^2+c_{24}\,k^4\bigr)\,\Gamma^2
 + \bigl(c_{12}\,k^2+c_{14}\,k^4\bigr)\,\Gamma
 + c_{04}\,k^4=0 }\,,
\end{align}
where $\Gamma=-\ii\,\omega$ and
\begin{align}
{\textstyle c_{60}=}
&{\textstyle \,
 \tau_\mathrm{s}^{-1}+\tau_{\sigma}^{-1}+\tau_{+}^{-1}+\tau_{-}^{-1}\,,
} \\
%%%% 
{\textstyle c_{62}=}
&{\textstyle \,
 \frac{\zeta_6}{T_{(0)}w_{(0)}}+\frac{2(D-1) \eta_3}{D T_{(0)}w_{(0)}} + {\bf A}_3\sigma_3\,,
} \\
%%%% 
{\textstyle c_{50}=}
&{\textstyle \,
\frac{\tau_\mathrm{s}\tau_{\sigma}+(\tau_\mathrm{s}+\tau_{\sigma})(\tau_{+}+\tau_{-})
+\tau_{+}\tau_{-}}
{\tau_\mathrm{s}\tau_{\sigma}\tau_{+}\tau_{-}} \,,
} \\
%%%% 
{\textstyle c_{52}=}
&{\textstyle \,
c_\mathrm{s}^2+\frac{\zeta_6(\tau_\mathrm{s}^{-1}+\tau_{\sigma}^{-1})}{T_{(0)}w_{(0)}}
+\frac{\zeta_{\NS}\tau_\mathrm{b}^{-1}}{w_{(0)}}
+\frac{2(D-1)\eta_3 (\hat{\tau}_\mathrm{s}^{-1}
+\tau_{\sigma}^{-1}+\tau_{+}^{-1}
+\tau_{-}^{-1})}{D w_{(0)}T_{(0)}}
} \nn\\
%%%% 
&{\textstyle \,
+{\bf A}_3\sigma_3 (\tau_\mathrm{s}^{-1}+\hat{\tau}_{\sigma}^{-1}
+\tau_{+}^{-1}+\tau_{-}^{-1})\,,
}\\
%%%% 
{\textstyle c_{54}=}
&{\textstyle \,
{\bf A}_3\sigma_3 \bigl(\frac{\zeta_6}{w_{(0)}T_{(0)}}
+\frac{2(D-1)\eta _3}{D w_{(0)}T_{(0)}}\bigr) \,,
} \\
%%%% 
{\textstyle c_{40}=}
&{\textstyle \,
\frac{\tau_\mathrm{s}
+\tau_{\sigma}+\tau_{+}+\tau_{-}}{\tau_\mathrm{s}\tau_{\sigma}\tau_{+}\tau_{-}}\,,
} \\
%%%% 
{\textstyle c_{42}=}
&{\textstyle \,
 c_\mathrm{s}^2(\tau_\mathrm{s}^{-1}+\tau_{\sigma}^{-1}+\tau_{+}^{-1}+\tau_{-}^{-1})
+\frac{\zeta_6}{w_{(0)}T_{(0)}\tau_\mathrm{s} \tau_{\sigma}}
+\frac{\zeta_{\NS}(\tau_\mathrm{s} \tau_{\sigma}
+(\tau_\mathrm{s}+\tau_{\sigma})\tau_\mathrm{b}^{-1}\tau_{+}\tau_{-})}
 {w_{(0)}\tau_\mathrm{s}\tau_{\sigma}\tau_{+}\tau_{-}}
}\nn\\
&{\textstyle \,
+\frac{2(D-1)\eta_{\NS}(\hat{\tau}_\mathrm{s}\tau_{\sigma}
+(\hat{\tau}_\mathrm{s}+\tau_{\sigma})(\tau_{+}+\tau_{-})+\tau_{+}\tau_{-})}
      {D w_{(0)}\tau_\mathrm{s}\tau_{\sigma}\tau_{+}\tau_{-}}
+\frac{{\bf A}_3\sigma_{\NS} (\tau_\mathrm{s}\hat{\tau}_{\sigma}
+(\tau_\mathrm{s}+\hat{\tau}_{\sigma})(\tau_{+}+\tau_{-})+\tau_{+}\tau_{-})}
{\tau_\mathrm{s} \tau_{\sigma}\tau_{+}\tau_{-}} \,,
} \\
%%%% 
{\textstyle c_{44}=}
&{\textstyle \,
{\bf A}_3\sigma_3\bigl(
\frac{\zeta_6(\hat{\tau}_{\sigma}^{-1}+\tau_\mathrm{s}^{-1})}{T_{(0)} w_{(0)}}
+\frac{\zeta_{\NS}\tau_\mathrm{b}^{-1}}{w_{(0)}}
+\frac{2(D-1)\eta_3(\hat{\tau}_\mathrm{s}^{-1}+\hat{\tau}_{\sigma}^{-1}
+\tau_{+}^{-1}+\tau_{-}^{-1})}{DT_{(0)}w_{(0)}}
+\frac{\det{\bf A} T_{(0)} w_{(0)}}{{\bf A}_3}\bigr)
\,,
} \\
%%%% 
{\textstyle c_{30}=}
&{\textstyle \,
\frac{1}{\tau_\mathrm{s}\tau_{\sigma}\tau_{+}\tau_{-}}\,,
} \\
%%%% 
{\textstyle c_{32}=}
&{\textstyle \,
\frac{c_\mathrm{s}^2(\tau_\mathrm{s}\tau_{\sigma}
+(\tau_\mathrm{s}+\tau_{\sigma})(\tau_{+}+\tau_{-})+\tau_{+}\tau_{-})}
{\tau_\mathrm{s}\tau_{\sigma}\tau_{+}\tau_{-}}
+\frac{\zeta_{\NS} (\tau_\mathrm{s}+\tau_{\sigma}
+\tau_\mathrm{b}^{-1}\tau_{+}\tau_{-})}{w_{(0)}\tau_\mathrm{s}\tau_{\sigma}\tau_{+}\tau_{-}}
}\nn\\
&{\textstyle \, 
+\frac{2(D-1)\eta_{\NS} (\hat{\tau}_\mathrm{s}+\tau_{\sigma}+\tau_{+}+\tau_{-})}
      {D w_{(0)}\tau_\mathrm{s}\tau_{\sigma}\tau_{+}\tau_{-}}
+\frac{{\bf A}_3\sigma_{\NS} (\tau_\mathrm{s}+\hat{\tau}_{\sigma}+\tau_{+}+\tau_{-})}
      {\tau_\mathrm{s}\tau_{\sigma}\tau_{+}\tau_{-}}
\,,
} \\
%%%% 
{\textstyle c_{34}=}
&{\textstyle \,
{\bf A}_3\sigma_{\NS}\Bigl(
\frac{\zeta_6}{T_{(0)} w_{(0)}\tau_\mathrm{s} \tau_{\sigma}}
+\frac{\zeta_{\NS}(\tau_\mathrm{s} \hat{\tau}_{\sigma}
+(\tau_\mathrm{s}+\hat{\tau}_{\sigma})\tau_\mathrm{b}^{-1}\tau_{+}\tau_{-})}
{w_{(0)}\tau_\mathrm{s}\tau_{\sigma}\tau_{+}\tau_{-}}
}\nn\\
&{\textstyle \,\qquad\quad 
+\frac{2(D-1)\eta_{\NS} (\hat{\tau}_\mathrm{s}\hat{\tau}_{\sigma}
+(\hat{\tau}_\mathrm{s}+\hat{\tau}_{\sigma})(\tau_{+}+\tau_{-})
+\tau_{+}\tau_{-})}
      {D w_{(0)} \tau_\mathrm{s} \tau_{\sigma}\tau_{+}\tau_{-}}
} \nn\\
%%%% 
&{\textstyle \,\qquad\quad 
+\frac{\det{\bf A}\sigma_{3} T_{(0)} w_{(0)} (\hat{\tau}_{\sigma}^{-1}
+\tau_\mathrm{s}^{-1}+\tau_{+}^{-1}+\tau_{-}^{-1})}{{\bf A}_3\sigma_{\NS}}\Bigr)
\,,
} \\
%%%% 
{\textstyle c_{22}=}
&{\textstyle \,
\frac{1}{\tau_\mathrm{s} \tau_{\sigma}\tau_{+}\tau_{-}}\bigl(c_\mathrm{s}^2(\tau_\mathrm{s}
+\tau_{\sigma}+\tau_{+}+\tau_{-})+\frac{\zeta_{\NS}}{w_{(0)}}
+\frac{2(D-1) \eta_{\NS}}{Dw_{(0)}}
+{\bf A}_3\sigma_{\NS}\bigr)
\,,
} \\
%%%% 
{\textstyle c_{24}=}
&{\textstyle \,
\frac{{\bf A}_3\sigma_{\NS}}{\tau_\mathrm{s}\tau_{\sigma}\tau_{+}\tau_{-}}\Bigl(
\frac{\zeta_{\NS} (\tau_\mathrm{s} +\hat{\tau}_{\sigma}
+\tau_\mathrm{b}^{-1}\tau_{+}\tau_{-})}{w_{(0)}}
+\frac{2(D-1)\eta_{\NS}(\hat{\tau}_\mathrm{s}+\hat{\tau}_{\sigma}
+\tau_{+}+\tau_{-})}{D w_{(0)}}
}\nn\\
&{\textstyle \,\qquad\qquad 
+\frac{\det{\bf A} w_{(0)}T_{(0)} 
(\tau_\mathrm{s}\hat{\tau}_{\sigma}+(\tau_\mathrm{s}+\hat{\tau}_{\sigma})(\tau_{+}+\tau_{-})
+\tau_{+}\tau_{-})}{{\bf A}_3}\Bigr)
\,,
} \\
%%%% 
{\textstyle c_{12}=}
&{\textstyle \,
\frac{c_\mathrm{s}^2}{\tau_\mathrm{s}\tau_{\sigma}\tau_{+}\tau_{-}} \,,
} \\
%%%% 
{\textstyle c_{14}=}
&{\textstyle \,
\frac{{\bf A}_3\sigma_{\NS}}{\tau_\mathrm{s}\tau_{\sigma}\tau_{+}\tau_{-}}
 \bigl(
 \frac{\zeta_{\NS}}{w_{(0)}}+\frac{2(D-1)\eta_{\NS}}{D w_{(0)}}
 +\frac{\det{\bf A} w_{(0)}T_{(0)}
          (\tau_\mathrm{s}+\hat{\tau}_{\sigma}+\tau_{+}+\tau_{-})}
       {{\bf A}_3}\bigr)
} \,,
\\
%%%% 
{\textstyle c_{04}=}
&{\textstyle \,
\frac{\det{\bf A} \sigma_{\NS} T_{(0)} w_{(0)}}
 {\tau_\mathrm{s} \tau_{\sigma}\tau_{+}\tau_{-}} \,.
}
\end{align}
Here we have defined non-negative constants
\begin{align}
 \hat{\tau}_\mathrm{s} \equiv r_\mathrm{s}\,\tau_\mathrm{s}
 =\frac{\eta_3}{\eta_{\NS}T_{(0)}}\,
 \tau_\mathrm{s}\,, \qquad 
 \hat{\tau}_{\sigma} \equiv \frac{\sigma_3}{\sigma_{\NS}}\,\tau_{\sigma}\,,
\end{align}
and redefined $\tau_{\rm b}$ as
\begin{align}
 \tau_\mathrm{b} &\equiv 
 \frac{\zeta_{\NS}\det{\boldsymbol\gamma}}{\cK^2\gamma_{+}+P_{\zeta\zeta\gamma}} 
 \label{tau_bulk}\,,\\
 \gamma_{+}           &\equiv a^2 \gamma_1+2a \gamma_2+\gamma_3\geq 0\,,\\
 P_{\zeta\zeta\gamma} &\equiv (\zeta_3\,\zeta_6-\zeta_5^2)\,\gamma_1
                           +2\,(\zeta_4\,\zeta_5-\zeta_2\,\zeta_6)\,\gamma_2
                           +(\zeta_1\,\zeta_6-\zeta_4^2)\,\gamma_3 \geq 0 \,,
\end{align}
which becomes $\gamma_1/(\zeta_1\,T)$ 
when the parameters are taken as in Sec.\ \ref{Israel_model}.
Note that complex parameters $\tau_\pm$ appear always 
through the combinations $\tau_{+}+\tau_{-}=2\,\mathrm{Re}\,\tau_{+}\,(\geq 0)$\,, 
$\tau_{+}\tau_{-}=|\tau_{+}|^2\,(\geq 0)$ or 
$\tau_{+}^{-1}+\tau_{-}^{-1}=2\,\mathrm{Re}\,\tau_{+}/|\tau_{+}|^2\,(\geq 0)$\,.
One can check that all the coefficients are positive, 
and thus at least the necessary condition for the stability is satisfied. 
For a full analysis to be performed, 
one should further check the Routh-Hurwitz stability criterion, 
which we have not carried out yet.

The dispersion relation around $k=0$ gives seven solutions, 
and four of the seven take the following form:
\begin{align}
 \omega = -\frac{\ii}{\tau_{\pm}} + \ord(k^2)\,, \quad
 \omega = -\frac{\ii}{\tau_\mathrm{s}} + \ord(k^2)\,,\quad
 \omega = -\frac{\ii}{\tau_{\sigma}} + \ord(k^2)\,.
\end{align}
They correspond to the relaxation modes, 
and as the observation time becomes much longer than 
the relaxation times $\mbox{Re\,}\tau_{\pm},\,\tau_\mathrm{s}$\,, and $\tau_{\sigma}$\,, 
these modes fade away in time and will not be observed eventually. 

The remaining three modes are hydrodynamic modes 
and have the following expansion in $k$\,:
\begin{align}
 \omega =
& c_\mathrm{s}\, |k| - \ii\, c_1\,k^2  
 + \Bigl(c_2 - \frac{c_1^2}{2\,c_\mathrm{s}}\Bigr)\, |k|^3 
 + \ord(k^4) \,,\label{sound_hydro_mode}\\
 \omega = 
 &-\ii\,\frac{\det{\bf A}\,\sigma_{\NS}\,T_{(0)}\,w_{(0)}}{c_\mathrm{s}^2}\, k^2
  + \ord(k^4)
\end{align}
with
\begin{align}
 c_1 = &~
 \frac{1}{2}\, \Bigl(\frac{\zeta_{\NS}}{w_{(0)}}+\frac{2(D-1)\,\eta_{\NS}}{D\,w_{(0)}}
+\frac{({\bf A}_3\,n_{(0)}+{\bf A}_2\, w_{(0)})^2\,\sigma_{\NS}\,T_{(0)}}
      {c_\mathrm{s}^2\,w_{(0)}}\Bigr) \,, 
\label{sound_c_1}\\
%%%% 
 c_2 = &~
 \Bigl(\frac{\zeta_{\NS}\,(\tau_{+}+\tau_{-}-\tau_\mathrm{b}^{-1}\tau_{+}\tau_{-})}{2w_{(0)}}
 +\frac{(D-1)\,\eta_{\NS}\,(1-r_\mathrm{s})\,\tau_\mathrm{s}}{D\,w_{(0)}}\Bigr)
  \,c_\mathrm{s} \nn\\
%%%% 
&~ \ 
+\frac{T_{(0)}\,\sigma_{\NS}\,({\bf A}_3\,n_{(0)}+{\bf A}_2\, w_{(0)})^2}
 {2c_\mathrm{s}^3\,w_{(0)}\,\tau_{\sigma}}\Bigl(c_\mathrm{s}^2
 +\frac{\zeta_{\NS}}{w_{(0)}\,\tau_{\sigma}}
 +\frac{2\,(D-1)\,\eta_{\NS}}{D\, w_{(0)}\,\tau_{\sigma}}
 -\frac{\det{\bf A}\, \sigma_{\NS}\, T_{(0)}\,w_{(0)}}{c_\mathrm{s}^2\,\tau_{\sigma}}\Bigr)
\,.
\label{sound_c_2}
\end{align}
In particular, if we neglect particle diffusions ($\cH=\sigma_{\NS}=0$), 
we have 
\begin{align}
 c_1&= \frac{\zeta_{\NS}}{2w_{(0)}}+\frac{(D-1)\,\eta_{\NS}}{D\,w_{(0)}} \,,\\
 c_2&= \Bigl(\frac{\zeta_{\NS}\,
 (\tau_{+}+\tau_{-}-\tau_\mathrm{b}^{-1}\tau_{+}\tau_{-})}{2w_{(0)}}
  +\frac{(D-1)\,\eta_{\NS}\,(1-r_\mathrm{s})\,\tau_\mathrm{s}}{D\,w_{(0)}}\Bigr)
  \,c_\mathrm{s} \,.
\end{align}
Up to $\ord(k^4)$\,, this dispersion relation coincides with that of the Israel-Stewart model 
if we identify $\tau_{+}+\tau_{-}-\tau_\mathrm{b}^{-1}\tau_{+}\tau_{-}$ 
and $(1-r_\mathrm{s})\,\tau_\mathrm{s}$ 
as the relaxation times $\tau_\Pi$ and $\tau_{\pi}$ of the Israel-Stewart model, respectively 
(see e.g., Eq.\ (47) in \cite{Romatschke:2009im}).%
\footnote{%%
In order for the correspondence to hold, 
we need to further choose the parameters 
such that $\tau_{+}+\tau_{-}-\tau_\mathrm{b}^{-1}\tau_{+}\tau_{-}$
and $(1-r_\mathrm{s})\,\tau_\mathrm{s}$ are both positive.
} %%%%%%%%%% 

%%%%%%%%%%%%%%%%%%%%%%%%%%%%%%%%%%%%%%%%%%%%%%%%%%%%%% 
\section{Conclusion and discussions}
%%%%%%%%%%%%%%%%%%%%%%%%%%%%%%%%%%%%%%%%%%%%%%%%%%%%%% 
\label{conclusion}

In this paper, we have studied the relativistic viscoelastic model \cite{FS} 
proposed recently on the basis of Onsager's linear regression theory 
on nonequilibrium thermodynamics. 
We first rederived the model using a local argument 
based on the current conservation laws 
and the positivity of entropy production rate. 
We then studied in detail the properties of the model 
and showed that our model universally reduces 
to the standard relativistic Navier-Stokes fluid mechanics  
if the observation time is sufficiently longer than the relaxation times. 

We also studied linear perturbations around a hydrostatic equilibrium 
in Minkowski spacetime. 
We showed that the wave equations for the propagation of disturbance 
become symmetric hyperbolic for some range of parameters, 
so that the model is free of acausality problems. 
This fact suggests that the relativistic viscoelastic model 
can be regarded as a causal completion of 
relativistic Navier-Stokes fluid mechanics, 
defining the latter as its long time limit.

Although the wave equations are not hyperbolic for generic values of parameters,
the problem of ill posedness 
in numerical simulations will be significantly remedied 
from the situations encountered in Navier-Stokes fluid mechanics. 
To see this, let us consider a shear mode as an example. 
As we saw in Sec.\ \ref{Shear_modes}, 
the dispersion relation in the long wavelength limit 
is given by Eq.\ \eq{shear_dispersion},
\begin{align}
 \omega = - \ii\, (\eta_{\NS}/w_{(0)})\,  k^2
 - \ii\, (\eta_{\NS}/w_{(0)})^2\,(1-r_\mathrm{s})\,\tau_\mathrm{s}\, k^4
 + \ord(k^6) 
\label{shear_omega}
\end{align}
and  has the same structure as that of the Israel-Stewart model 
up to $\ord(k^6)$ so long as $(0\leq)\,r_\mathrm{s} < 1$\,. 
This implies that, even for a parameter region 
where the wave equations are not hyperbolic, 
the behaviors at short wavelength scales 
are still remedied to an extent similar to that of the Israel-Stewart model, 
and thus the problems associated with the causality violation 
are expected to occur less likely in numerical simulations.  
It should be interesting to check this statement with a direct numerical simulation.

As discussed in Sec.\ \ref{hyperbolic_dispersion}, 
the dispersion relations for linear perturbations 
with generic parameters exhibit two kinds of branches. 
One is the ``hydrodynamic branch,'' where $\omega\to 0$ as $k\to 0$\,, 
and corresponds to the poles in retarded Green's function 
in the Kubo formula for dissipative fluid mechanics. 
If we neglect the effect of particle diffusion ($\cH=\sigma_{\NS}=0$), 
these poles in the relativistic theory of viscoelasticity 
coincide with the poles of the Israel-Stewart model 
up to $\ord(k^6)$ for shear modes and $\ord(k^4)$ for sound modes 
[see Eqs.\ \eq{shear_dispersion}, \eq{sound_c_1} and \eq{sound_c_2}] 
by identifying 
$\tau_{+}+\tau_{-}-\tau_\mathrm{b}^{-1}\tau_{+}\tau_{-}$ and 
$(1-r_\mathrm{s})\,\tau_\mathrm{s}$ 
with the relaxation times $\tau_\Pi$ and $\tau_\pi$\,, respectively, 
in the Israel-Stewart model.
In the so-called fluid/gravity correspondence \cite{Kovtun:2005ev,Baier:2007ix}, 
such poles are actually found in retarded Green's functions 
calculated at the boundary of an asymptotically AdS geometry, 
and the relaxation time is obtained to have the value $\tau_\pi=(2-\ln 2)/(2\pi T)$ 
for strongly coupled $\mathcal{N}=4$ Super Yang-Mills theory. 
This suggests that we should set 
$(1-r_\mathrm{s})\,\tau_\mathrm{s}=(2-\ln 2)/(2\pi T)$ 
if we want to establish a mapping between 
the fluids described by strongly-coupled Yang-Mills theory 
and those described by our viscoelastic model.

The other branch (``rheological branch'') gives a behavior 
that $\omega$ converges to a nonvanishing, pure imaginary value,  
$\omega \to -\ii/\tau_\mathrm{s}+\ord(k^2)$\,, as $k\to 0$\,,
and thus corresponds to the relaxation of strains. 
These relaxation poles are usually discarded in the discussion of viscous fluids, 
because the observation time for fluids is much longer than the relaxation times 
and the relaxation modes disappear at such time scales. 
However, if such poles can also be found in retarded Green's function 
at the boundary theory, 
then the fluid/gravity correspondence 
may be understood within a more general framework of 
the ``viscoelasticity/gravity correspondence.''% 
\footnote{%%
To establish this, one first would need to investigate 
whether the parameters $r_\mathrm{s}$ and $\tau_\mathrm{s}$ can be 
obtained consistently for sound and shear modes. 
} %%%%%%%%%%
It would be interesting to pursue the study in this direction. 
It should also be interesting to investigate 
the viscoelasticity/gravity correspondence 
along the line of the recent study 
relating the solutions of the Navier-Stokes equations 
to those of the Einstein equations \cite{Bredberg:2011jq,Compere:2011dx}.

As other future directions to be pursued, 
it should be important to extend the model 
such that one can treat more complicated systems 
like multicomponent viscoelastic materials. 
Such extension is actually straightforward and is under investigation. 
Another interesting direction is to extract 
the transport coefficients from kinetic theory 
or to extend the theory such as to include higher-derivative corrections.

%%%%%%%%%%%%%%%%%%%%%%%%%%%%%%%%%%%%%%%%%%%%%%%%%%%%%% 
\section*{Acknowledgments}
%%%%%%%%%%%%%%%%%%%%%%%%%%%%%%%%%%%%%%%%%%%%%%%%%%%%%% 

The authors thank Tatsuo Azeyanagi, Hikaru Kawai, Teiji Kunihiro, 
Shin-ichi Sasa and Kentaroh Yoshida for useful discussions. 
This work was supported by the Grant-in-Aid for the Global COE program 
``The Next Generation of Physics, Spun from Universality and
Emergence" from the Ministry of Education, Culture, Sports, 
Science and Technology (MEXT) of Japan. 
This work was also supported by the Japan Society for the Promotion of Science 
(JSPS) (Grant No.\,21$\cdot$1105) and by MEXT (Grant No.\,19540288).

\appendix

%%%%%%%%%%%%%%%%%%%%%%%%%%%%%%%%%%%%%%%%%%%%%%%%%%%%%% 
\section{Entropic formulation of relativistic viscoelastic fluid mechanics}
%%%%%%%%%%%%%%%%%%%%%%%%%%%%%%%%%%%%%%%%%%%%%%%%%%%%%% 
\label{appendix:Review}

In this Appendix we give a brief review on how the fundamental equations 
[Eqs.\ \eq{conservation1}--\eq{rheology3}] are obtained 
from the relativistic theory of viscoelasticity \cite{FS}
constructed on the basis of Onsager's linear regression theory 
\cite{Onsager:I,Onsager:II,Casimir,LL_stat}. 
We use the same geometrical setup and the same definition of viscoelastic materials 
as those given in Sec.\ \ref{Definitions}.
See \cite{FS} for a more detailed description.

We assume that the local thermodynamic properties 
of the material particle at $x$ (already in its local equilibrium) 
are specified by the set of local quantities $\bigl(b^A(x),\,c^I(x),\,d^P(x)\bigr)$\,. 
Here $c^I(x)$ denote the densities of the existing additive conserved quantities $C^I$\,. 
$b^A(x)$ denote the ``intrinsic'' intensive variables 
possessed by each material particle (such as strains), 
and $d^P(x)$ denote the remaining ``external'' intensive variables 
which further need to be introduced 
to characterize each subsystem thermodynamically 
(such as the background electromagnetic or gravitational fields). 
We distinguish density quantities from other intensive quantities, 
and by multiplying them with the spatial volume element $\sqrt{h}$\,, 
we construct new quantities which are spatial densities on each timeslice.
For example, the entropy density $s$ and the densities $c^I$ of conserved charges 
are density quantities, 
and for them we construct the following spatial densities: 
$\ts\equiv \sqrt{h}\,s$\,, $\tc^I\equiv \sqrt{h}\,c^I$\,.
The local equilibrium hypothesis implies that 
the local entropy $\ts(x)$ is already maximized 
at each spacetime point $x$ 
and is given as a function of the above local variables; 
$\ts(x)=\ts\bigl(b^A(x),\,\tc^I(x),\,d^P(x)\bigr)$\,.
If we denote by $(\epsilon_{\mathrm{s}},\,\epsilon_{\mathrm{t}})$ 
the spacetime scale where the local equilibrium is realized,
then at spacetime scales larger than $(\epsilon_{\mathrm{s}},\,\epsilon_{\mathrm{t}})$\,, 
we need to take into account the effect 
that the material particles communicate with each other 
by exchanging conserved quantities 
(such as energy-momentum and particle number). 
The second law of thermodynamics tells us that, 
if boundary effects can be neglected, 
this should proceed such that the total entropy of the larger region gets increased. 

\begin{figure}[htbp]
\begin{quote}
\begin{center}
\includegraphics[scale=0.55]{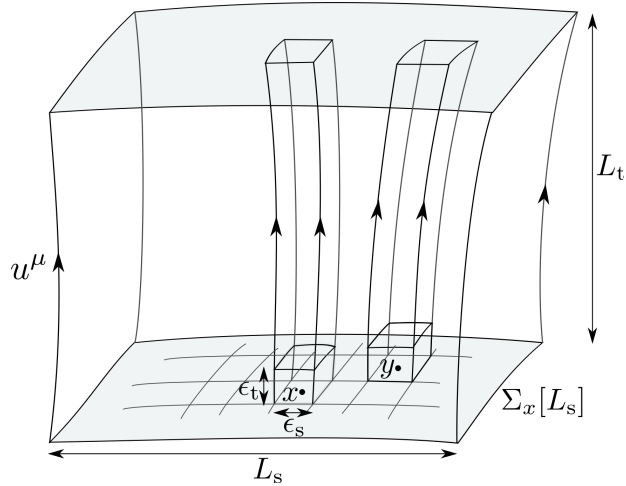}
\caption{Time evolution of material particles 
in the large region $\Sigma_x[L_{\mathrm{s}}]$ \cite{FS}.
\label{large_region}}
\end{center}
\end{quote}
\vspace{-2ex}
\end{figure}

In order to describe such dynamics mathematically, 
we first introduce the spacetime scale $(L_{\mathrm{s}},\,L_{\mathrm{t}})$ 
which is much larger than the spacetime scale 
$(\epsilon_{\mathrm{s}},\,\epsilon_{\mathrm{t}})$ 
and assign to each spacetime point $x=(x^0=t,\bx)$ on timeslice $\Sigma_t$ 
a spatial region $\Sigma_x[L_{\mathrm{s}}]$ of linear size $L_{\mathrm{s}}$ 
(see Fig.\ \ref{large_region}). 
%%% 
%%% 
We then consider the total entropy of the region $\Sigma_x[L_{\mathrm{s}}]$\,: 
\begin{align}
 \Stot(t;\,\Sigma_x[L_{\mathrm{s}}]) 
   \equiv 
   \int_{\Sigma_x[L_{\mathrm{s}}]} \rmd^D\by\, 
             \ts\bigl(b^A(t,\by),\,\tc^I(t,\by),\,d^P(t,\by)\bigr) \,. 
\end{align}
The irreversible evolutions of intrinsic variables 
$a^r(x)\equiv \bigl(b^A(x),\,\tc^I(x)\bigr)$ at $x$ 
will proceed toward an equilibrium of the region $\Sigma_x[L_{\mathrm{s}}]$\,. 
Due to the condition $L_{\mathrm{s}} \gg \epsilon_{\mathrm{s}}$\,, 
we can assume that the influence from the surroundings of 
the region $\Sigma_x[L_{\mathrm{s}}]$ is not relevant 
to the dynamics of $a^r(x)$ because $x$ is well inside the region. 
An equilibrium state of the region $\Sigma_x[L_{\mathrm{s}}]$ 
will be realized when the observation is made 
for a long period of time, $L_{\mathrm{t}}$\,, 
and can be characterized by the condition
\begin{align}
 \frac{\delta \Stot(t;\,\Sigma_x[L_{\mathrm{s}}])}{\delta a^r(x)}=0\,. 
\label{equilib_value}
\end{align}
Note that the functional derivative is taken 
only with respect to a spatial, $D$-dimensional unit in the functional. 
We denote the values of $a^r(x)$ at the equilibrium by 
$a^r_0(x;\,L_{\mathrm{s}}) 
 \equiv \bigl(b^A_0(x;\,L_{\mathrm{s}}),\,\tc^I_0(x;\,L_{\mathrm{s}})\bigr)$\,. 
One should note here that, 
since $\tc^I(t,\by)$ are conserved quantities, 
the variations \eq{equilib_value} with respect to $\tc^I$-type variables 
should be taken with total charges kept fixed at prescribed values: 
\begin{align}
 \int_{\Sigma_x[L_{\mathrm{s}}]}\rmd^D\by \,\tc^I(t,\by) 
  \equiv C^I\bigl(\Sigma_x[L_{\mathrm{s}}]\bigr) \,. 
\label{conservation_law}
\end{align}
A simple analysis using the Lagrange multipliers shows 
that the condition of global equilibrium is expressed locally as 
\begin{align}
 \frac{\partial \ts}{\partial b^A}(x)=0 \qquad \text{and}\qquad 
  h_\mu^{~\nu}(x)\,\nabla_\nu \beta_I(x) = 0 \,,
\label{equilibrium_condition}
\end{align}
where $\beta_I$ is the thermodynamic variable conjugate to $\tc^I$ 
that is defined by
\begin{align}
 \beta_I(x) \equiv \frac{\partial \ts}{\partial \tc^I}(x) \,.
\end{align}

The total entropy of the region $\Sigma_x[L_{\mathrm{s}}]$ at an equilibrium 
is given by
\begin{align}
 \Stot_0(t;\,\Sigma_x[L_{\mathrm{s}}]) 
   \equiv 
   \int_{\Sigma^0_x[L_{\mathrm{s}}]} \rmd^D\by\, 
             \ts\bigl(b^A_0(t,\by),\,\tc^I_0(t,\by),\,d^P(t,\by)\bigr) \,, 
\end{align}
where $\Sigma^0_x[L_{\mathrm{s}}]$ is a hypersurface orthogonal to 
the velocity field at the equilibrium, $u_0^\mu \equiv p^\mu_0/e_0$\,. 
When the material can be regarded as being at an equilibrium at spatial infinity,
we can fix the labeling $s$ of the new timeslices $\{\Sigma^0_{s}\}$ at the equilibrium
with the labeling $t$ of the original timeslices $\{\Sigma_t\}$ 
by setting $s=t$ if $\Sigma^0_s$ conforms with $\Sigma_t$ at spatial infinity. 
If we denote coordinates corresponding to the new foliation $\{\Sigma^0_t\}$ 
by $(x^{\prime\,\mu})=(x^{\prime\,0}=t\,,\,x^{\prime\,i})$, 
then the velocity field $u_0=u_0^\mu\,\partial^{\,\prime}_\mu$ 
will be expressed in the following form:
\begin{align}
 u_0 = \frac{1}{N_0}\,\partial_t + \frac{1}{N^i_0}\,\partial^{\,\prime}_i\,.
\label{new_lapse}
\end{align}
This expression defines the new lapse $N_0$ and the new shifts $N_0^i$ 
at the equilibrium that are realized at spacetime scale $(L_{\rm s},L_{\rm t})$.
For configurations other than the equilibrium, 
the total entropy $\Stot(t;\,\Sigma_x[L_{\mathrm{s}}])$ 
is smaller than that of the equilibrium $\Stot_0(t;\,\Sigma_x[L_{\mathrm{s}}])$\,, 
so that if we denote their difference by
\begin{align}
 \Delta\Stot(t;\,\Sigma_x[L_{\mathrm{s}}])
  \equiv \Stot(t;\,\Sigma_x[L_{\mathrm{s}}])
  -\Stot_0(t;\,\Sigma_x[L_{\mathrm{s}}])\,,
\end{align}
$\Delta \Stot$ is always nonpositive.

In the previous paper \cite{FS}, 
it is proposed that the difference $\Delta\Stot$ can be effectively written in 
the following form at the lowest order in the derivative expansion 
for linear nonequilibrium thermodynamics:
\begin{align}
 &\Delta\Stot(t;\,\Sigma_x[L_{\mathrm{s}}])\nn\\
 &= - \,\frac{1}{2} \,\int_{\Sigma^0_x[L_{\mathrm{s}}]} \rmd^D \by \,N_0^{-1}\,\sqrt{-g}\,
 \begin{pmatrix} (b-b_0)^A & 
  \nabla_\mu \beta_I
 \end{pmatrix} 
  \begin{pmatrix}
  \ell_{AB}  & \ell_A^{\,\nu J}\cr
  \ell_B^{\,\mu I} & \ell^{\,\mu I,\, \nu J}
  \end{pmatrix}
 \begin{pmatrix} (b-b_0)^B \cr 
  \nabla_\nu \beta_J
 \end{pmatrix} \,.
\label{total_entropy}
\end{align}
Here the scalar function $N_0$ is the lapse at the equilibrium 
defined in Eq.\ \eq{new_lapse},
the coefficient $\Bigl(\begin{smallmatrix}
  \ell_{AB}  & \ell_A^{\,\nu J}\cr
  \ell_B^{\,\mu I} & \ell^{\,\mu I,\, \nu J}
  \end{smallmatrix}\Bigr)$ 
is a symmetric, positive semidefinite matrix, 
and all the elements are spatial tensors, 
$\ell_A^{\,\mu I}\,u_\mu = 0 = \ell^{\,\mu I,\, \nu J}\,u_\nu$\,. 
The integral region can be replaced by $\Sigma_x[L_{\mathrm{s}}]$ 
because the difference is of higher orders in the derivative expansion. 
See the appendix in \cite{FS} for a derivation of \eq{total_entropy} for simple cases. 
The functional form of the total entropy,
$\Stot(t;\,\Sigma_x[L_{\mathrm{s}}])
 =\Stot_0(t;\,\Sigma_x[L_{\mathrm{s}}])+ \Delta\Stot(t;\,\Sigma_x[L_{\mathrm{s}}])$\,, 
is called the \textit{entropy functional} in \cite{FS}.

We now consider Onsager's linear regression theory \cite{Onsager:I,Onsager:II,Casimir,LL_stat}
assuming that the total entropy is given with this entropy functional. 
In Onsager's treatment the irreversible evolutions of 
thermodynamic variables $a^r(x)$ are given by
\begin{align}
 [\dot{a}^r(x)]_{\mathrm{irr}} = L^{rs}\,f_s(x)\,.
\end{align}
Here $f_s(x)$ is the thermodynamic force defined by
\begin{align}
 f_s(x) = \frac{\delta\,\Delta\Stot(t;\,\Sigma_x[L_{\mathrm{s}}])}{\delta a^s(x)} \,,
\end{align}
and in the relativistic nonlinear thermodynamics, 
the dot should be defined as $\dot{a}^r\equiv N\,\Lie_u a^r$ \cite{FS}, 
where $\Lie_u$ is the Lie derivative 
with respect to the velocity $u=u^\mu(x)\,\partial_\mu$\,.
$L^{rs}$ are the so-called phenomenological coefficients 
and can be shown to satisfy 
Onsager's reciprocal relation \cite{Onsager:I,Onsager:II,Casimir}
\begin{align}
 L^{rs} = (-1)^{|a^r| + |a^s|}\,L^{sr}\,,
\end{align}
where the index $|a^r|$ expresses how the variables transform 
under time reversal, 
$a^r(x)\to (-1)^{|a^r|}\,a^r(x)$\,.%
\footnote{%%% 
When the background fields $d^P$ change as $d^P\to d^P_{\,\mathrm{T}}$ under time reversal, 
the reciprocal relation is expressed as
$L^{rs}(d^P) = (-1)^{|a^r| + |a^s|}\,L^{sr}(d^P_{\,\mathrm{T}})$\,.
}  %%%%%%%%%% 
The Curie principle says that $L^{rs}$ can be block diagonalized 
with respect to the transformation properties of the indices $(r,s)$ 
under spatial rotations and the parity transformation \cite{de_Groot-Mazur}, 
that is, under the subgroup $\mathrm{O}(D)$ of the local Lorentz group $\mathrm{O}(D,1)$ 
in local inertial frames. 
For example, when $a^r$ constitute a contravariant vector, $(a^r)\equiv(a^\mu)$\,, 
the equations of linear regression 
should be set for each of the normal and tangential components 
to the timeslice through $x$\,:
\begin{align}
  [\dot{a}(x)]_{{\mathrm{irr}}\,\bot}^{\mu}(x) 
  &= L_\bot^{\mu\nu}\,\biggl[\frac{\delta\,\Delta\Stot}
  {\delta a^\nu(x)}\biggr]_{\mbox{\raisebox{4pt}{\scriptsize{$\bot$}}}}\,,\\
 [\dot{a}(x)]_{{\mathrm{irr}}\,\|}^{\mu}(x) 
  &= L_\|^{\mu\nu}\,\biggl[\frac{\delta\,\Delta\Stot}
  {\delta a^\nu(x)}\biggr]_{\mbox{\raisebox{4pt}{\scriptsize{$\|$}}}}\,,
\end{align}
where for a contravariant vector $v^\mu$ 
we define $v_\bot^\mu\equiv (-u^\mu u_\nu)\,v^\nu$ 
and $v_\|^\mu\equiv h^\mu_{~\nu}\,v^\nu$ 
(and similarly for covariant vectors).
Covariance and positivity further impose the condition 
that $L_\bot^{\mu\nu}$ and $L_\|^{\mu\nu}$ should be expressed as
$L_\bot^{\mu\nu} = L_\bot\,u^\mu u^\nu$ $(L_\bot>0)$ and 
$L_\|^{\mu\nu} = L_\|\,h^{\mu\nu}$ $(L_\|>0)$\,, respectively.

If we further know the reversible evolutions of thermodynamic variables, 
$[\dot{a}^r(x)]_{\mathrm{rev}}$\,, 
which are not accompanied by entropy productions, 
then the dynamics of the system can be determined as
\begin{align}
 \dot{a}^r(x) = [\dot{a}^r(x)]_{\mathrm{rev}} + [\dot{a}^r(x)]_{\mathrm{irr}}
              = [\dot{a}^r(x)]_{\mathrm{rev}} 
                + L^{rs}\,\frac{\delta\,\Delta\Stot(t;\,\Sigma_x[L_{\mathrm{s}}])}
                               {\delta a^s(x)} \,.
\end{align}

For viscoelastic materials, 
the relevant thermodynamic variables are the following:
\begin{center}
\begin{tabular}{|c|c||c|c||c|c|}
\hline
 $b^A$ & $b^A_0$ & 
 $\tc^I$  & $\beta_I$ & $d^P$ & $(\partial\ts/\partial d^P)$ \\ \hline\hline
%%%%%%%%%%%%%% 
 $\varepsilon_{\mu\nu}$& $0$ & 
 $\tp_\mu$ & $ - u^\mu/T$ & $g_{\mu\nu}$ & $\sqrt{h}\,T_{\mathrm{(q)}}^{\mu\nu}/2T$ \\
%%%%%%%%%%%%%% 
 $\varepsilon_\mu$& $0$ & 
 $\tn$ & $-\mu/T$ & & \\
%%%%%%%%%%%%%% 
 $\theta$& $0$ & & & & \\
\hline
\end{tabular}
\end{center}
where $T^{\mu\nu}_{\mathrm{(q)}}\equiv e\,u^\mu\,u^\nu+\tau_{\mathrm{(q)}}^{\mu\nu}$
is the quasiconservative energy-momentum tensor  
with $\tau_{\mathrm{(q)}}^{\mu\nu}$ the quasiconservative stress tensor.
The entropy functional is then written as
\begin{align}
 \,\Delta\Stot&(t;\,\Sigma_x[L_{\mathrm{s}}])\nn\\
 &= -\frac{1}{2}\int_{\Sigma^0_x[L_{\mathrm{s}}]}\!\!\!\rmd^D\by\,
 \sqrt{-g}\,N_0^{-1}\,\times\nn\\
&\qquad\qquad\times \Biggl[
 \begin{pmatrix}
  \varepsilon_{\langle\mu\nu\rangle} & 
  \nabla^{\langle \mu} \bigl(\partial\ts/\partial \tp_{\nu\rangle}\bigr)
 \end{pmatrix} 
  \begin{pmatrix}
  \ell_1^{\langle\mu\nu\rangle,\langle\rho\sigma\rangle} 
 &\ell^{\langle\mu\nu\rangle,}_{2~~~\langle\rho\sigma\rangle} \cr
  \ell_{2\,\langle\mu\nu\rangle,}^{~~~~~\langle\rho\sigma\rangle}
 & \ell_{3\,\langle\mu\nu\rangle,\langle\rho\sigma\rangle}
  \end{pmatrix}
 \begin{pmatrix}
  \varepsilon_{\langle\rho\sigma\rangle} \cr 
  \nabla^{\langle \rho} \bigl(\partial\ts/\partial \tp_{\sigma\rangle}\bigr)
 \end{pmatrix} \nn\\
 &\qquad\qquad\qquad 
 +\begin{pmatrix}
  \varepsilon_{\mu} & 
  \partial_\mu \bigl(\partial\ts/\partial\tn\bigr)
 \end{pmatrix} 
  \begin{pmatrix}
  \ell_1^{\mu\nu} & \ell_2^{\mu\nu}\cr
  \ell_2^{\mu\nu} & \ell_3^{\mu\nu}
  \end{pmatrix}
 \begin{pmatrix}
  \varepsilon_{\nu} \cr 
  \partial_\nu \bigl(\partial\ts/\partial\tn\bigr)
 \end{pmatrix} \nn\\
 &\qquad\qquad\qquad 
 +\begin{pmatrix}
    \tr \varepsilon & \theta & \nabla_\mu\bigl(\partial\ts/\partial\tp_\mu\bigr)
 \end{pmatrix} 
  \begin{pmatrix}
  \hat{\ell}^{\mathrm{s}}_1 & \hat{\ell}^{\mathrm{s}}_2 & \hat{\ell}^{\mathrm{s}}_4 \cr
  \hat{\ell}^{\mathrm{s}}_2 & \hat{\ell}^{\mathrm{s}}_3 & \hat{\ell}^{\mathrm{s}}_5 \cr
  \hat{\ell}^{\mathrm{s}}_4 & \hat{\ell}^{\mathrm{s}}_5 & \hat{\ell}^{\mathrm{s}}_6
  \end{pmatrix}
 \begin{pmatrix}
    \tr \varepsilon \cr \theta \cr \nabla_\mu\bigl(\partial\ts/\partial\tp_\mu\bigr)
 \end{pmatrix} \Biggr]  \,,\nn\\
 &= -\frac{1}{2}\int_{\Sigma^0_x[L_{\mathrm{s}}]}\!\!\!\rmd^D\by\,
 \sqrt{-g}\,N_0^{-1}\,\times\nn\\
&\qquad\qquad\times \Biggl[
 \begin{pmatrix}
  \varepsilon_{\langle\mu\nu\rangle} & 
  (-1/T)\,K_{\langle\mu\nu\rangle}
 \end{pmatrix} 
  \begin{pmatrix}
  \ell_1^{\langle\mu\nu\rangle,\langle\rho\sigma\rangle} 
 &\ell^{\langle\mu\nu\rangle,}_{2~~~\langle\rho\sigma\rangle} \cr
  \ell_{2\,\langle\mu\nu\rangle,}^{~~~~~\langle\rho\sigma\rangle}
 & \ell_{3\,\langle\mu\nu\rangle,\langle\rho\sigma\rangle}
  \end{pmatrix}
 \begin{pmatrix}
  \varepsilon_{\langle\rho\sigma\rangle} \cr 
  (-1/T)\,K_{\langle\rho\sigma\rangle}
 \end{pmatrix} \nn\\
 &\qquad\qquad\qquad 
 +\begin{pmatrix}
  \varepsilon_{\mu} & 
  \partial_\mu (-\mu/T)
 \end{pmatrix} 
  \begin{pmatrix}
  \ell_1^{\mu\nu} & \ell_2^{\mu\nu}\cr
  \ell_2^{\mu\nu} & \ell_3^{\mu\nu}
  \end{pmatrix}
 \begin{pmatrix}
  \varepsilon_{\nu} \cr 
  \partial_\nu (-\mu/T)
 \end{pmatrix} \nn\\
 &\qquad\qquad\qquad 
 +\begin{pmatrix}
    \tr \varepsilon & \theta & (-1/T)\,\tr K
 \end{pmatrix} 
  \begin{pmatrix}
  \hat{\ell}^{\mathrm{s}}_1 & \hat{\ell}^{\mathrm{s}}_2 & \hat{\ell}^{\mathrm{s}}_4 \cr
  \hat{\ell}^{\mathrm{s}}_2 & \hat{\ell}^{\mathrm{s}}_3 & \hat{\ell}^{\mathrm{s}}_5 \cr
  \hat{\ell}^{\mathrm{s}}_4 & \hat{\ell}^{\mathrm{s}}_5 & \hat{\ell}^{\mathrm{s}}_6
  \end{pmatrix}
 \begin{pmatrix}
    \tr \varepsilon \cr \theta \cr (-1/T)\,\tr K
 \end{pmatrix} \Biggr]  \,,
\end{align}
where the coefficient matrices are symmetric and positive semidefinite, 
and their indices are all orthogonal to $u^\mu$\,.%
\footnote{%%
We actually need to impose the latter condition in the Landau-Lifshitz frame.
One can show that if this condition is relaxed, 
the energy-momentum tensor comes to have terms related to a heat flux, 
which should not appear in the Landau-Lifshitz frame.
} %%%%%%%%%% 
Note that for this parametrization, the contributions from 
the rotation $\nabla_{[\mu}u_{\nu]}$ are discarded.
Since the matrices must be invariant tensors, 
we can assume that they take the following form:%%%%%%
\footnote{%%%%%%
Note that $\ell_k^{\langle\mu\nu\rangle,\langle\rho\sigma\rangle}\,
\varepsilon_{\langle\rho\sigma\rangle}
 = 2\, \ell^\mathrm{t}_k\,\varepsilon^{\langle\mu\nu\rangle}$ $(k=1,2,3)$\,.
} %%%%%%
\begin{align}
  \begin{pmatrix}
  \ell_1^{\langle\mu\nu\rangle,\langle\rho\sigma\rangle} 
 &\ell_2^{\langle\mu\nu\rangle,\langle\rho\sigma\rangle}\cr
  \ell_2^{\langle\mu\nu\rangle,\langle\rho\sigma\rangle} 
 & \ell_3^{\langle\mu\nu\rangle,\langle\rho\sigma\rangle}
  \end{pmatrix}
 &= 2\,
  \begin{pmatrix}
  \ell_1^\mathrm{t} 
 &\ell_2^\mathrm{t} \cr
  \ell_2^\mathrm{t} 
 & \ell_3^\mathrm{t} 
  \end{pmatrix}
  \,N_0\,h^{\langle\mu}_{\mu'}\, h^{\nu\rangle}_{\nu'}\, h^{\mu'\rho}\,h^{\nu'\sigma} \,, \\ 
  \begin{pmatrix}
  \ell_1^{\mu\nu} & \ell_2^{\mu\nu}\cr
  \ell_2^{\mu\nu} & \ell_3^{\mu\nu}
  \end{pmatrix}
 &=
    \begin{pmatrix}
  \ell_1^\mathrm{v} & \ell_2^\mathrm{v}\cr
  \ell_2^\mathrm{v} & \ell_3^\mathrm{v}
  \end{pmatrix} \,N_0\, h^{\mu\nu}\,,\\
  \begin{pmatrix}
  \hat{\ell}^{\mathrm{s}}_1 & \hat{\ell}^{\mathrm{s}}_2 & \hat{\ell}^{\mathrm{s}}_4 \cr
  \hat{\ell}^{\mathrm{s}}_2 & \hat{\ell}^{\mathrm{s}}_3 & \hat{\ell}^{\mathrm{s}}_5 \cr
  \hat{\ell}^{\mathrm{s}}_4 & \hat{\ell}^{\mathrm{s}}_5 & \hat{\ell}^{\mathrm{s}}_6
  \end{pmatrix}
 &= 
  \begin{pmatrix}
  \ell^{\mathrm{s}}_1 & \ell^{\mathrm{s}}_2 & \ell^{\mathrm{s}}_4 \cr
  \ell^{\mathrm{s}}_2 & \ell^{\mathrm{s}}_3 & \ell^{\mathrm{s}}_5 \cr
  \ell^{\mathrm{s}}_4 & \ell^{\mathrm{s}}_5 & \ell^{\mathrm{s}}_6
  \end{pmatrix}\,N_0
\,,
\end{align}
where 
$\Bigl(\begin{smallmatrix}
  \ell_1^\mathrm{t} 
 &\ell_2^\mathrm{t} \cr
  \ell_2^\mathrm{t} 
 & \ell_3^\mathrm{t} 
\end{smallmatrix}\Bigr)$\,, 
$\Bigl(\begin{smallmatrix}
  \ell_1^\mathrm{v} & \ell_2^\mathrm{v}\cr
  \ell_2^\mathrm{v} & \ell_3^\mathrm{v}
\end{smallmatrix}\Bigr)$
and 
$\Bigl(\begin{smallmatrix}
  \ell^{\mathrm{s}}_1 & \ell^{\mathrm{s}}_2 & \ell^{\mathrm{s}}_4 \cr
  \ell^{\mathrm{s}}_2 & \ell^{\mathrm{s}}_3 & \ell^{\mathrm{s}}_5 \cr
  \ell^{\mathrm{s}}_4 & \ell^{\mathrm{s}}_5 & \ell^{\mathrm{s}}_6
\end{smallmatrix}\Bigr)$
are positive semidefinite.
The irreversible evolutions of thermodynamic variables then become
\begin{align}
 \bigl[\dot\varepsilon_{\langle\mu\nu\rangle}\bigr]_\mathrm{irr}
  &\equiv \frac{1}{\sqrt{h}}\,
  L^{\varepsilon_{\langle\mu\nu\rangle}\varepsilon_{\langle\rho\sigma\rangle}}\,
  \frac{\delta\Delta\Stot}{\delta\varepsilon_{\rho\sigma}}\,,
\label{eq_varepsilon_munu}\\
 \bigl[\dot\varepsilon_{\mu}\bigr]_{\mathrm{irr}\,\bot}
  &\equiv \frac{1}{\sqrt{h}}\,
  L_\bot^{\varepsilon_\mu \varepsilon_\nu}\,
  \biggl[\frac{\delta\Delta\Stot}{\delta\varepsilon_\nu}
  \biggr]_{\mbox{\raisebox{3pt}{\scriptsize{$\bot$}}}} \equiv 0\,,
\label{eq_varepsilon_mu_perp}\\
 \bigl[\dot\varepsilon_{\mu}\bigr]_{\mathrm{irr}\,\|}
  &\equiv \frac{1}{\sqrt{h}}\,
  L_\|^{\varepsilon_\mu \varepsilon_\nu}\,
  \biggl[\frac{\delta\Delta\Stot}{\delta\varepsilon_\nu}
  \biggr]_{\mbox{\raisebox{3pt}{\scriptsize{$\|$}}}}\,,
\label{eq_varepsilon_mu_para}\\
 \begin{pmatrix}
  \bigl[(\tr\varepsilon)^\cdot\bigr]_\mathrm{irr} \cr
  \bigl[\,\dot\theta\,\bigr]_\mathrm{irr}
  \end{pmatrix}
  &\equiv \frac{1}{\sqrt{h}}\,
  \begin{pmatrix}
   L^{\tr\varepsilon\,\tr\varepsilon} & L^{\tr\varepsilon\,\theta}\cr
   L^{\tr\varepsilon\,\theta} & L^{\theta\,\theta}\
  \end{pmatrix}
  \begin{pmatrix}
   \delta\Delta\Stot/\delta (\tr\varepsilon) \cr
   \delta\Delta\Stot/\delta \theta
  \end{pmatrix}\,,
\label{eq_scalar}\\
 \bigl[\,\dot\tp_{\mu}\bigr]_{\mathrm{irr}\,\bot}
  &\equiv 
  \sqrt{h}\,L_\bot^{\tp_\mu \tp_\nu}\,
  \biggl[\frac{\delta\Delta\Stot}{\delta\tp_\nu}
  \biggr]_{\mbox{\raisebox{3pt}{\scriptsize{$\bot$}}}}\,,
\label{eq_tp_mu_perp}\\
 \bigl[\,\dot\tp_{\mu}\bigr]_{\mathrm{irr}\,\|}
  &\equiv 
  \sqrt{h}\,L_\|^{\tp_\mu \tp_\nu}\,
  \biggl[\frac{\delta\Delta\Stot}{\delta\tp_\nu}
  \biggr]_{\mbox{\raisebox{3pt}{\scriptsize{$\|$}}}}\,,
\label{eq_tp_mu_para}\\
 [\dot\tn]_\mathrm{irr} 
  &\equiv \sqrt{h}\,L^{\tn \tn}\, \frac{\delta\Delta\Stot}{\delta \tn}\,.
\label{eq_tn}
\end{align}
We now make the following irreducible decompositions of the phenomenological constants
under the group $\mathrm{O}(D)$ in a local inertial frame:
\begin{align}
   L^{\varepsilon_{\langle\mu\nu\rangle}\varepsilon_{\langle\rho\sigma\rangle}} 
   &\equiv  L^\mathrm{t}\,
  h_{\langle\mu}^{\mu'} h_{\nu\rangle}^{\nu'} h_{\mu'\rho}h_{\nu'\sigma} \,,\qquad 
 L_\|^{\varepsilon_{\mu}\varepsilon_{\nu}}  \equiv L^\mathrm{v}\,h_{\mu\nu} \,,\\
  L^{\tr\varepsilon\,\tr\varepsilon} &\equiv L^\mathrm{s}_1\,, \qquad
 L^{\tr\varepsilon\,\theta}  \equiv L^\mathrm{s}_2\,, \qquad 
 L^{\theta\,\theta}  \equiv L^\mathrm{s}_3\,, \\
  L_\bot^{\tp_\mu \tp_\nu} &\equiv L_\bot\,u_\mu u_\nu\,,\quad
 L_\|^{\tp_\mu \tp_\nu} \equiv L_\|\,h_{\mu\nu}\,,\quad
 L^{\tn\tn} \equiv M\,.
\end{align}
Then the irreversible evolutions of thermodynamic variables can be written as \cite{FS}
\begin{align}
 N^{-1}[\dot{\varepsilon}_{\langle\mu\nu\rangle}]_\mathrm{irr}
   &= - 2L^\mathrm{t}\ell_1^\mathrm{t}\,\varepsilon_{\langle\mu\nu\rangle} 
   +(2L^\mathrm{t}\ell_2^\mathrm{t}/T)\,K_{\langle\mu\nu\rangle}  \,, 
\label{app_rheology1_irr}\\
 N^{-1} \bigl[\dot{\varepsilon}_\mu\bigr]_{\mathrm{irr}} 
  &= - L^\mathrm{v}\, h_\mu^{~\nu}\,
   \bigl[\ell_1^\mathrm{v}\,\varepsilon_{\nu}
          +\ell_2^\mathrm{v}\, \partial_\nu (-\mu/T)\bigr] \,, 
\label{app_rheology2_irr}\\
 N^{-1}\bigl[(\tr\varepsilon)^\cdot\bigr]_\mathrm{irr} 
   &=
  - (L^\mathrm{s}_1\,\ell^\mathrm{s}_1 + L^\mathrm{s}_2\,\ell^\mathrm{s}_2)\,\tr\varepsilon
  - (L^\mathrm{s}_1\,\ell^\mathrm{s}_2 + L^\mathrm{s}_2\,\ell^\mathrm{s}_3)\,\theta
  + (L^\mathrm{s}_1\,\ell^\mathrm{s}_4 + L^\mathrm{s}_2\,\ell^\mathrm{s}_5)\,
  \frac{1}{T}\tr K\,,
\label{app_rheology3_irr}\\
 N^{-1} \bigl[\,\dot{\theta}\,\bigr]_\mathrm{irr} 
  &= -\, (L^\mathrm{s}_2\,\ell^\mathrm{s}_1
  + L^\mathrm{s}_3\,\ell^\mathrm{s}_2)\,\tr\varepsilon
  - (L^\mathrm{s}_2\,\ell^\mathrm{s}_2 + L^\mathrm{s}_3\,\ell^\mathrm{s}_3)\,\theta
  + (L^\mathrm{s}_2\,\ell^\mathrm{s}_4 + L^\mathrm{s}_3\,\ell^\mathrm{s}_5)\,
  \frac{1}{T}\tr K\,,
\label{app_rheology4_irr}\\
 \frac{1}{\sqrt{-g}}\,\bigl[\,\dot\tp_{\nu}\bigr]_{\mathrm{irr}\,\bot}
  &= -\,c_\bot\,L_\bot\,(-u_\nu u_\lambda)\,\nabla_\mu\,
  \bigl[\, 2\,\ell^\mathrm{t}_2\,\varepsilon^{\langle\mu\lambda\rangle}
  -(2/T)\,\ell^\mathrm{t}_3\,K^{\langle\mu\lambda\rangle} \nn\\
 &\qquad\qquad\qquad\qquad\qquad\quad 
  + \bigl( \ell^\mathrm{s}_4\,\tr\varepsilon + \ell^\mathrm{s}_5\,\theta
  - (1/T)\,\ell^\mathrm{s}_6\,\tr K \bigr)\,h^{\mu\lambda}\bigr]\,,
\label{app_tp1_irr}\\
 \frac{1}{\sqrt{-g}}\,\bigl[\,\dot\tp_{\nu}\bigr]_{\mathrm{irr}\,\|}
  &= -\,c_\|\,L_\|\,h_{\nu\lambda}\,\nabla_\mu\,
  \bigl[\, 2\,\ell^\mathrm{t}_2\,\varepsilon^{\langle\mu\lambda\rangle}
  -(2/T)\,\ell^\mathrm{t}_3\,K^{\langle\mu\lambda\rangle} \nn\\
 &\qquad\qquad\qquad\qquad\qquad\quad 
  + \bigl( \ell^\mathrm{s}_4\,\tr\varepsilon + \ell^\mathrm{s}_5\,\theta
  - (1/T)\,\ell^\mathrm{s}_6\,\tr K \bigr)\,h^{\mu\lambda}\bigr]\,,
\label{app_tp2_irr}\\
 \frac{1}{\sqrt{-g}}\,[\dot\tn]_\mathrm{irr} 
 &= -\,\sqrt{h}\,(-\partial^2\ts/\partial\tn^2)\,M\,
  \nabla_\mu\, \bigl[\,\ell^\mathrm{v}_2\,h^{\mu\nu}\,\varepsilon_\nu
  + \ell^\mathrm{v}_3\,h^{\mu\nu}\,\partial_\nu(-\mu/T)\bigr]\,.
\label{app_tn_irr}
\end{align}
Here, in order to evaluate $\delta\Delta\Stot/\delta\tp_\mu$\,, 
we have used the decomposition of the matrix 
$\partial^2\ts/\partial\tp_\mu\partial\tp_\nu$ 
(negative-definite for each irreducible component) as%
\begin{align}
 \sqrt{h}\, \frac{\partial^2 \ts}{\partial\tp_\mu \partial\tp_\nu}
  = -c_{\bot}\,u^\mu u^\nu - c_{\|}\,h^{\mu\nu}
\label{s_pp-decomp}
\end{align}
with positive quantities $c_{\bot}$ and $c_{\|}$\,.

If we assume that $L_\tp\equiv c_\bot L_\bot= c_\| L_\|$ and 
$L_\tn\equiv\sqrt{h}\,(-\partial^2\ts/\partial\tn^2)\,M$ are constant, 
then Eqs.\ \eq{app_tp1_irr}--\eq{app_tn_irr} are rewritten as
\begin{align}
 [\dot{\tp}_\nu]_\mathrm{irr}=-\,\sqrt{-g}\,\nabla^\mu \tau^\mathrm{(d)}_{\mu\nu}\,,
  \qquad [\dot{\tn}]_\mathrm{irr}=-\,\sqrt{-g}\,\nabla^\mu \nu^\mathrm{(d)}_\mu\,,
\label{app_current_irr}
\end{align}
where the dissipation currents are given by
\begin{align}
 \tau^\mathrm{(d)}_{\mu\nu}
  &\equiv L_\tp\,\bigl[\, 2\,\ell^\mathrm{t}_2\,\varepsilon_{\langle\mu\nu\rangle}
  -(2/T)\,\ell^\mathrm{t}_3\,K_{\langle\mu\nu\rangle} 
  + \bigl( \ell^\mathrm{s}_4\,\tr\varepsilon + \ell^\mathrm{s}_5\,\theta
  - (1/T)\,\ell^\mathrm{s}_6\,\tr K \bigr)\,h_{\mu\nu}\bigr]\,,
\label{app_tau_d}\\
 \nu^\mathrm{(d)}_\mu
  &\equiv L_\tn\,\bigl[\,\ell^\mathrm{v}_2\,\varepsilon_\mu
  + \ell^\mathrm{v}_3\,h_\mu^{~\nu}\,\partial_\nu(-\mu/T)\bigr]\,.
\label{app_nu_d}
\end{align}

On the other hand, as for the isentropic evolutions,
we assume that the evolutions of the densities of conserved quantities 
are given by
\begin{align}
 [\dot{\tp}_\nu]_\mathrm{rev}=-\,\sqrt{-g}\,\nabla^\mu \tau^\mathrm{(r)}_{\mu\nu}\,,
  \qquad [\dot{\tn}]_\mathrm{rev}=-\,\sqrt{-g}\,\nabla^\mu \nu^\mathrm{(r)}_\mu
\label{app_current_rev}
\end{align}
with the reversible currents of the following form:%
\footnote{%%
At the end of this appendix, we comment on 
how these reversible parts are determined in the entropic formulation.
} %%%%%%%%%%
\begin{align}
 \tau^\mathrm{(r)}_{\mu\nu}
  &\equiv \tau^\mathrm{(q)}_{\mu\nu}-\,2\cG\,\varepsilon_{\langle\mu\nu\rangle} 
     -\cK\,\bigl(\tr\varepsilon-a\,\theta)\,h_{\mu\nu}\,, 
\label{app_tau_r}\\
 \nu^\mathrm{(r)}_\mu
  &\equiv -\cH\,\varepsilon_\mu\,.    
\label{app_nu_r}
\end{align}
As for the evolutions of the strains $E_{\mu\nu}=(\varepsilon_{\mu\nu}\,,\,
\varepsilon_\mu\,,\,\theta)$\,, 
we set them to be in the most generic form:
\begin{align}
 N^{-1}\bigl[\dot{\varepsilon}_{\langle\mu\nu\rangle}\bigr]_{\mathrm{rev}}
  &= (2\cG L^\mathrm{t}/L_\tp\,T)\,K_{\langle\mu\nu\rangle}\,,
\label{app_rheology1_rev}\\ 
 N^{-1}\bigl[\dot{\varepsilon}_{\mu}\bigr]_{\mathrm{rev}}
  &=-\,(L^\mathrm{v}\cH/L_\tn)\,h_\mu^{~\nu}\,\partial_\nu(-\mu/T)\,,
\label{app_rheology2_rev}\\ 
 \begin{pmatrix}
  N^{-1}\bigl[(\tr\varepsilon)^\cdot\bigr]_{\mathrm{rev}} \cr
   N^{-1}\bigl[\,\dot{\theta}\,\bigr]_{\mathrm{rev}}
 \end{pmatrix}
  &= -L_\tp^{-1}\,
   \begin{pmatrix} 
    L^\mathrm{s}_1 & L^\mathrm{s}_2 \cr 
    L^\mathrm{s}_2 & L^\mathrm{s}_3
   \end{pmatrix}
   \begin{pmatrix}
    \cK' \,\theta - (\cK/T)\,\tr K \cr -\cK' \tr\varepsilon + (\cK/T)\,\tr K
   \end{pmatrix}\,.
\label{app_rheology3_rev}
\end{align}

Combining Eqs.\ \eq{app_current_irr}--\eq{app_nu_d} 
and Eqs.\ \eq{app_current_rev}--\eq{app_nu_r}, 
and using the formulas 
$\dot\tp_\nu=\sqrt{-g}\,\nabla_\mu (u^\mu\,p_\nu)
=\sqrt{-g}\,\nabla^\mu(e \,u_\mu u_\nu)$
and
$\dot\tn=\sqrt{-g}\,\nabla^\mu (n\, u_\mu)$\,,
we obtain 
\begin{align}
 \sqrt{-g}\,\nabla^\mu (e\,u_\mu u_\nu)
  &=\dot{\tp}_\nu 
  =\bigl[\dot{\tp}_\nu\bigr]_{\mathrm{rev}}
  +\bigl[\dot{\tp}_\nu\bigr]_{\mathrm{irr}}
  =-\sqrt{-g}\,\nabla^\mu\bigl(\tau_{\mu\nu}^\mathrm{(r)}+\tau_{\mu\nu}^\mathrm{(d)}\bigr)\,,\\
 \sqrt{-g}\,\nabla^\mu (n\,u_\mu)
  &= \dot{\tn}
  =\bigl[\dot{\tn}\bigr]_{\mathrm{rev}}
  +\bigl[\dot{\tn}\bigr]_{\mathrm{irr}}
  =-\sqrt{-g}\,\nabla^\mu\bigl(\nu_{\mu}^\mathrm{(r)}+\nu_{\mu}^\mathrm{(d)}\bigr)\,.
\end{align}
We thus find that \eq{app_tp1_irr}--\eq{app_tn_irr} 
[or Eqs.\ \eq{app_current_irr}--\eq{app_nu_d}] 
and Eqs.\ \eq{app_current_rev}--\eq{app_nu_r}
can be summarized as current conservations:
\begin{align}
  \nabla_\mu T^{\mu\nu}=0\,,\qquad \nabla_\mu n^\mu =0\,,
\label{current_conservation_viscoelastic}
\end{align}
where each of the conserved currents, 
\begin{align}
 T^{\mu\nu} \equiv e\,u^\mu u^\nu + \tau^{\mu\nu} \,,\qquad
 n^\mu \equiv n\,u^\mu + \nu^\mu \,,
\end{align}
consists of the convective current
($p^\nu u^\mu=e\,u^\mu u^\nu$ or $n u^\mu$) 
and the additional current ($\tau^{\mu\nu}$ or $\nu^\mu$), 
the latter being further decomposed 
into the reversible and the dissipative currents:
\begin{align}
 \tau^{\mu\nu}\equiv\tau^{\mu\nu}_\mathrm{(r)}+\tau^{\mu\nu}_\mathrm{(d)}\,,\qquad
  \nu^{\mu}\equiv\nu^{\mu}_\mathrm{(r)}+\nu^{\mu}_\mathrm{(d)}\,.
\end{align}

Furthermore, one can easily show that
the evolution equations on $E_{\mu\nu}=(\varepsilon_{\mu\nu}\,,\,
\varepsilon_\mu\,,\,\theta)$ 
[Eqs.\ \eq{app_rheology1_irr}--\eq{app_rheology4_irr}]
together with the explicit form 
of the reversible and the dissipative currents 
[Eqs.\ \eq{app_tau_d}, \eq{app_nu_d}, \eq{app_tau_r}, and \eq{app_nu_r}]
can be rewritten into the following set of equations:
\begin{align}
 \begin{pmatrix} 
     -(2\lambda_1/T)\,\Lie_u\varepsilon_{\langle\mu\nu\rangle} \cr
    \tau_{\langle\mu\nu\rangle}-\tau^{(q)}_{\langle\mu\nu\rangle}
     \end{pmatrix} 
  &= 2\,({\boldsymbol\cG}+{\boldsymbol\eta})
     \begin{pmatrix} 
      \varepsilon_{\langle\mu\nu\rangle} \cr
     -(1/T)\,K_{\langle\mu\nu\rangle} 
     \end{pmatrix} 
\label{pre_rheology1}
  \,, \\
 \begin{pmatrix} 
   -(\lambda_2/T)\,\Lie_u\varepsilon_\mu \cr
         \nu_\mu
  \end{pmatrix} 
 &=  ({\boldsymbol\cH}+{\boldsymbol\sigma})
     \begin{pmatrix} 
      \varepsilon_\mu \cr
      h_\mu^{~\nu} \partial_\nu(-\mu/T) 
     \end{pmatrix} 
\label{pre_rheology2}
\,,\\
 \begin{pmatrix} 
   -(1/T)\, {\boldsymbol\gamma}
   \begin{pmatrix} 
   \Lie_u(\tr\varepsilon) \cr 
   \Lie_u\theta 
   \end{pmatrix}
   \cr
   (1/D)\,(\tr\tau-\tr\tau_\mathrm{(q)})
 \end{pmatrix} 
  &= ({\boldsymbol\cK}+{\boldsymbol\zeta})
     \begin{pmatrix} 
       \tr \varepsilon \cr \theta \cr (-1/T)\,\tr K 
\label{pre_rheology3}
 \end{pmatrix} \,,
\end{align}
where
\begin{align}
 \lambda_1&\equiv \frac{L_{\tp}\,T}{2L^\mathrm{t}} \,,\qquad  
  \lambda_2\equiv \frac{L_{\tn}\,T}{L^\mathrm{v}}\,,\qquad
 {\boldsymbol\gamma}\equiv \begin{pmatrix} 
 \gamma_1 & \gamma_2 \cr 
 \gamma_2 & \gamma_3 
 \end{pmatrix}
 \equiv 
 \Biggl[\frac{1}{L_{\tp}\,T}\,
 \begin{pmatrix} 
   L^\mathrm{s}_1 & L^\mathrm{s}_2 \cr 
   L^\mathrm{s}_2 & L^\mathrm{s}_3 
\end{pmatrix}\Biggr]^{-1} \,,
\label{pre_rheology4}\\
{\boldsymbol\eta}
 &\equiv L_{\tp}\,
   \begin{pmatrix} 
   \ell^\mathrm{t}_1 & \ell^\mathrm{t}_2 \cr 
   \ell^\mathrm{t}_2 & \ell^\mathrm{t}_3 
   \end{pmatrix}\,,\quad 
{\boldsymbol\sigma}
 \equiv L_{\tn}\,
   \begin{pmatrix} 
   \ell^\mathrm{v}_1 & \ell^\mathrm{v}_2 \cr 
   \ell^\mathrm{v}_2 & \ell^\mathrm{v}_3 
   \end{pmatrix}\,,\quad 
{\boldsymbol\zeta} 
 \equiv L_{\tp}\,
   \begin{pmatrix} 
   \ell^\mathrm{s}_1 & \ell^\mathrm{s}_2 & \ell^\mathrm{s}_4\cr 
   \ell^\mathrm{s}_2 & \ell^\mathrm{s}_3 & \ell^\mathrm{s}_5\cr
   \ell^\mathrm{s}_4 & \ell^\mathrm{s}_5 & \ell^\mathrm{s}_6\cr
   \end{pmatrix}\,,
\label{pre_rheology5}\\
{\boldsymbol\cG}&\equiv
   \begin{pmatrix} 
   0 & \cG \cr 
   -\cG & 0 
   \end{pmatrix}\,,\quad 
{\boldsymbol\cH}\equiv
   \begin{pmatrix} 
   0 & \cH \cr 
   -\cH & 0 
   \end{pmatrix}\,,\quad 
{\boldsymbol\cK}\equiv
   \begin{pmatrix} 
   0 & \cK' & \cK   \cr 
   -\cK' & 0 & -\cK a \cr
   -\cK &  \cK a & 0\cr
   \end{pmatrix}\,.
\label{pre_rheology6}
\end{align}
Equations \eq{pre_rheology1}--\eq{pre_rheology6} 
totally agree with Eqs.\ \eq{2nd_law_1}--\eq{2nd_law_5}, 
% with $\cG=\lambda_1$\, $\cH=0$\,, 
% $\cK=\gamma_3/\det{\boldsymbol\gamma}$ and $a=\gamma_2/\gamma_3$\,,
from which Eqs.\ \eq{conservation1}--\eq{rheology3} follow, 
as we see in Sec.\ \ref{Fundamental_equations}.
This is what we promised to show at the beginning of this Appendix.

We close this appendix with a comment on 
how the reversible evolutions are determined. 
They are actually determined by the requirement 
that the reversible evolutions do not produce entropy 
and the final form of the total evolutions (reversible ones plus irreversible ones) 
should be given as in Eqs.\ \eq{pre_rheology1}--\eq{pre_rheology6}.
As an example, let us consider 
the irreversible evolution of $\varepsilon_{\langle\mu\nu\rangle}$ 
and the quantity $\tau_{\langle\mu\nu\rangle}-\tau^\mathrm{(q)}_{\langle\mu\nu\rangle}$\,: 
\begin{align}
 N^{-1}\,\bigl[\dot\varepsilon_{\langle\mu\nu\rangle}\bigr]_\mathrm{irr}
  &= -2 L^\mathrm{t}\,\ell^\mathrm{t}_1\,\varepsilon_{\langle\mu\nu\rangle}
  + \frac{2L^\mathrm{t}\,\ell^\mathrm{t}_2}{T}\,K_{\langle\mu\nu\rangle}\,,
\label{incomp1}\\
 \tau_{\langle\mu\nu\rangle}-\tau^\mathrm{(q)}_{\langle\mu\nu\rangle}
  &=-2\cG\,\varepsilon_{\langle\mu\nu\rangle}
  +2L_\tp\,\ell^\mathrm{t}_2\,\varepsilon_{\langle\mu\nu\rangle}
  -\frac{2L_\tp\,\ell^\mathrm{t}_3}{T}\,K_{\langle\mu\nu\rangle}\,.
\label{incomp2}
\end{align}
By multiplying the first equation by a factor $-L_\tp/L^\mathrm{t}$\,,
the equations can be rewritten with a symmetric matrix as
\begin{align}
 \begin{pmatrix}
  -(L_\tp/L^\mathrm{t})\,N^{-1}\,\bigl[
  \dot\varepsilon_{\langle\mu\nu\rangle}\bigr]_\mathrm{irr}
  \cr \tau_{\langle\mu\nu\rangle}-\tau^\mathrm{(q)}_{\langle\mu\nu\rangle}
 \end{pmatrix}
 =\left[
  \begin{pmatrix} 0 & 0 \cr -2\cG & 0 \end{pmatrix} 
  +2 L_\tp
  \begin{pmatrix} \ell^\mathrm{t}_1 & \ell^\mathrm{t}_2 \cr  
   \ell^\mathrm{t}_2 & \ell^\mathrm{t}_3 
  \end{pmatrix} 
  \right]
  \begin{pmatrix} \varepsilon_{\langle\mu\nu\rangle} \cr 
   -(1/T)\,K_{\langle\mu\nu\rangle}
  \end{pmatrix}\,.
\end{align}
The second term with a symmetric positive-semidefinite matrix 
represents irreversible processes with entropy production. 
Thus, in order for the first term \emph{not} to produce entropy, 
we need to introduce the reversible part 
in $\dot\varepsilon_{\langle\mu\nu\rangle}$ 
such that the resulting form can be written with an antisymmetric matrix. 
This consideration determines the reversible evolution uniquely as 
\begin{align}
 \begin{pmatrix}
  -(L_\tp/L^\mathrm{t})\,N^{-1}\,
  \bigl[\dot\varepsilon_{\langle\mu\nu\rangle}\bigr]_\mathrm{rev}
  \cr 0
 \end{pmatrix}
 = \begin{pmatrix} 0 & 2\cG \cr 0 & 0 \end{pmatrix} 
   \begin{pmatrix} \varepsilon_{\langle\mu\nu\rangle} \cr 
   -(1/T)\,K_{\langle\mu\nu\rangle}
   \end{pmatrix}\,.
\label{incomp3}
\end{align}
Noting that $N^{-1}\,\dot\varepsilon_{\langle\mu\nu\rangle}
=\Lie_u\varepsilon_{\langle\mu\nu\rangle}$\,,
we see that the total evolution is actually given as in \eq{pre_rheology1}.
The remaining equations can be obtained in a similar way.

%%%%%%%%%%%%%%%%%%%%%%%%%%%%%%%%%%%%%%%%%%%%%%%%%%%%%% 
\section{Constitutive equations in rheological models}
%%%%%%%%%%%%%%%%%%%%%%%%%%%%%%%%%%%%%%%%%%%%%%%%%%%%%% 
\label{constitutive_eq}

The theory of elasticity is based on Hooke's law 
which states that that stresses are proportional to strains in elastic materials. 
On the other hand, the theory of viscous fluids 
is based on Newton's law
which states that viscous stresses are proportional to velocity gradients in fluids,
and is described by the Navier-Stokes equations. 
However, for more general materials 
these theories are not applicable, 
and a class of such materials is called \emph{viscoelastic materials} 
and studied in the area of rheology.
The relation between stresses and strains for a given material
is called the \emph{constitutive equations}, 
which play a fundamental role in the study of rheology.
In this Appendix, we list a few well-known materials 
with their constitutive equations 
and compare them with the viscoelastic materials 
discussed in the bulk of the present paper.

%%%%%%%%%%%%%%%%%%%%%%%%%%%%%%%%%%%%%%%%%%%%%%%%%%%%%% 
\subsection*{Hookean materials}
%%%%%%%%%%%%%%%%%%%%%%%%%%%%%%%%%%%%%%%%%%%%%%%%%%%%%% 

The simplest constitutive equations constitute Hooke's law.
We first assume that, on each timeslice $\Sigma_t$\,, 
every material particle knows its own natural shape 
described by the reference metric $\barh_{\mu\nu}$\,, 
which measures distances in a material when it is free of elastic strains. 
This metric has the same meaning as the intrinsic metric in the main text, 
though it is not dynamical here ($\Lie_u\barh_{\mu\nu}=0$)\,.
When we discuss nonrelativistic dynamics, 
we will set it to be $\barh_{\mu\nu}=\mathrm{diag} (0,1,\dotsc,1)$ 
in a laboratory frame, 
as is taken in standard textbooks (e.g., \cite{LL_elasticity}).
Although we consider the strain tensor $E_{\mu\nu}$ in the main text, 
we here assume that elastic strains are purely spatial, 
and only consider the elastic strain tensor defined by 
$\varepsilon_{\mu\nu}\equiv (1/2)\,(h_{\mu\nu}-\barh_{\mu\nu})$\,.

Hooke's law can then be expressed as
\begin{align}
 \tau^{\mu\nu} = -\cK^{(\mu\nu)(\rho\sigma)}\, \varepsilon_{\rho\sigma}\qquad 
  (\cK^{(\mu\nu)(\rho\sigma)}\geq 0)\,,
\end{align}
where $\cK^{(\mu\nu)(\rho\sigma)}$ is a constant tensor, 
and $(~)$ denotes the symmetrization of indices 
with the normalization $((~))=(~)$\,.
For isotropic elastic materials which locally has no preferred direction, 
the coefficient $\cK^{(\mu\nu)(\rho\sigma)}$ 
can be expressed as the sum of the irreducible components  
$h^{\mu\nu}\, h^{\rho\sigma}$ and 
$(1/2)\, \bigl(h^{\mu\rho}\, h^{\nu\sigma}+ h^{\mu\sigma}\, h^{\nu\rho}
-(2/D)\, h^{\mu\nu}\, h^{\rho\sigma}\bigr)$\,, 
and we have
\begin{align}
 \tau_{\mu\nu} = -\cK\, (\tr \varepsilon)\, h_{\mu\nu}
  - 2\cG\, \varepsilon_{\langle\mu\nu\rangle}\,,
\end{align}
where $\cK$ and $\cG$ are the bulk and the shear modulus, respectively.
Relativistic motions of such elastic materials 
in gravitational fields are discussed in, e.g., \cite{Carter:1977qf}.

Since we are considering the linear approximation in $\varepsilon_{\mu\nu}$\,,
this stress tensor can also be written as
\begin{align}
 \tau_{\mu\nu} = -\cK\, (\bartr \varepsilon)\, \barh_{\mu\nu}
  - 2\cG\, \varepsilon_{\overline{\langle\mu\nu\rangle}}\,,
\label{Hooke_stress}
\end{align}
where $\bartr$ and $\varepsilon_{\overline{\langle\mu\nu\rangle}}$ 
are defined by
\begin{align}
 \bartr \varepsilon = \varepsilon_{\mu\nu}\,\barh^{\mu\nu}\,,\qquad 
 \varepsilon_{\overline{\langle\mu\nu\rangle}}= \varepsilon_{\mu\nu}
 - \frac{1}{D}\, (\bartr \varepsilon)\, \barh_{\mu\nu}\,.
\end{align}
Then, if we take the nonrelativistic approximation 
with $\barh_{\mu\nu}=\mathrm{diag} (0,1,\dotsc,1)$\,, 
we reproduce the standard Hookean stress tensor
\begin{align}
 \tau_{ij} = -\cK\, \bigl(\delta^{kl}\, \varepsilon_{kl}\bigr)\, \delta_{ij}
  - 2\cG\, \Bigl(\varepsilon_{ij}
                 - \frac{1}{D}\, \bigl(
  \delta^{kl}\, \varepsilon_{kl}\bigr) \delta_{ij}\Bigr)\,.
\end{align}

\begin{figure}[htbp]
\begin{quote}
\begin{center}
\includegraphics[scale=0.5]{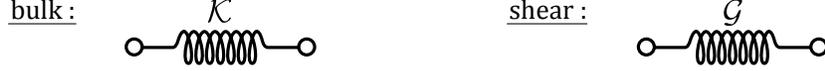}
\caption{The bulk part (left) and shear part (right) for a Hookean material.
\label{Hooke}}
\end{center}
\end{quote}
\vspace{-2ex}
\end{figure}

\begin{figure}[htbp]
\begin{quote}
\begin{center}
\includegraphics[scale=0.5]{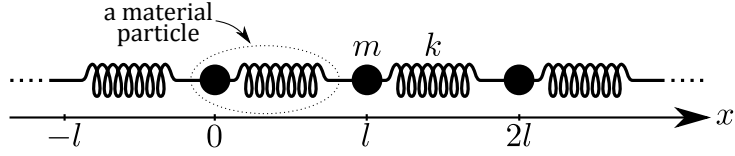}
\caption{Weights of mass $m$ are connected to the spring with spring constant $k$\,.
\label{Hooke_multi}}
\end{center}
\end{quote}
\vspace{-2ex}
\end{figure}

The constitutive equations for a Hookean material 
are schematically represented by a spring, as depicted in Fig.\ \ref{Hooke}.
%%%% 
%%%% 
To understand the diagram, 
we consider a Hookean material in $D=1$ spatial dimension. 
The material can be obtained 
by connecting in series tiny springs with a weight of mass $m$ 
at each end (see Fig.\ \ref{Hooke_multi}). 
%%%% 
%%%% 
Since the actual length between two adjacent weights 
at $x=x_n\equiv n\,l$ ($n\in \mathbb{Z}$) and $x=x_{n+1}$ is 
given by $\sqrt{h_{11}(x_{n})}\,l$\,, 
and since the natural length is given by
$\sqrt{\barh_{11}(x_{n})}\,l$\,, 
the stretch of the spring is given by
$(\sqrt{h_{11}}-\sqrt{\barh_{11}})\,l \simeq 
(\varepsilon_{11}/\sqrt{\barh_{11}})\,(x_{n})$\,,
where $\varepsilon_{11}=(1/2)\,(h_{11}-\bar{h}_{11})$.
Then the equation of motion for the weight at $x=x_n$ can be written as
\begin{align}
 m\,a_1(x_n)  = - k\,l\,\frac{\varepsilon_{11}}{\sqrt{\barh_{11}}}(x_{n-1}) 
                + k\,l\,\frac{\varepsilon_{11}}{\sqrt{\barh_{11}}}(x_{n}) \,,
\end{align}
where $a_1(x_n)$ is the acceleration of the weight at $x=x_n$ 
in the $x^1$-direction.
Then in the continuum limit $l\to 0$ 
with $e_0\equiv m/\sqrt{\barh_{11}}\,l$ and 
$\cK\equiv k\,l$ kept fixed at finite values, 
the equation becomes
\begin{align}
  e_0\,a_1(x)
  = -\partial_1 \bigl[-\cK\, \bigl(\bartr\varepsilon(x)\bigl)\,\delta^1_{~1}\bigr]\,,
\end{align}
so long as we take a coordinate system in which the intrinsic metric 
$\barh_{11}$ is spatially constant. 
If we define the energy density $e(x)\equiv m/\sqrt{h_{11}}\,l$ 
and neglect the difference $(e(x)-e_0)\,a_1(x)$\,, 
which is of higher orders in $\varepsilon_{11}$\,, 
we obtain the Euler equation
\begin{align}
  e(x)\,a_1(x) = -\partial_1 \tau^1_{~1}
\end{align}
with the stress tensor $\tau^1_{~1} = -\cK\, (\bartr\varepsilon)\,\delta^1_{~1}$\,.
This stress tensor coincides with \eq{Hooke_stress} in $D=1$ dimension,
and in this sense the left diagram in Fig.\ \ref{Hooke} represents 
(the bulk part of) the constitutive equations for a Hookean material. 
On the other hand, the right diagram in Fig.\ \ref{Hooke} 
is simply a schematic generalization for the shear part 
and does not have any physical meaning other than the information 
that the shear part of the stress tensor is given by
$-2\cG\,\varepsilon_{\langle\mu\nu\rangle}$\,.

%%%%%%%%%%%%%%%%%%%%%%%%%%%%%%%%%%%%%%%%%%%%%%%%%%%%%% 
\subsection*{Navier-Stokes (Newtonian) fluids}
%%%%%%%%%%%%%%%%%%%%%%%%%%%%%%%%%%%%%%%%%%%%%%%%%%%%%% 

\begin{figure}[htbp]
\begin{quote}
\begin{center}
\includegraphics[scale=0.5]{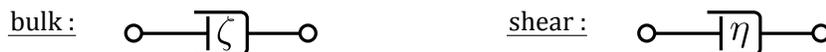}
\caption{The bulk part (left) and the shear part (right) 
for a Navier-Stokes (or Newtonian) fluid.
A dashpot yields a viscous stress 
proportional to the time derivative of the induced metric.\label{dashpot}}
\end{center}
\end{quote}
\vspace{-2ex}
\end{figure}

Newton's viscosity law says that 
the viscous stress tensor is proportional to velocity gradients, 
and in our notations this can be written as
\begin{align}
 \tau_\mathrm{(d)}^{\mu\nu} = -\zeta^{(\mu\nu)(\rho\sigma)} K_{\rho\sigma}\qquad 
  (\zeta^{(\mu\nu)(\rho\sigma)}\geq 0)\,,
\end{align}
because the extrinsic curvature $K_{\mu\nu}=(1/2)\,\Lie_u h_{\mu\nu}$ 
can also be expressed as velocity gradients, 
$K_{\mu\nu}=(1/2)\,h_\mu^{~\rho}\, h_\nu^{~\sigma}\,(\nabla_\rho u_\sigma
+\nabla_\sigma u_\rho )$\,.
In particular, for simple fluids (that do not have any specific directions locally) 
we have
\begin{align}
 \tau_\mathrm{(d)}^{\mu\nu} = - \zeta\, (\tr K)\, h^{\mu\nu}
  - 2\eta\, K^{\langle\mu\nu\rangle} \,,
\end{align}
where $\zeta\,(\geq 0)$ and $\eta\,(\geq 0)$ are 
the bulk and the shear viscosity, respectively.
These constitutive equations can be interpreted as representing 
the resistance due to the time derivative of the induced metric, 
$K_{\mu\nu}=(1/2)\,\Lie_u h_{\mu\nu}$\,, 
and are schematically represented by a dashpot as in Fig.\ \ref{dashpot}.
%%%% 
%%%% 

For simple fluids,
the reversible part of 
the stress tensor, $\tau_\mathrm{(r)}^{\mu\nu}$\,,
should be proportional to $h^{\mu\nu}$ by definition, 
and we write it as $\tau_\mathrm{(r)}^{\mu\nu}=P\,h^{\mu\nu}$\,.
Then the total stress tensor for simple viscous fluids is given by
\begin{align}
 \tau^{\mu\nu}=\tau_\mathrm{(r)}^{\mu\nu}
  +\tau_\mathrm{(d)}^{\mu\nu}
 =P\,h^{\mu\nu}
  - \zeta\, (\tr K)\, h^{\mu\nu} - 2\eta\, K^{\langle\mu\nu\rangle}\,.
\end{align}
Materials with the constitutive equations of this form 
are called Navier-Stokes (or Newtonian) fluids.

%%%%%%%%%%%%%%%%%%%%%%%%%%%%%%%%%%%%%%%%%%%%%%%%%%%%%% 
\subsection*{Kelvin-Voigt materials}
%%%%%%%%%%%%%%%%%%%%%%%%%%%%%%%%%%%%%%%%%%%%%%%%%%%%%% 

\begin{figure}[htbp]
\begin{quote}
\begin{center}
\includegraphics[scale=0.5]{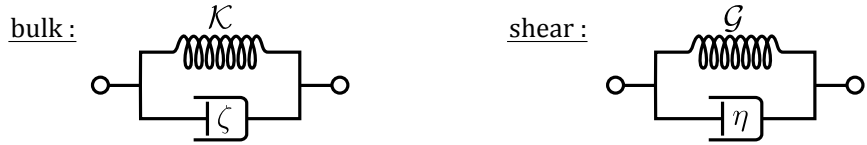}
\caption{The bulk part (left) and shear part (right) for a Kelvin-Voigt material.
\label{Kelvin-Voigt}}
\end{center}
\end{quote}
\vspace{-2ex}
\end{figure}

If an elastic material (so that $\Lie_u \bar{h}_{\mu\nu}=0$) 
further obeys Newton's viscosity law, 
the stress tensor is given in the following form:
\begin{align}
 \tau_{\mu\nu}
  = -\bigl(\cK\, \tr \varepsilon + \zeta\, \tr K\bigr)\, h_{\mu\nu} 
     - 2\cG\, \varepsilon_{\langle\mu\nu\rangle} - 2\eta\, K_{\langle\mu\nu\rangle} \,.
\end{align}
Such materials are called Kelvin-Voigt materials 
and are sometimes used to explain creep phenomena in viscoelastic materials. 
Relativistic motions of such materials are 
discussed in, e.g., \cite{Kranys:1977}.
Since Kelvin-Voigt materials have fixed intrinsic metric 
($0\equiv\Lie_u\barh_{\mu\nu}=K_{\mu\nu}-\Lie_u\varepsilon_{\mu\nu}$), 
we have $K_{\mu\nu}=\Lie_u \varepsilon_{\mu\nu}$ 
and the stress tensor can be rewritten in the following form:
\begin{align}
 \tau_{\mu\nu}
  = -\bigl(\cK\,  \tr \varepsilon +  \zeta\, \Lie_u\tr\varepsilon \bigr)\, h_{\mu\nu}
  - 2\cG\, \varepsilon_{\langle\mu\nu\rangle}
  - 2\eta\, \Lie_u\varepsilon_{\langle\mu\nu\rangle} \,.
\end{align}
The constitutive equations for a Kelvin-Voigt material thus can be represented 
by the diagrams in Fig.\ \ref{Kelvin-Voigt}. 
Since a spring and a dashpot are connected in parallel in each diagram, 
the total stress is given as the sum of the stress of each component. 
%%%% 
%%%% 

Unlike Hookean materials, the stress-strain relation is process-dependent.
However, for Kelvin-Voigt materials the stress tensor at each moment 
can be determined only by measuring the induced metric $h_{\mu\nu}$ and 
its temporal derivative $K_{\mu\nu}$ at the moment, 
and we do not need to know the preceding history of the strains.

For more generic materials, the stress tensor indeed depends 
on the whole preceding history of the strains. 
The simplest among such materials are Maxwell materials, described below.

%%%%%%%%%%%%%%%%%%%%%%%%%%%%%%%%%%%%%%%%%%%%%%%%%%%%%% 
\subsection*{Maxwell materials}
%%%%%%%%%%%%%%%%%%%%%%%%%%%%%%%%%%%%%%%%%%%%%%%%%%%%%% 

\begin{figure}[htbp]
\begin{quote}
\begin{center}
\includegraphics[scale=0.5]{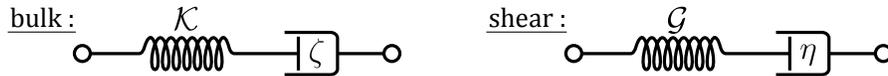}
\caption{The bulk part (left) and the shear part (right) for a Maxwell material.
\label{Maxwell}}
\end{center}
\end{quote}
\vspace{-2ex}
\end{figure}

The constitutive equations for a Maxwell material 
are depicted in Fig.\ \ref{Maxwell}.
%%%% 
%%%% 
Since a spring and a dashpot are connected in series, 
the stress of the spring and the stress of the dashpot should be equal. 
As is already explained, the stress of the spring is given by
\begin{align}
 \tau_{\mu\nu}= 
 -\cK\, (\tr \varepsilon)\, h_{\mu\nu} - 2\cG\, \varepsilon_{\langle\mu\nu\rangle} \,.
\label{Maxwell_model_tau_1}
\end{align}
Recall that the induced metric $h_{\mu\nu}$ measures the actual shape of 
each material particle (i.e., the total length of the diagram),
while the intrinsic metric $\barh_{\mu\nu}$ measures 
the natural shape of each material particle 
(i.e., the length of the dashpot plus the natural length of the spring).
Thus, the stress of the dashpot, 
which is proportional to the temporal derivative of $\barh_{\mu}$ 
(i.e., the temporal derivative of the length of the dashpot), 
is given by
\begin{align}
 \tau_{\mu\nu}
  = -\zeta\, (\tr \barK)\, h_{\mu\nu} - 2\eta\, \barK_{\langle\mu\nu\rangle}\,.
\label{Maxwell_model_tau_2}
\end{align}
Since these stresses are equal, from Eqs.\ \eq{Maxwell_model_tau_1} and 
\eq{Maxwell_model_tau_2}, we obtain the equations
\begin{align}
 \tr \barK = (\cK/\zeta)\, \tr \varepsilon \,,\qquad 
 \barK_{\langle\mu\nu\rangle} = (\cG/\eta)\, \varepsilon_{\langle\mu\nu\rangle} \,.
\end{align}
Since $\barK_{\mu\nu}$ is the temporal derivative of $\barh_{\mu\nu}$\,, 
$\barK_{\mu\nu}=(1/2)\,\Lie_u \barh_{\mu\nu}$\,, 
these equations describe the dynamics of $\barh_{\mu\nu}$ 
and are called the rheology equations in the main text.
Note that the structure of a Maxwell material is critically different from 
that of a Kelvin-Voigt material 
in that the intrinsic metric $\barh_{\mu\nu}$ of the former is dynamical.

We should also emphasize that 
even if we measure the shape of a viscoelastic material, $h_{\mu\nu}$\,, 
and its derivative $K_{\mu\nu}$ at a given moment, 
we cannot readily determine the value of the stress tensor $\tau_{\mu\nu}$ 
at the moment  
because there is no way to know the values of strains $\varepsilon_{\mu\nu}$ 
when $\barh_{\mu\nu}$ is dynamical.
However, if we observe the evolution of $h_{\mu\nu}$ 
during a finite interval of time, 
the initial value of $\barh_{\mu\nu}$ can be obtained, 
and by solving the rheology equations 
we can determine the value of the intrinsic metric $\barh_{\mu\nu}$ 
at each moment. 

%%%%%%%%%%%%%%%%%%%%%%%%%%%%%%%%%%%%%%%%%%%%%%%%%%%%%% 
\subsection*{Zener materials}
%%%%%%%%%%%%%%%%%%%%%%%%%%%%%%%%%%%%%%%%%%%%%%%%%%%%%% 

\begin{figure}[htbp]
\begin{quote}
\begin{center}
\includegraphics[scale=0.5]{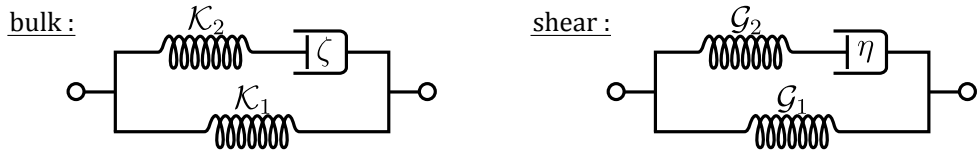}
\caption{The bulk part (left) and the shear part (right) for a Zener material.
\label{SLS}}
\end{center}
\end{quote}
\vspace{-2ex}
\end{figure}

We next consider Zener materials or the standard linear solid model 
whose constitutive equations are given by the diagrams in Fig.\ \ref{SLS}.
%%%% 
%%%% 
This model includes Kelvin-Voigt materials and Maxwell materials 
as limiting cases ($\cK_2=\cG_2=0$ and $\cK_1=\cG_1=0$\,, respectively). 
However, as is clear from Fig.\ \ref{SLS}, 
if a Zener material is left intact after an initial deformation, 
it will get back to its original natural shape.
In other words, this kind of material does not posses permanent strains 
unlike Maxwell materials, 
and in this sense Zener materials are said to be solid-like.
%%\footnote{
%%The Poynting-Thomson materials, which correspond to a diagram a spring 
%%and the Kelvin-Voigt material connected in series, also includes the Kelvin-Voigt model 
%%and the Maxwell model as limiting cases but they are also solid-like materials.}.

If we want to describe the relativistic dynamics of a Zener material 
using our theory of viscoelasticity, 
we need to extend the framework, 
introducing another nondynamical intrinsic metric $\barh^{(2)}_{\mu\nu}$ 
in addition to the original dynamical intrinsic metric $\barh_{\mu\nu}$\,.
Here $\barh^{(2)}_{\mu\nu}$ measures the natural length 
of the lower spring in Fig.\ \ref{SLS}, 
while $\barh_{\mu\nu}$ measures the length of the dashpot 
plus the natural length of the upper spring.

If we consider more generic materials, 
we accordingly should introduce more additional intrinsic metrics 
(dynamical or nondynamical). 
Such generalizations correspond to considering multielement models 
(such as the generalized Maxwell model) known in the study of rheology.
In this paper we only consider the cases with a single intrinsic metric, 
and such generalizations will be discussed elsewhere. 

%%%%%%%%%%%%%%%%%%%%%%%%%%%%%%%%%%%%%%%%%%%%%%%%%%%%%% 
\subsection*{Viscoelastic materials considered in this paper}
%%%%%%%%%%%%%%%%%%%%%%%%%%%%%%%%%%%%%%%%%%%%%%%%%%%%%% 

\begin{figure}[htbp]
\vspace{3ex}
\begin{center}
\includegraphics[scale=0.6]{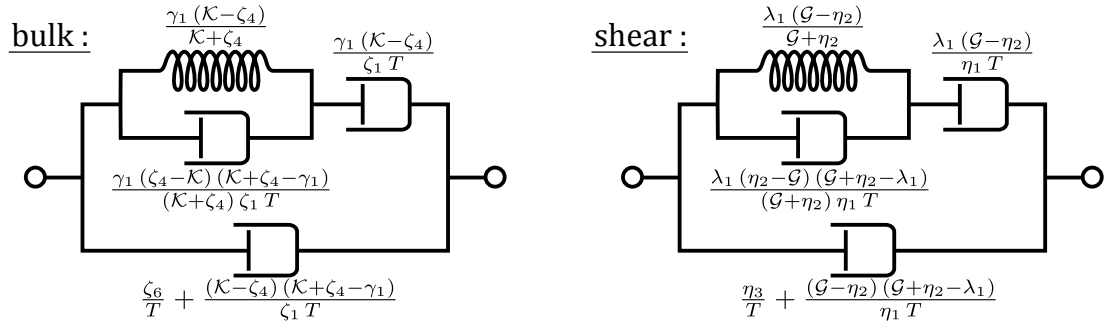}
\begin{quote}
\caption{Schematic structure of the bulk part (left) and the shear part (right). 
Note that the contribution of $\tau_{\mu\nu}^\mathrm{(q)}$ is omitted for simplicity.
\label{fig:bulk_shear}}
\end{quote}
\end{center}
\vspace{-6ex}
\end{figure}

\begin{figure}[htbp]
\begin{center}
\vspace*{0.5cm}
\includegraphics[scale=0.6]{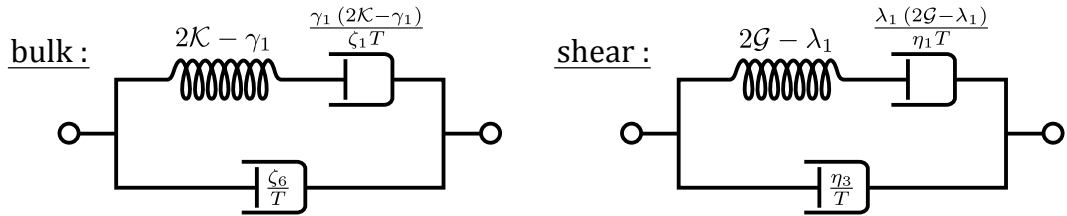}
\begin{quote}
\caption{A three-element model where a dashpot is connected 
in parallel with a Maxwell material.
\label{fig:AFKY}}
\end{quote}
\end{center}
\vspace{-7ex}
\end{figure}

As for the rheological model discussed in this paper, 
we here consider for brevity 
the case when the effects of thermal expansion can be neglected 
($\zeta_2=\zeta_5=a=\gamma_2=\cK'=0$). 
Then the stress tensor and the rheology equations are given by
\begin{align}
 \tau_{\mu\nu}
 =&~ \tau_{\mu\nu}^\mathrm{(q)}-2\,(\cG-\eta_2)\,\varepsilon_{\langle\mu\nu\rangle} 
      - (2\eta_3/T)\,K_{\langle\mu\nu\rangle} \nn\\
  &~ -\bigl[\bigl(\cK-\zeta_4)\, \tr \varepsilon
  + (\zeta_6/T)\,\tr K\bigr]\, h_{\mu\nu} \,,\\
 \Lie_u\varepsilon_{\langle\mu\nu\rangle}
  &= -\frac{\eta_1\,T}{\lambda_1}\, \varepsilon_{\langle\mu\nu\rangle}
     +\frac{\cG + \eta_2}{\lambda_1} \, K_{\langle\mu\nu\rangle} \,,\\
  \Lie_u(\tr\varepsilon) 
  &=   -\frac{\zeta_1\,T}{\gamma_1}\, \tr \varepsilon
  +\frac{\cK +\zeta_4}{\gamma_1}\,\tr K \,.
\end{align}
These equations can be summarized as in Fig.\ \ref{fig:bulk_shear}.
%%%% 
%%%% 
In particular, one can show that
Maxwell's original definition is realized 
if we set $\cK=\gamma_1-\zeta_4$ and $\cG=\lambda_1-\eta_2$ 
(see Sec.\ \ref{short_elastic}).
The corresponding diagrams are given in Fig.\ \ref{fig:AFKY}.
%%%% 
%%%% 
The Maxwell model can be obtained if we additionally set $\zeta_6=\eta_3=0$\,, 
which is the case where the simplified Israel-Stewart model 
is obtained, as shown in Sec.\ \ref{Israel_model}.

%%%%%%%%%%%%%%%%%%%%%%%%%%%%%%%%%%%%%%%%%%%%%%%%%%%%%% 
\section{Euler and Gibbs-Duhem relations}
%%%%%%%%%%%%%%%%%%%%%%%%%%%%%%%%%%%%%%%%%%%%%%%%%%%%%% 
\label{Euler-Gibbs-Duhem}

In this appendix we consider the case $\tau^{\mu\nu}_{\mathrm{(q)}}=P\,h^{\mu\nu}$\,.
Then the variation equation of entropy, Eq.\ \eq{delta_s_visc}, is given by
\begin{align}
 \delta \ts =&~ 
  - \frac{u^\nu}{T} \,\delta \tp_\nu
  - \frac{\mu}{T}\,\delta \tn
  + \frac{P}{T}\, \delta \sqrt{h}
  + \frac{\te}{2T}\, u^{\mu}u^{\nu} \,\delta g_{\mu\nu} \nn\\
  &~ - \frac{\sqrt{h}}{T} \, 2\lambda_1\,\varepsilon^{\langle\mu\nu\rangle}
                                 \,\delta \varepsilon_{\langle\mu\nu\rangle}
  - \frac{\sqrt{h}}{T}\,\bigl(
  \gamma_1\,\tr\varepsilon+\gamma_2\,\theta\bigr)\,\delta (\tr\varepsilon) 
  \nn\\
  &~ - \frac{\sqrt{h}}{T} \,\lambda_2\,\varepsilon^\mu\,\delta\varepsilon_\mu
  - \frac{\sqrt{h}}{T}\,\bigl(
 \gamma_3\,\theta +\gamma_2\,\tr\varepsilon\bigr)\, \delta\theta \,.
\label{delta_s_visc_2}
\end{align}
Here, if we consider the variation $\delta = u^\mu \nabla_\mu$\,, we obtain
\begin{align}
 \sqrt{h}\,u^\mu\, \partial_\mu s =&~ 
    \frac{\sqrt{h}}{T} \,u^\mu (\partial_\mu e - \mu\,\partial n) \nn\\
  &~ - \frac{\sqrt{h}}{T} \, 2\lambda_1\,\varepsilon^{\langle\mu\nu\rangle}
                                 \,u^\rho \nabla_\rho \varepsilon_{\langle\mu\nu\rangle}
  - \frac{\sqrt{h}}{T}\,\bigl(\gamma_1\,\tr\varepsilon+\gamma_2\,\theta\bigr)\,
             u^\mu \partial_\mu (\tr\varepsilon) \nn\\
  &~ - \frac{\sqrt{h}}{T} \,\lambda_2\,\varepsilon^\mu\,u^\rho \nabla_\rho \varepsilon_\mu
  - \frac{\sqrt{h}}{T}\,\bigl(\gamma_3\,\theta +\gamma_2\,\tr\varepsilon\bigr)\, 
             u^\mu \partial_\mu\theta \,.
\label{delta_1}
\end{align}
On the other hand, if we consider the variation $\delta = \Lie_u$\,, we obtain
\begin{align}
 \sqrt{h}\, \nabla_\mu (su^\mu) =&~ 
  - \frac{\sqrt{h}\,u^\nu}{T} \,\nabla_\mu (p_\nu u^\mu)
  - \frac{\sqrt{h}\,\mu}{T}\, \nabla_\mu (nu^\mu)
  + \frac{\sqrt{h}\,P}{T}\, \nabla_\mu u^\mu \nn\\
  &~ - \frac{\sqrt{h}}{T} \, 2\lambda_1\,\varepsilon^{\langle\mu\nu\rangle}
                                 \,\Lie_u \varepsilon_{\langle\mu\nu\rangle}
  - \frac{\sqrt{h}}{T}\,\bigl(\gamma_1\,\tr\varepsilon+\gamma_2\,\theta\bigr)\,
             u^\mu \partial_\mu (\tr\varepsilon) \nn\\
  &~ - \frac{\sqrt{h}}{T} \,\lambda_2\,\varepsilon^\mu\,\Lie_u \varepsilon_\mu
  - \frac{\sqrt{h}}{T}\,\bigl(\gamma_3\,\theta +\gamma_2\,\tr\varepsilon\bigr)\, 
             u^\mu \partial_\mu\theta \,.
\label{delta_2}
\end{align}
Subtracting Eq.\ \eq{delta_1} from Eq.\ \eq{delta_2},
we obtain the following equation:
\begin{align}
 \ts \, \tr K  =&~  \frac{\te-\mu \ts+\sqrt{h}\,P}{T}\, \tr K \nn\\
  &~ - \frac{\sqrt{h}}{T} \, 2\lambda_1\,\bigl(\tr(\varepsilon^2K)
  -(4/D)\,(\tr\varepsilon)\,\tr(\varepsilon K)
  +(4/D^2)\, (\tr K)\,(\tr\varepsilon)^2\bigr) \nn\\
 &~ - \frac{\sqrt{h}}{T} \,\lambda_2\,\varepsilon^\mu\, \varepsilon^\nu K_{\mu\nu} \,.
\end{align}
We can neglect the terms in the second and third lines 
because they are of higher orders,
and thus we obtain the equation
\begin{align}
  (e+P-Ts-\mu\, n) \tr K = 0\,.
\end{align}
Since this should hold for any processes in our linear approximations, 
the following relation must hold:
\begin{align}
 e+P-Ts-\mu\, n = 0\,.
\label{Euler}
\end{align}
This has the same form with  the standard Euler relation although 
the energy density $e$ and the entropy density $s$ here 
contain contributions from the strain tensor 
$E_{\mu\nu}=(\varepsilon_{\mu\nu}\,,\,\varepsilon_\mu\,,\,\theta)$\,.
From this and Eq.\ \eq{delta_s_visc_2}, we can derive the Gibbs-Duhem-like equation:
\begin{align}
 & s\,\delta T+ n \,\delta \mu - \delta P \nn\\
 &= - 2\lambda_1\,\varepsilon^{\langle\mu\nu\rangle}
                                 \,\delta \varepsilon_{\langle\mu\nu\rangle}
  - \bigl(\gamma_1\,\tr\varepsilon+\gamma_2\,\theta\bigr)\,\delta (\tr\varepsilon) 
  - \lambda_2\,\varepsilon^\mu\,\delta\varepsilon_\mu
  - \bigl(\gamma_3\,\theta +\gamma_2\,\tr\varepsilon\bigr)\, \delta\theta \,.
\label{Gibbs-Duhem}
\end{align}
In the limit where the strains relax completely ($E_{\mu\nu}\to 0$), 
this reduces to the standard Gibbs-Duhem equation for simple fluids.

%%%%%%%%%%%%%%%%%%%%%%%%%%%%%%%%%%%%%%%
\baselineskip=0.85\normalbaselineskip
%%%%%%%%%%%%%%%%%%%%%%%%%%%%%%%%%%%%%%%

\end{document}